\documentclass[goettingen, gauss, print, additionalreferee]{thesis}
\title{Exploring the solar paradigm to explain stellar variability}
\author{Nina-Elisabeth N\`{e}mec}
\town{Amstetten, \"{O}sterreich}%insert your place of birth, the country only appearing, if it is not Germany, with the name of the country in German as in "Paris, Frankreich" 

\thesisadvisorycommitteea{Dr. Alexander I. Shapiro}
\instthesisadvisorycommitteea{Max-Planck-Institut f\"ur Sonnensystemforschung, G\"ottingen, Germany}
\thesisadvisorycommitteeb{Dr. Natalie A. Krivova}
\instthesisadvisorycommitteeb{Max-Planck-Institut f\"ur Sonnensystemforschung, G\"ottingen, Germany}

\thesisadvisorycommitteec{Prof. Dr. Stefan Dreizler}
\instthesisadvisorycommitteec{Institut f\"ur Astrophysik, Georg-August-Universit\"at G\"ottingen, G\"ottingen, Germany}

\refereea{Prof. Dr. Stefan Dreizler}
\instrefereea{Institut f\"ur Astrophysik, Georg-August-Universit\"at G\"ottingen, G\"ottingen, Germany}
\refereeb{Dr. Alexander I. Shapiro}
\instrefereeb{Max-Planck-Institut f\"ur Sonnensystemforschung, G\"ottingen, Germany}
\refereec{Prof. Dr. Gibor Basri}%Only appears with option %"additionalreferee"
\instrefereec{University of California, Berkeley, United States of America}

\commissiona{Prof. Dr. Sami K. Solanki}
\instcommissiona{Max-Planck-Institut f\"ur Sonnensystemforschung, G\"ottingen, Germany}
\commissionb{Prof. Dr. Ariane Frey}
\instcommissionb{II. Physikalisches Institut, Georg-August-Universit\"at G\"ottingen, G\"ottingen, Germany}
\commissionc{Prof. Dr. W. Kollatschny}
\instcommissionc{Institut f\"ur Astrophysik, Georg-August-Universit\"at G\"ottingen, G\"ottingen, Germany}
\commissiond{Prof. Dr. Hardi Peter}
\instcommissiond{Max-Planck-Institut f\"ur Sonnensystemforschung, G\"ottingen, Germany}

\submittedyear{2020}
\publicationyear{2021}
\isbn{}

\newcommand{\comment}[1]{}
\usepackage[german,english]{babel}
\usepackage[OT2, T1]{fontenc}
\usepackage{natbib}
\usepackage{printlen}
\usepackage{color}
\usepackage{mathtools}
\usepackage{gensymb}
\usepackage{subcaption}
\usepackage{textcomp}
\usepackage{sidecap}
\usepackage[bottom]{footmisc}

\newcommand{\rotrate}{$\Omega_{\odot}$}

\newcommand{\rvar}{$R_{var}$}

\defcitealias{Nina1}{N20b}
\defcitealias{Isik2020}{I20}
\defcitealias{Isik2018}{I18}
\defcitealias{Shapiro2014}{S14}

\begin{document}
\selectlanguage{english}
\maketitle

\tableofcontents
%\listofigures
%\listoftables

\chapter*{Abstract\markboth{Summary}{Summary}}
\addcontentsline{toc}{chapter}{Abstract}

The unprecedented precision of broadband stellar photometry achieved with the planet-hunting missions CoRoT and \textit{Kepler} initiated a new era in examining the magnetically-driven brightness variations of hundreds of thousands of stars. Such brightness variations are well studied and understood for the Sun. The plethora of data allows to accurately compare solar and stellar brightness variations. An intriguing question is whether the observed trends in the stellar photometric variability (e.g. the dependence of the variability on the stellar rotation period) can be explained by utilising the solar paradigm, in particular the physical concepts of brightness variations learnt from the Sun. The goal of this work is to find out, through comparison of observational and simulated data, if any physical concepts of solar brightness variability have to be altered to reproduce the distribution of Sun-like stars variabilities.

Comparisons between solar and stellar variability suffer from several observational biases. Stellar brightness variations are routinely measured in various spectral passbands and direct measurements of solar variability in these passbands do not exist.
Therefore, measurements of stellar variability are often compared to measurements of the Total Solar Irradiance variability (i.e. the spectrally integrated solar radiative flux at 1 AU from the Sun), introducing potential biases. Additionally, observations of solar variability are made from the equatorial plane, corresponding to a right angle between the Sun's rotation axis and the line-of-sight (the 7.25$^{\circ}$ tilt between the solar rotation axis and the ecliptic plane can be neglected). In this thesis I build a model based on a surface flux transport model (SFTM) and the Spectral And Total Irradiance REconstruction (SATIRE) approach to calculate  the effect of the inclination and different passbands on the solar variability on both the activity cycle (11 years) and the rotational (27 days) timescales. This  model is presented in Sect. \ref{sec:paper_1}. We quantify the rotational variability of the Sun as it would be observed by different space missions and the effect of the inclination in Sect. \ref{sec:paper_2}. In the next step we extend our model to stars that are more active than the Sun. This extension is based on the observation that the solar disk coverage by spots increases faster with the activity than that by faculae. Until now such a behaviour has not been explained. I demonstrate in Sect. \ref{sec:paper_3} that the cancellation of small magnetic flux concentrations, which are associated with faculae, is able to explain this behaviour. In Sect.~\ref{sec:paper_4} I present calculations of brightness variations for fast-rotating stars. I conclude that in order to model the observed dependence of the stellar variability on the rotation period, the degree of nesting (i.e. the tendency of active regions to emerge in the vicinity of previous emergences) of active regions should increase with decreasing rotation periods.

 \chapter*{Zusammenfassung\markboth{Summary}{Summary}}
\addcontentsline{toc}{chapter}{Zusammenfassung}

Die beispiellose Pr\"azision der Breitband-Sternfotometrie, die mit den Planetensuchmissionen CoRoT und \textit{Kepler} erreicht wurde, leitete eine neue \"{A}ra bei der Untersuchung der magnetisch bedingten Helligkeitsschwankungen von Hunderttausenden von Sternen ein. Solche Helligkeitsvariationen sind f\"ur die Sonne gut untersucht und verstanden. Die F\"{u}lle der Daten erm\"{o}glicht einen genauen Vergleich der solaren und stellaren Helligkeitsvariationen. Eine faszinierende Frage ist, ob die beobachteten Trends in der photometrischen Variabilit\"{a}t der Sterne (z.B. die Abh\"{a}ngigkeit der Variabilit\"{a}t von der Rotationsperiode) mit Hilfe des Sonnenparadigmas erkl\"art werden k\"{o}nnen, insbesondere der physikalischen Konzepte der Helligkeitsvariationen, die von der Sonne abgeleitet wurden. Das Ziel dieser Arbeit ist es, durch Vergleich von Beobachtungs- und Simulationsdaten herauszufinden, welche physikalischen Konzepte der solaren Helligkeitsvariabilit\"{a}t ge\"{a}ndert werden m\"{u}ssen, um die Verteilung der sonnen\"{a}hnlichen Sternvariabilit\"{a}ten zu reproduzieren.

Vergleiche zwischen Sonnen- und Sternvariabilit\"at leiden unter mehreren Beobachtungsverzerrungen. Stellare Helligkeitsschwankungen werden routinem\"a\ss ig in verschiedenen spektralen Filtersystemen gemessen, und direkte Messungen der Sonnenvariabilität in diesen Filtersystemen gibt es nicht.
Daher werden Messungen der stellaren Variabilit\"at oft mit Messungen der Variabilit\"at der gesamten Sonneneinstrahlung (d.h. der wellenl\"angenintegrierten solaren Strahlungsintensit\"at im Abstand von 1 AE von der Sonne) verglichen, was zu Verzerrungen f\"uhren kann. Zus\"atzlich werden Beobachtungen der solaren Variabilit\"at von der \"Aquatorebene aus gemacht, die einem rechten Winkel zwischen der Sonnenrotationsachse und der Sichtlinie entspricht (die Neigung von 7,25$^{\circ}$ zwischen der Sonnenrotationsachse und der Ekliptikebene kann vernachl\"assigt werden). In dieser Arbeit baue ich ein Modell auf, das auf einem Oberfl\"achenfluss-Transportmodell (SFTM) und dem Ansatz der Spektralen und totalen Strahlungsrekonstruktion (SATIRE) basiert, um die Auswirkung der Neigung und verschiedener Filtersysteme auf die solare Variabilit\"at sowohl auf den Aktivit\"atszyklus (11 Jahre) als auch auf die Rotationszeitskalen (27 Tage) zu berechnen. Dieses Modell wird in Sect. \ref{sec:paper_1} vorgestellt. Ich quantifiziere die Rotationsvariabilit\"at der Sonne, wie sie von verschiedenen Raumfahrtmissionen beobachtet werden w\"urden, und den Einfluss der Inklination in Sect. \ref{sec:paper_2}. Im n\"achsten Schritt erweitere ich das Modell auf Sterne, die aktiver sind als die Sonne. Diese Erweiterung beruht auf der Beobachtung, dass die Bedeckung der Sonnenscheibe durch Flecken mit der Aktivit\"at schneller zunimmt als die durch Fackeln. Ein solches Verhalten ist bisher nicht erkl\"art worden. Ich demonstriere in Sect. \ref{sec:paper_3}, dass die gegenseitige Ausl\"oschung kleiner magnetischer Flusskonzentrationen, die mit Fackeln verbunden sind, dieses Verhalten erkl\"aren kann. In \ref{sec:paper_4} stelle ich Berechnungen von Helligkeitsschwankungen f\"ur Sterne mit kurzen Rotationsperioden vor. Aktive Regionen tendieren dazu, sich in der N\"ahe des Entstehungsorts anderer aktiver Regionen zu bilden. Ich komme zu dem Schluss, dass zur Modellierung der beobachteten Abh\"angigkeit der stellaren Variabilit\"at von der Rotationsperiode diese Tendenz mit sinkender Rotationsperiode verst\"arkt werden muss.

\chapter{Introduction}\label{intro}

\section{Solar variability}

\subsection{Magnetic activity and sunspots}

The Sun is only one of the many magnetically active stars known to exist in the Universe. It holds a special position among those stars for us, as it makes the Earth habitable. The magnetic activity of the Sun is quite dynamic and its activity directly influences life on Earth. While phenomena like the Northern and Southern polar lights are a beautiful manifestation of the interaction between the solar magnetic activity and the Earth (a connection normally referred to as space weather), the same interaction occasionally makes our electrical grids and satellite systems susceptible to power outages.

\begin{figure*}
\includegraphics[width=\textwidth]{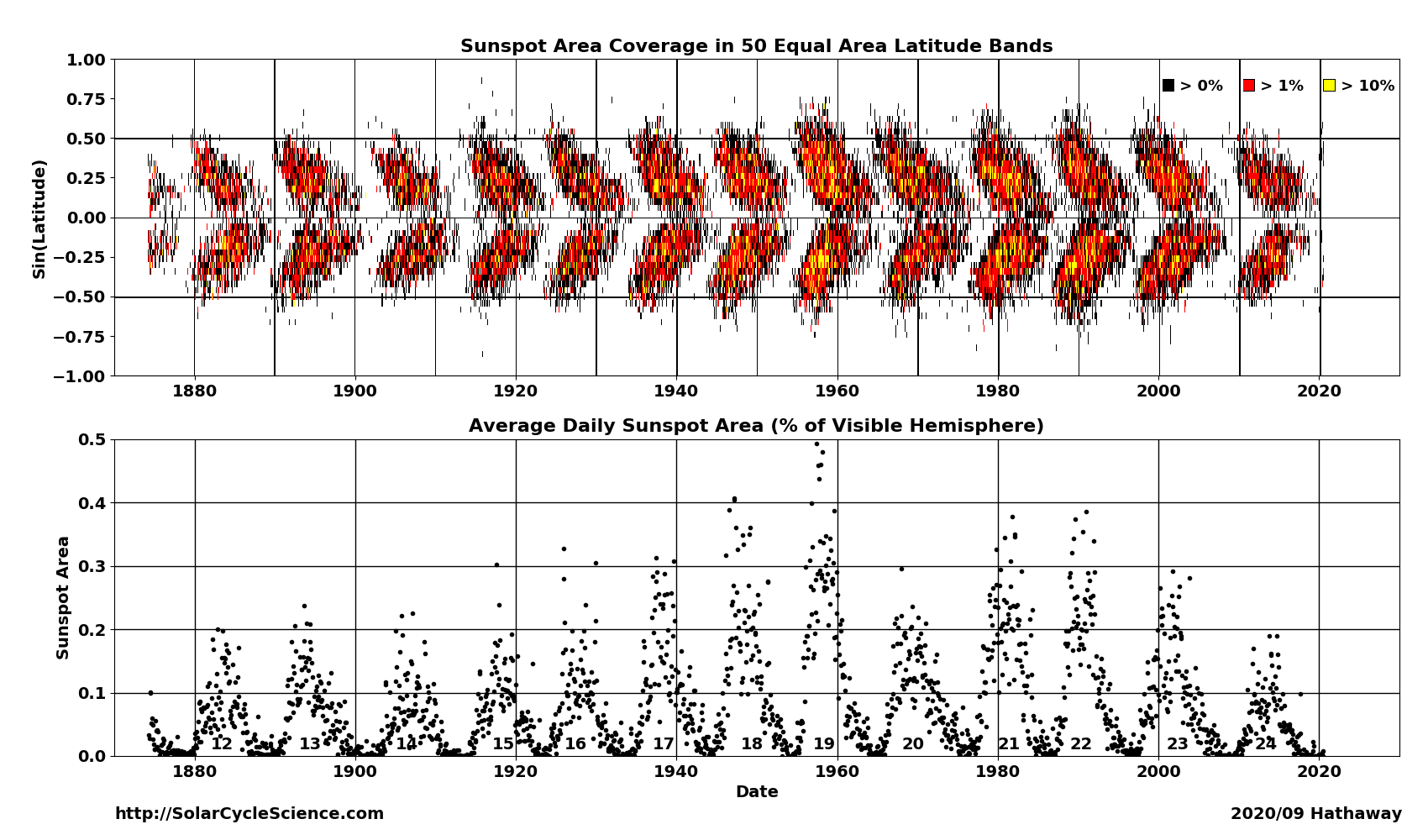}
\caption{Sunspot areas as a function of time. Top panel: Latitudinal dependence of the rotationally averaged sunspot area. Bottom panel: rotationally averaged daily sunspot number as a function of time. Courtesy of David Hathaway. http://solarcyclescience.com/solarcycle.html.}%\protect\footnotemark.} 
\label{fig:sun_cycle}
\end{figure*}

The magnetic activity of the Sun reveals itself in many shapes and forms. Most prominent of those are the so-called sunspots that appear on the solar surface from time to time. Sunspot observations have a long history, with the earliest observations made in ancient China. With the invention of the telescope in the early 1700s, the Western world started more detailed studies of sunspots. While there have been many observers, the first one to publish sunspot observations was \cite{Fabricius1611}. These observations already revealed that the Sun is rotating, as he tracked the movement of the sunspots across the solar disc.
%\cite{Carrington1858} and \cite{Spoerer1883} 
The top panel in Fig. \ref{fig:sun_cycle} gives the so-called \textit{butterfly} diagram. Each wing on this diagram has a time-span of about 11--years. That the latitudinal position of the sunspot varies throughout the cycle was already pointed out by \cite{Carrington1858} and \cite{Spoerer1883}. They found that spots emerge at higher latitudes at the beginning of the cycle and then migrating towards the equator at the end of the cycle. Such a behaviour gives rise to the characteristic butterfly wing-like shape. Spots also only emerge up to latitudes of $\pm$30 degree around the equatorial region. The bottom panel in Fig. \ref{fig:sun_cycle} gives the average daily sunspot number and the 11-year periodicity is even more prominently visible here. The solar cycle is also sometimes referred to as \textit{Schwabe's cycle}, after its discoverer Heinrich Schwabe \citep{Schwabe1844}. When \cite{Hale1938} unraveled that sunspots are magnetic phenomena, more questions opened up regarding the underlying dynamo action that produces the observed patters in the sunspots.
However, the exact way the solar dynamo operates is still unclear even after the more than 80 years of active research.

%\footnotetext{http://solarcyclescience.com/solarcycle.html}

\subsection{Solar brightness variability}

Solar variability has various manifestations that are related to the change in the solar activity. One of the most exciting of them is the variability of the solar brightness, which is the main focus of this thesis.
 The solar irradiance in a broad sense is the power per unit area received from the Sun at a distance of 1 astronomical unit (AU). The amount of irradiance is wavelength and time dependent. One common measure is the Total Solar Irradiance (TSI), which is the measure of the spectrally integrated solar radiative flux at a distance of 1 AU. The TSI is measured by spaceborne radiometers,  which contain an absorptive blackened cavity, which is maintained in thermal equilibrium. The incident radiation is absorbed in the cavity and  changes its temperature. By accurate measurements of the cavity temperature the intensity of the incident radiation can thus be determined. 

\begin{figure*}
\includegraphics[width=\textwidth]{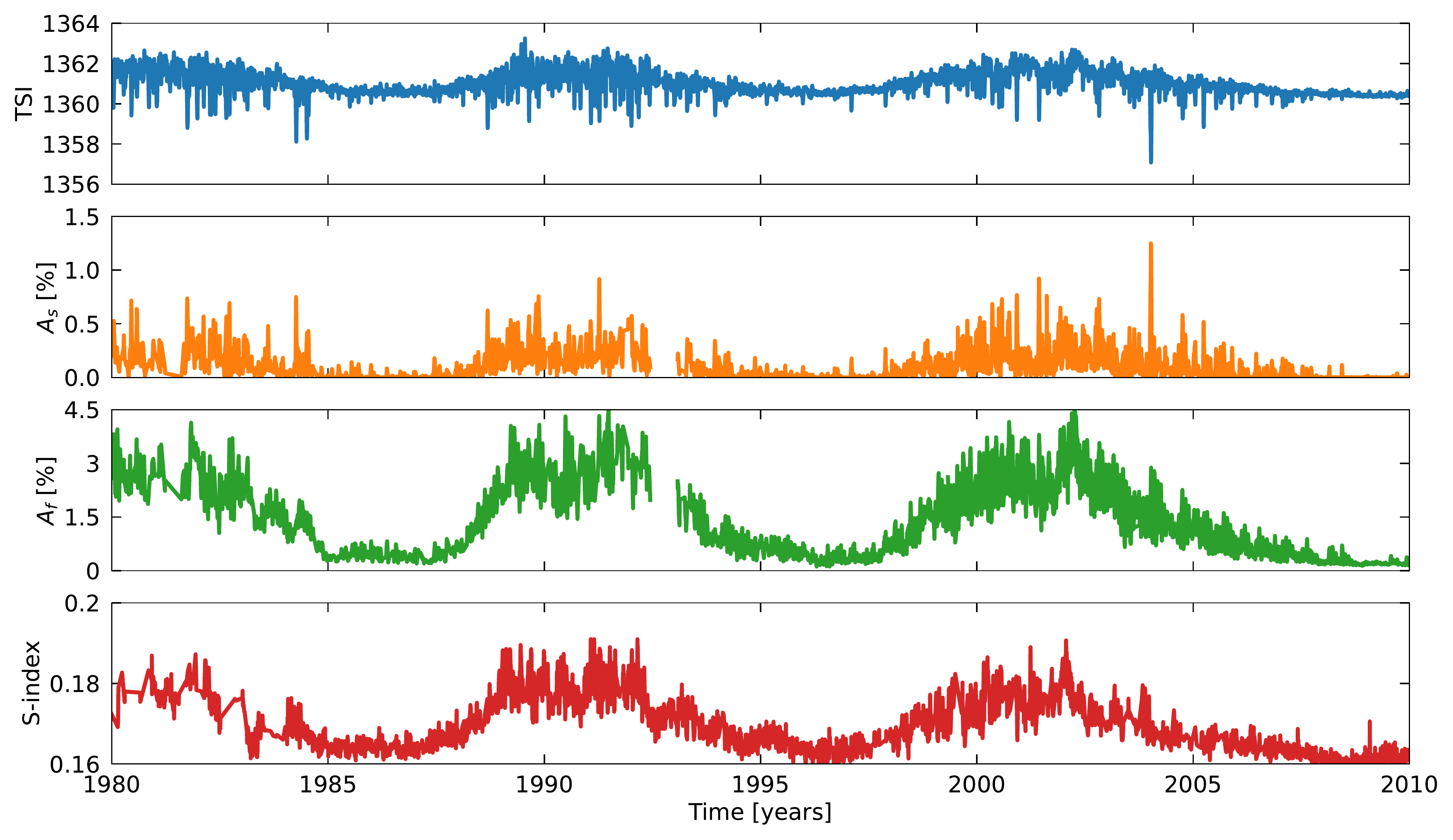}
\caption{Disk area coverages of different solar magnetic features and activity proxies as a function of time. The spot ($A_s$) and facular ($A_f$) disk area coverages  are from \cite{Yeo2014}. The TSI values are from the PMOD \citep{Froehlich2006} composite and the S-index values are from \cite{Bertello2016}.}
\label{fig:sun_activity}
\end{figure*}

Already the first spaceborne TSI measurements in the late 1970s revealed that the amount of irradiance that the Earth receives is not constant, but varies on multiple
time-scales, including variations over the  11-year activity cycle. We show the 
Physikalisch- Meterologisches Observatorium Davos (PMOD) composite \citep{Froehlich2006}\protect\footnotemark in orange lines Fig. \ref{fig:sun_activity}, and the daily sunspot area as returned by the Spectral And Total Irradiance REconstruction \citep[SATIRE,][]{Yeo2014} model in blue lines for comparison. \footnotetext{version42\_65\_1709, ftp://ftp.pmodwrc.ch/pub/data}
Evidently, the rise and fall in the TSI and sunspot area happen at the same time, counterintuitively, the Sun is the brightest at sunspot maximum. This is because of the presence of a second manifestation of magnetic fields on the solar surface: bright faculae (green lines in  Fig. \ref{fig:sun_activity}). Faculae are more diffuse in nature than the dark spots and their area is generally larger. At sunspot maximum, the 
irradiance change brought about by faculae over weights that of spots so that TSI increases. The difference in amplitude between cycle maximum and cycle minimum is around 0.1\%, whereas day-to-day variations can be considerably larger (up to 0.3\%).
%\footnotetext{ftp://ftp.pmodwrc.ch/pub/data}

While sunspots and faculae are photospheric phenomena, the magnetic fields also expand throughout the whole solar atmosphere. The layer directly above the photosphere is called chromosphere. Temperatures rise from about 5770K to 50k  - 100k K, and because of this high temperatures chromospheric spectral lines appear in emission. One of the most famous chromospheric spectral features  are the Ca II H\&K lines at 3933.66\,\AA{} and 3968.47\,\AA{}, respectively. Monitoring of the Ca II H\&K lines started in the late 1960s when Olin Wilson founded the Mount-Wilson HK program, establishing the Mount-Wilson Observatory (MWO) S-index \citep[S-index for short,][]{Vaughan1980} of chromospheric activity. Measurements of the solar Ca II H\&K lines were not made directly, but by using the reflected light of the Moon. However, solar data from the Mount-Wilson observatory are sparse, hence other observations from the NSO Sacramento Peak, the Kodaikanal Observatory and the Lowell Observatory Solar-Stellar Spectrograph have to be combined to obtain a composite that stretches from the 1960s to now \citep{Egeland2017}. The solar S-index variation is shown in panel d of Fig. \ref{fig:sun_activity}. What is clear from Fig. \ref{fig:sun_activity} is that the Sun is the brightest, when its chromospheric emission is the highest.
%What is clear from Fig. \ref{fig:sun_activity} is that both, photometric and chromospheric, activity variations are in-phase and that the Sun is the brightest at activity maximum.

\section{Stellar variability}

The umbrella-term \textit{Sun-like} covers cool, main-sequence dwarf stars with convective envelopes and of spectral classes F, G, and M. The term \textit{solar-analogues} however narrows it down to cool stars with similar fundamental parameters (effective temperature and metallicity as the Sun.

\subsection{Chromospheric and photometric brightness variations}

%chromospheric studies
The MWO program not only observed the chromosperic variability of the Sun, but is most-well known for its monitoring of roughly 2,000 Sun-like stars. These long-term observations revealed that stars exhibit variations in their chromospheric emissions as well, with some stars even displaying cyclic variations \citep{Baliunas1995,Saar1999}. This variations can be related to the rotation period \citep{Skumanich1972, Noyes1984}, with young stars rotating faster and having a tendency for irregular or shorter cycles, whereas their older peers rotate slower and exhibit more pronounced, longer cycles. \cite{Vaughan1980} reported that there is an apparent gap (now called the Vaughan-Preston Gap), between highly active stars and stars with lower activity levels, later confirmed by \cite{Brandenburg1998}, who split the stars further into an active and inactive branch. While the existence of the active branch has come under scrutiny in more recent studies, the clustering of more inactive stars has shown to be robust even with longer data-sets and different methods \citep[see e.g.][]{Distefano2017,BoroSaikia2018, Olspert2018}.

%photometric studies
Complimentary ground-based photometric studies of the long-term variability of solar-type stars have confirmed the existence of stellar activity cycles as observed from the Mount-Wilson Survey.
 \cite{Lockwood1997} found, that with increasing chromospheric variability,the photometric variability increases. 
 Additionally, as noted by \cite{Radick1998}, younger, fast rotating stars with higher chromosperic emission are the dimmest in terms of photometry at high chromospheric emission levels, whereas older, slower rotating stars have their photometric brightness maximum at the time of maximum chromospheric activity, as observed from the Sun. 
Observational studies of the Sun \citep[e.g.][]{Foukal1993,Chapman1997} have shown that with increasing spot area, the ratio between facular and spot areas decreases. \cite{Shapiro2014} have used the dependence of spot and facular areas on the solar S-index to extrapolate from the solar case to more active stars and found that the anti-correlation between photometric and chromospheric variability can be explained by the transition from faculae to spot dominated on the activity timescale. \cite{Reinhold2019} later found that this transition coincides with the Vaughan-Preston gap.

%shorter timescales
These studies of the photometric activity were mostly long-term studies conducted on the activity cycle timescale and data was only available for some thousands of stars. Additionally, ground-based observations are also challenging for various reasons (seeing, day-night-cycle, weather, only to name a few). 
The field of stellar variability studies in terms of photometric observations was truly revolutionized with the introduction of space born observatories, such as the Convection, Rotation and planetary Transits \citep[CoRoT][]{COROT2,COROT}, the \textit{Kepler} telescope \citep{KEPLER} and the Transiting Exoplanet Survey Satellite \citep[TESS,][]{TESS}. While the primary goal of these missions was to detect exoplanets with the transit method, their high precision and high cadence also opened up possibilities to study stellar variability on the rotation period and below \citep[e.g.][]{Basri2013} and retrieve the rotation periods of stars to further explore the rotation period--stellar age relation \citep[][]{Walkowicz2013}.
The rotational variability of a sample of stars with detected rotation periods and effective temperatures between 5500 and 6000 K from \cite{McQuillan2014} is given as a function of rotation period in Fig. \ref{fig:rot_var}.
This figure reveals how sophisticated the picture of stellar variability is. While there is a dependence of the variability on the rotation periods for stars with periods lager than ten days, this not the case for the faster rotators. 

\subsection{The Sun in the stellar context}

With some thousands of stars observed by ground-based, long-term photometric and chromospheric activity monitoring programs and hundreds of thousands of stars observed by space missions, we finally address an important question: How does the Sun compare to Sun-like stars in terms of its variability?

We distinguish between the long-term variability on the cycle time-scale and the short-term variability on the rotational timescale.
When \cite{Vaughan1980} and \cite{Noyes1984} compared the solar chromosperic emission to the chromosperic emission of Sun-like stars from the MWO survey, they found that the Sun lies somewhat between the active and inactive branch, suggesting some transition phase between those branches. \cite{Radick2018} narrowed the analysis down with roughly 70 Sun-like stars, of which almost 80\% are actually even solar-analogues, they found that the solar-chromospheric variability is even higher than the linear dependence of chromospheric emission of the S-index as a function of the latter suggests. In terms of the long-term photometric variability, stars with similar chromospheric activity as the Sun showed considerably larger variations in brightness amplitude than 0.1\%, by factors of 10 to 30. Importantly, \cite{Radick2018} acknowledged, that the chosen solar cycle for this type of comparisons is of importance and that also the inclination (that is the angle between a star's rotation axis and the line-of-sight of the observer) plays a role.

\begin{figure*}
\centering
\includegraphics[width=0.7\textwidth]{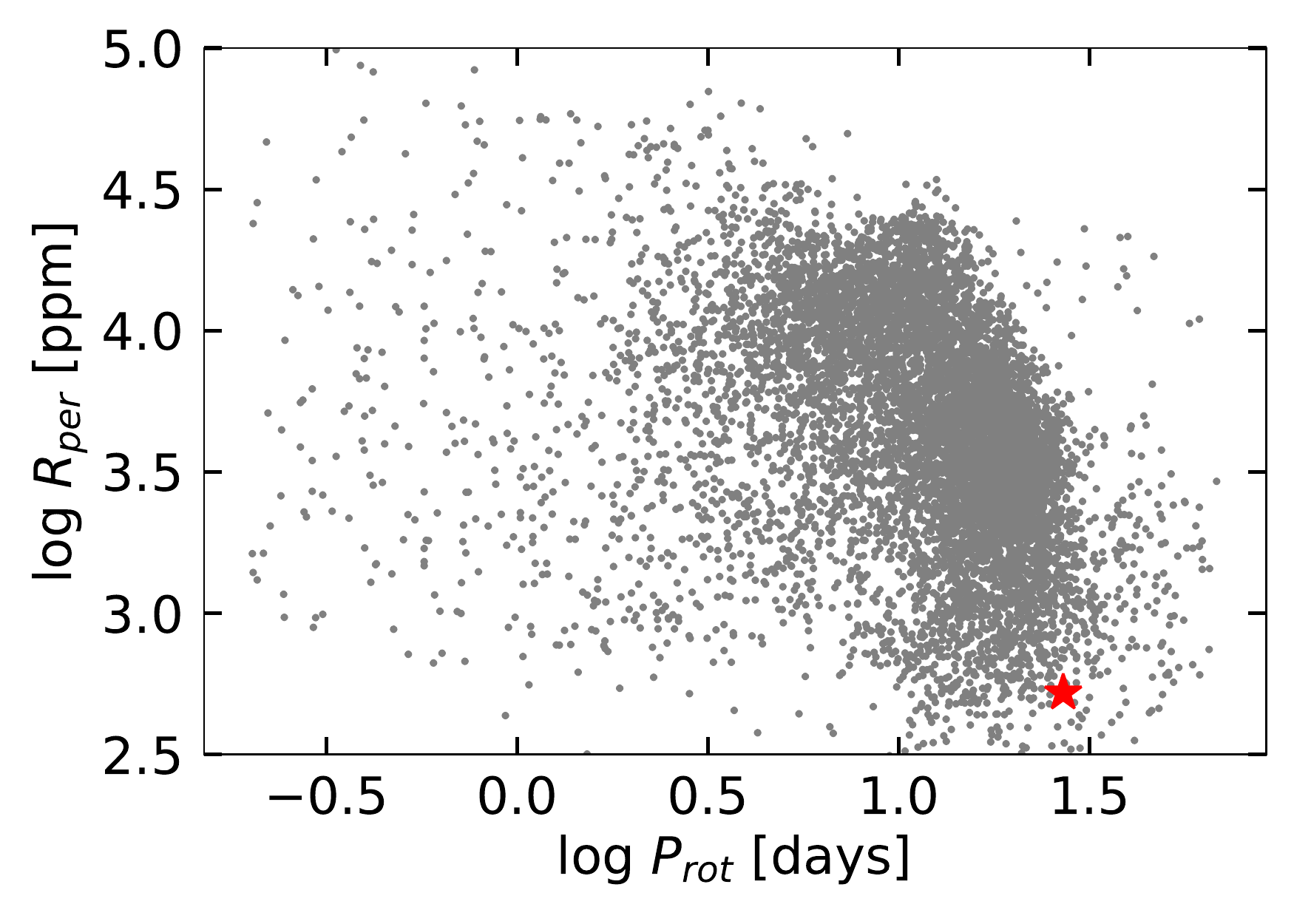}
\caption{Rotational variability as a function of the rotation rate of \textit{Kepler} stars with effective temperatures between 5500 and 6000 K with known rotation periods. Grey points are the data from \cite{McQuillan2014}, and the  red star represents the median value of the solar rotational variability calculated using the SATIRE-S TSI for cycles 21--24.}
\label{fig:rot_var}
\end{figure*}

%rotational variability
In order to compare the solar rotational variability to stars observed by \textit{Kepler}, we calculated the rotational variability of the Sun using the TSI of the SATIRE-S model from solar cycles 21--24 (1976--2019). The rotational variability of the Sun is represented by the red star in Fig. \ref{fig:rot_var}. Two things are uncanny. Firstly, there is large spread of variabilities on the timescales of the solar rotation and secondly the solar variability is unambiguously low compared to its peers. 
\cite{Eliana2020} have argued, that algorithms like the auto-correlation function used by e.g. \cite{McQuillan2014} have difficulties detecting the rotation period even of the Sun, as the light-curves of those stars are rather irregular. Other factors complicate the comparison in this picture as well. The inclination might be different for the stars in the sample and in this instance also the passband of the instruments used introduces hurdles that need to be overcome. Detection biases are one side of the story, however \cite{Timo2020} found that the light curves of stars for which they could detect rotation periods are vastly different from the solar paradigm. These stars show more regular patterns in their variability, which raises the question, if we can model the observed distribution of brightness variations of Sun-like stars by simply extrapolating from the solar paradigm, or if any physical concepts, such as the distribution of the magnetic field on the surfaces of stars have to be altered to reproduce the observed photometric trends.

\section{Motivation and thesis outline}

The aforementioned observational findings have lead to several questions, one of the most intriguing ones being if the lessons learnt from the solar paradigm can be applied to explain stellar variabilities.
One of the main goals of this thesis is to find out through comparison of observational and simulated data, which, if any physical concepts of solar brightness variability have to be altered to reproduce the distribution of variabilities of Sun-like stars. Central to this approach is the model I present in Sect. \ref{sec:paper_1}. In Sect. \ref{sec:paper_2} I directly apply this model to quantify the effect of the inclination and spectral passbands used by past and current space missions to facilitate the solar-stellar comparison. I present a physics-based explanation for the transition between faculae and spot domination of more active stars using the model in Sect. \ref{sec:paper_3}. In Sect. \ref{sec:paper_4} I aim to  aim to model the observed distribution of variability across rotation rates from 4 days down to 30 days by employing the SFTM and changing the characteristics of the distribution of the magnetic features. In Sect. \ref{thesis_conclusio} I present a summary of the most important findings of this thesis and provide examples of further applications of our model.

\chapter{Power spectra of solar brightness variations at various inclinations}\label{sec:paper_1}

The contents of this section are identical to the printed version of N\`{e}mec, N.-E. et al., A\&A, Vol. 636, A42, 2020, reproduced with permission \textcopyright ESO. DOI: 10.1051/0004-6361/202037588\\
\noindent
\textbf{Contributions to the paper}: I developed the model, produced the results and provided the main scientific interpretation.  %, together with R.V. Tagirov she also wrote the masking code. R. H. Cameron ran the surface flux transport simulations. The manuscript has been read and approved by all authors.

\section{Introduction}

Recent planet-hunting missions such as CNES' Convection, Rotation and planetary Transit \citep[CoRoT,][]{COROT,COROT2}, NASA's \textit{Kepler} \citep{KEPLER}, and the Transiting Exoplanet Survey Satellite  \citep[TESS,][]{TESS} have opened up new possibilities for studying stellar variability up to timescales of the rotational period and, in some cases, even beyond. \citep{Timo2017,Montet2017}. 
The plethora of data obtained by these missions underlines the need for a better understanding and modelling of stellar brightness variations. 
One of the possible approaches for such an approach to modelling is to rely on the solar paradigm; that is, to take a model which reproduces the observed variability of solar brightness and extend it to other stars. For example, such an approach has been used by \cite{witzkeetal2018} , who extended the Spectral And Total Irradiance REconstruction  \cite[SATIRE,][]{Fligge2000,Krivova2003} model of solar brightness variability to calculate brightness variations over the timescale of the activity cycle in stars with different metallicities and effective temperatures. 
 Later,  \cite{Veronika_rot} utilised a similar model to investigate how the amplitude of the rotational stellar brightness variability as well as the detectability of stellar rotation periods depend on the metallicity.  Here, we perform one more extension of the SATIRE model to study how the amplitude of solar brightness variability depends on the angle between solar rotation axis and directions to the observer (hereafter, the inclination). 

The brightness variability of the Sun is brought about by magnetic features (such as dark spots and bright faculae) on its surface \citep[see, e.g. reviews by][]{Solanki2013, Ermolli2013}. The visibility of the magnetic features and their brightness contrasts depend on the position of the observer relative to the solar rotation axis. This causes the solar brightness variability to depend on the  inclination. A quantitative assessment of such a dependence is of particular importance in attempting to answer the question of how solar photometric variability compares to that of other stars. To properly address this question, we need to take into account that the Sun is observed from its near-equatorial plane (i.e. at inclinations close to $90^{\circ}$), while stars are observed at random, mostly unknown, inclinations. 

The effect of the inclination on solar variability can only be assessed with models since
solar brightness has never been measured out of the ecliptic. For example, to account for possible long-term climate response to the change of the Earth’s orbital inclination in relation to the solar equator, \cite{Vieira2012}  developed a model based on combining synoptic maps and disk images obtained from the Helioseismic and Magnetic Imager \citep[HMI,][]{HMI} data. They found  that on timescales of several thousands of years, the total solar irradiance (TSI) variability due to the change in the  Earth’s orbital inclination is negligibly small.

A number of studies have modelled the dependence of solar brightness variability on the inclination over the timescale of the 11-year activity cycle. These studies have been motivated by ground-based observations of Sun-like stars that have revealed the Sun exhibits lower photometric variability on the activity cycle timescale than most Sun-like  stars with near-solar levels of magnetic activity \citep[][]{Lockwood1990,Lockwood2007,Radick2018}.  \cite{Schatten1993} proposed that this enigmatic behaviour of the Sun is due to its equator-on view from the Earth. He found that the amplitude of the activity cycle in solar brightness significantly increases with decreasing inclination.
Later, \cite{Knaack2001} and \cite{Shapiro2014} employed a more accurate model and also found an increase of the variability for the out-of-ecliptic observer, but the effect of the inclination appeared to be considerably weaker than that reported by \cite{Schatten1993}. All in all, the current consensus is that the effect of inclination cannot explain the low variability of the Sun on the activity cycle timescale and, consequently, other explanations for this have been proposed \citep{Shapiro2016,witzkeetal2018, Karoff2018}.

\cite{Schatten1993},  \cite{Knaack2001}, and \cite{Shapiro2014} assumed an axisymmetric band-like distribution of faculae and spots. Such an assumption is justifiable for modelling solar brightness variations on the activity cycle timescale but it does not allow modelling brightness variability on the solar rotational timescale. Indeed, the activity cycle variability is caused by the overall modulation with regard to the solar surface coverage by magnetic features from activity minimum to maximum and it depends only on the time-averaged surface distribution of magnetic features (which can be approximated by the axisymmetric band-like structure rather well). In contrast, rotational variability is caused by the evolution of individual magnetic features and their transits across the visible solar disc as the Sun rotates. Consequently, it depends on the exact distribution of magnetic features.

An attempt to model the effect of the inclination on the rotational solar brightness variability was recently carried out by \cite{Shapiro2016}. They  used distribution of magnetic features on the visible solar disk provided by \cite{Yeo2014} and obtained the distribution of magnetic features on the far-side of the Sun (part of which would become visible for the observer not bound to the Earth) assuming that the near and far sides of the Sun are point-symmetric with respect to each other through the centre of the Sun.  They found that an observer bound to the ecliptic plane witnesses the Sun to be spot-dominated on the rotational timescale, but with decreasing inclination the amplitude of the rotational variability decreases (in contrast to the brightness variability on the activity cycle timescale, which increases with decreasing inclination) and the facular contribution becomes dominant. Despite being more advanced relative to previous studies, the assumption of the point-symmetric distribution of solar magnetic features employed in \cite{Shapiro2016}  does not account for the appearance and disappearance of magnetic features which rotate in and out of the visible
solar disc. This has led to the contribution of a number of artefacts which did not allow  for a study of the effects of the inclination on the detectability of stellar rotation periods. These effects might play, however, an important role in understanding the observed distribution of rotation periods in Kepler stars \citep{Reinhold2019, VanSaders2019}. Also, these artefacts hindered the accurate assessment of the inclination effect on the timescale of solar rotation.  Such an assessment is, in turn, needed for the interpretation of the data from the planet-hunting missions. For example, the \textit{Kepler} data indicated that also solar brightness variability on the timescale of solar rotation appears to be lower than that of most of the stars with known near-solar fundamental parameters and rotation periods \cite{Timo2020}. 

Here we take a different approach than \cite{Shapiro2016} and we utilise a surface flux transport model \citep[SFTM,][]{Cameron2010} to obtain the distribution of solar magnetic features over the entire solar surface (i.e. on both near and far sides of the Sun).  This distribution is then fed into the SATIRE model to calculate the solar brightness variability  for different solar activity levels, various photometric filter system used in stellar observations, and at different inclinations. In particular, we show how the change of the inclination affects the power spectrum of solar brightness variations. This allows for a study of the impact of the inclination on brightness variability depending on the timescale of the variability. 
In Sect.~\ref{Methods}, we describe how we compute the solar disc area coverages by magnetic features from the SFTM and then calculate the brightness variations following the SATIRE model.  We also list the main parameters of the model and explore their impact on the brightness variations. In Sect.~\ref{ecliptic}, we show how the strength of an individual cycle affects the solar photometric variability in different passbands before we move to different inclinations in Sect.~\ref{inclination}. In Sect.~\ref{inclination}, we also decompose the solar brightness variability into components arising from the evolution of magnetic features and from the solar rotation. We present our main conclusions in Sect.~\ref{conclusion}.

\section{Methods}\label{Methods}

\subsection{Calculating brightness variations}\label{sec:SATIRE}

We built our method based on the SATIRE model, in which  brightness variations on timescales longer than a day are attributed to the emergence and evolution of magnetic field on the surface of the Sun, as well as on solar rotation \citep[][]{Fligge2000,Krivova2003}.
The photospheric magnetic features are divided into three main classes: sunspot umbra ($u$), sunspot penumbra ($p$), and faculae ($f$). 
The intensities of these features and that of the quiet Sun ($q$) depend on the wavelength and the cosine of heliocentric angle $\theta$ ($\mu = \mathrm{cos}\theta$), but they are also time-independent. The intensities were computed
by \cite{Unruh1999} \citep[following][]{Castelli1994} with the use of the spectral synthesis
code ATLAS9 \citep{Kurucz1992}.
 The 1D atmospheric structures of umbra, penumbra, and quiet Sun were calculated using radiative equilibrium models, while the facular model is a modified version of FAL-P by \cite{Fontenla1993}.

The spectral solar irradiance $S(t,\lambda_w)$ (i.e. spectral radiative flux from the Sun, normalised to one AU), where $t$ is the time and $\lambda_w$ the wavelength (which should not be confused with $\lambda $ used for the latitude later in this paper), is calculated by summing the intensities weighted by the corresponding fractional disc area coverages of the magnetic features (designated with the index $k$) as given by

\begin{equation}
    S(t,\lambda_w) = S^{q}(\lambda_w) +\sum_{mn} \sum_{k}(I_{mn}^{k}(\lambda_w)-I_{mn}^{q}(\lambda_w)) \, \alpha_{mn}^{k}(t)
    \Delta\Omega_{mn}.
\label{SSI}
\end{equation}

\noindent Here the summation is done over the pixels of the magnetograms and the $m$ and $n$ indexes are the pixel coordinates (longitude and latitude, respectively), $\alpha_{mn}^{k}$ is the fraction of pixel ($m$,$n$) covered by the magnetic feature $k$, $\Delta \Omega_{mn}$ is the solid angle of the area on the solar  disc corresponding to one pixel, as seen from the distance of 1 AU, and S$^{q}$ is the quiet Sun irradiance, defined as

\begin{equation}
   S^{q}(\lambda_w) = \sum_{mn} I_{mn}^{q}(\lambda_w)\Delta\Omega_{mn}.
\end{equation}

\noindent 
The solid angles of pixels as well as corresponding intensity values depend on the vantage point of the observer. Consequently, the solar irradiance values $S(t,\lambda_w)$ given by Eq. (\ref{SSI}) also depend on the vantage point of the observer and, in particular, on the inclination.

\subsection{Surface flux transport model}\label{sec:SFTM}

To simulate the full surface distribution of magnetic features, we use the SFTM in the form presented in \cite{Cameron2010}. The SFTM describes the passive transport of the radial component of the magnetic field B, considering the effects of differential rotation $\Omega(\lambda)$ (with $\lambda$ being the latitude), meridional flow $\nu(\lambda)$ at the solar surface, and a horizontal surface diffusion  thanks to a non-zero diffusivity $\eta_H$. The emerged active regions gradually disperse due to the radial diffusion $\eta_r$, with the flux finally decaying after cancellation between opposite polarities, where they overlap. The governing equation is

%\begin{multline}
\begin{equation}
\begin{split}
  \frac{\partial B}{\partial t} &= - \Omega(\lambda)\frac{\partial B}{\partial \phi} - \frac{1}{R_{\odot}\cos\lambda}\frac{\partial}{\partial \lambda}(\nu(\lambda)B \cos(\lambda))\\&+\eta_H\left ( \frac{1}{R_{\odot}^{2}\cos\lambda}\frac{\partial}{\partial \lambda}\left ( \cos(\lambda)\frac{\partial B}{\partial \lambda} \right ) + \frac{1}{R_{\odot}^{2}\cos^{2}\lambda} \frac{\partial ^{2}B}{\partial\phi^{2}}\right ) \\ &+D(\eta_r) + S(\lambda,\phi,t),
%\end{multline}
\end{split}
\label{flux}
\end{equation}

\noindent where $R_{\odot}$ is the solar radius, $\phi$ is the longitude of the 
active region, and $D$ is a linear operator that describes the decay due to radial
diffusion with the radial surface diffusivity $\eta_r$. For the linear operator $D$ the form of \cite{Baumann2006} was used. 
The horizontal diffusivity $\eta_H$ was taken to be 250 km$^{2}$s$^{-1}$ as in \cite{Cameron2010}  and the radial surface diffusivity $\eta_r$ was set to  25 km$^{2}$s$^{-1}$ according to \cite{Jiang2011_2}. The time average (synodic) differential rotation profile was taken from \cite{Snodgrass1983} and is given as (in degree per day):
\begin{equation}
\Omega(\lambda) = 13.38 -2.3 \cdot \sin^{2}\lambda -1.62\cdot \sin^{4}\lambda. 
\label{diff_rot}
\end{equation}

 The time-averaged meridional flow is expressed following \cite{vanBallegooijen1998}, namely,
\begin{equation}
\nu(\lambda)=
  \begin{cases}
    11\cdot \sin(2.4\lambda) \quad \text{m/s,} & \text{where} \quad \lambda \leq 75 ^\circ \\
        0,         & \text{otherwise}.
  \end{cases}
 \label{merflow}
\end{equation}

The source term $S(\lambda,\Phi,t)$ in Eq. (\ref{flux}) describes the magnetic flux, which is prescribed to be in the form of  two patches with opposite polarities \citep[][]{vanBallegooijen1998,Baumann2004}. The patches are centred at $\lambda_{+}$ and $\phi_{+}$ for the positive polarity patch and $\lambda_{-}$ and $\phi_{-}$ for the negative polarity patch. The field of each patch is given by

\begin{equation}
B^{\pm}(\lambda,\phi) = B_{\mathrm{max}}\left(\frac{0.4 \Delta \beta}{\delta}\right)^{2} e^{-2[1-\cos(\beta_{\pm}(\lambda,\phi))]/\delta^{2}} ,
\label{source}
\end{equation}

\noindent where $B^{\pm}$ is the flux density of the positive and negative polarity,
$\beta_{\pm}$($\lambda$,$\phi$) are the heliocentric angles between point ($\lambda$, $\phi$) and the centres of the polarity patches, $\Delta \beta$ is the separation between the two polarities and $\delta$ is the size of the individual polarity patches, taken to be 4$^\circ$. $B_{\mathrm{max}}$ is a scaling factor introduced by \cite{Cameron2010} and \cite{Jiang2011_2} and was fixed to 374 G. This value was found by forcing the total unsigned flux to match the measurements from the Mount Wilson and Wilcox Solar Observatories.

\cite{Jiang2011_1} constructed a semi-empirical source term $S(\lambda,\Phi,t)$ for the 1700--2010 period so that its statistical properties reflect those of the Royal Greenwich Observatory sunspot record. Here we adopt the $S(\lambda,\Phi,t)$ term from \cite{Jiang2011_1}  but with one important modification. As an observer stationed at a vantage point outside the ecliptic sees both the near- and far-sides of the Sun (as defined by the Earth-bound observer), it is crucial to avoid any systematic differences between the active region distributions on the two sides. To this purpose we have modified  $S(\lambda,\Phi,t)$  so that the emergence of active regions happens at random longitudes, whereas the butterfly-like shape of their latitudinal emergence, as well as the number of emergences and the tilt-angle distributions, over the course of the cycle is preserved.

All in all, the adapted source term describes the emergence of active regions on the solar surface in a  statistical way. We stress that the goal of this study is not to reproduce the exact solar light curve as it would be seen from outside the ecliptic, but to study the effect of the inclination on the power spectrum of solar brightness variations at different levels of solar activity. The statistical representation of the source term is fully sufficient for this purpose.

\subsection{From magnetic fluxes to area coverages}\label{sec:ff}

The SFTM returns simulated magnetograms, with a pixel-size of 1$^\circ \rm \times$ 1$^\circ$. We follow the approach of \cite{DasiEspuig2014} and divide each pixel ($m$,$n$) into 100 sub-pixels, with a size of 0.1$^\circ \rm \times$ 0.1$^\circ$ each.

To calculate the brightness variations, we need to distinguish between spots and faculae.
The spot areas and positions at the day of emergence have been provided by \cite{Jiang2011_1} together with the source term $S(\lambda,\Phi,t)$. After spots  emerge, their positions  on the solar surface are affected by the differential rotation described by Eq.~(\ref{diff_rot}) and the meridional flow described by Eq.~(\ref{merflow}). The spot sizes are calculated by following a decay law during their evolution.
 We have found studies in the literature that support linear and parabolic decay laws and different values for the decay rate \citep[][]{MorenoInsertis1988, MartinezPillet1993,Petrovay1997,Baumann2005,Hathaway2008}. As \cite{Baumann2005} found, it is not possible to distinguish between a linear and parabolic decay law from, for example, the area distribution of sunspots.
 For simplicity, we chose a linear decay law of:
\begin{equation}
    A(t) = A_0 - R_d \cdot (t - t_0),
    \label{eq:spot_decay}
\end{equation}
\noindent where $A(t)$ is the area on a given day $t$ and $t_0$ is the day on which the spot has its maximum area $A_0$  (provided in the input). The decay rate $R_d$ is measured in microsemi-hemispheres (MSH) per day and is a semi-free parameter of the model, which will be discussed in more detail in Sect. \ref{sec:params}. The decay rate $R_d$ has been studied extensively before. In particular, \cite{MartinezPillet1993} have reported several values of the decay rate, ranging from 25 to 47 MSH day$^{-1}$. The value we found to be most optimal for our model is 80 MSH day$^{-1}$ (see a detailed description of the procedure used to determine $R_d$ in Sect.~\ref{sec:params}). The slightly higher value, compared with observational estimates, which we obtained for our modelling can be explained by the low spatial resolution of the source term in Eq.~(\ref{source}). A group of spots might be represented by one large spot (due to the resolution of the source term), which then decay with a rate that is equal to the sum of the decay rates of the individual spots.

Having established the spatial and temporal spot distribution, we can then correct the simulated magnetograms for the spot magnetic flux, which is important for the masking of the faculae. The correction is done on the original 1$^\circ \rm \times$ 1$^\circ$ grid corresponding to the SFTM output since, in contrast to the spot distribution which is calculated on the 0.1$^\circ \rm \times$ 0.1$^\circ$ grid, we calculate more diffuse facular distribution on the original grid. If a  1$^\circ \rm \times$ 1$^\circ$ pixel is found to be free of spots, the correction is equal to 0 and the magnetic field in the pixel is directly taken from the SFTM.  If a given pixel is found to be partially covered by spots the magnetic field in the pixel is corrected as,

\begin{equation}
B'_{\mathrm{(m,n)}} = B_{\mathrm{m,n}} - B_{\mathrm{spot}} \cdot a^s_{\mathrm{m,n}},
 %\label{eq:fac_f}
\end{equation}

\noindent where $B_{\mathrm{m,n}}$ is the pixel field returned by the SFTM, $B_{\mathrm{spot}}$ is the mean magnetic field of a spot, and $a^s_{\rm mn}$ is the fractional coverage of the pixel (m,n) by spots.
The value of $B_{\rm spot}$ is taken from observations. \cite{Keppens1996} have measured the umbral and penumbral field strength of solar sunspots. We do not distinguish between umbral and penumbral regions and we use an area-weighted average of the values of 800 G reported in \cite{Keppens1996}.

The remaining magnetic field $B'_{\mathrm{(m,n)}}$ (with $B'_{\mathrm{(m,n)}}=B_{\mathrm{(m,n)}}$ for pixels free of spots) is then attributed to faculae and is calculated following the SATIRE approach:
\begin{equation}
\alpha^f_{\rm m,n}= 
  \begin{cases}
    \frac{B'_{\rm m,n}}{B_{\rm sat}} & \text{if B$_{\mathrm{mn}}$ $<$  B$_{\mathrm{sat}}$} \\
        1         & \text{if B$_{\mathrm{mn}}$ $\ge$ B$_{\mathrm{sat}}$},
  \end{cases}
 \label{eq:fac_f_f}
\end{equation}
\noindent where B$_{\mathrm{sat}}$ is the saturation threshold, in accordance to the SATIRE-S model \citep{Krivova2003,Wenzler2004,Ball2012}. In this model, the facular filling factor increases linearly with the magnetic field strength, eventually reaching unity at a saturation.
Given that the SFTM provides information only at time of the maximum area and during the subsequent decay of the active regions, we need to additionally consider the growth phase of the spots (i.e. take into account that they do not emerge instantaneously). We employ a linear growth law with a constant rate $R_g$ similar to the decay law given by Eq.~(\ref{eq:spot_decay}). For $R_g$ we have not found any appropriate studies so that it is treated as  a free parameter (see the next section).

\subsection{Model parameters}\label{sec:params}

\begin{figure*}
\includegraphics[width=\textwidth]{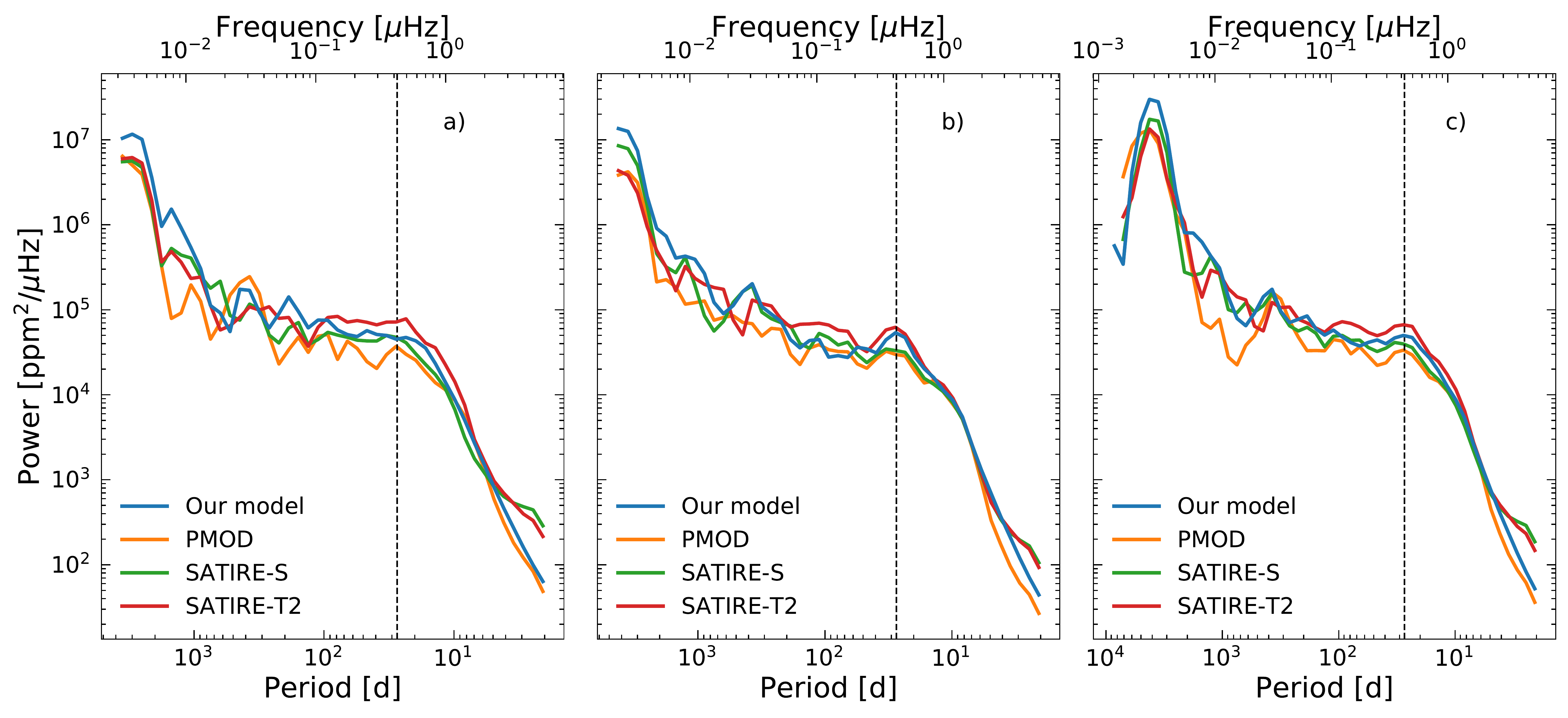}
\caption{Comparison of the power spectra of solar brightness variations produced by our model to those given by the PMOD-composite as well as SATIRE-S and SATIRE-T2 models for cycle 22 (panel a) and cycle 23 (panel b) and the combined timeseries (panel c). The vertical dashed black line indicates the synodic solar rotation period of 27.3 days.} 
\label{fig:Comparison_model_data}
\end{figure*}

\begin{table}
\caption{List of the parameters used in our model}              % title of Table
\label{table:param_descricption}      % is used to refer this table in the text
\centering                                      % used for centering table
\begin{tabular}{|l| l| l|}          % centered columns (4 columns)
\hline\hline                        % inserts double horizontal lines
Parameters & Description & Best value \\    % table heading
\hline                                   % inserts single horizontal line
    R$_d$ & decay rate spots  & 80 MSH day$^{-1}$\\      % inserting body of the table
    R$_g$ & growth rate spots & 600 MSH day$^{-1}$ \\
    B$_{sat}$ & saturation threshold faculae & 500 G\\
\hline   
%inserts single line
\label{paper_1_best_params}
\end{tabular}
\end{table}

To find the best set of model parameters, we compare the power spectra of the computed TSI time series to the power spectra of TSI from other sources. We use the Physikalisch-Meterologisches Observatorium Davos (PMOD)
composite \\
\citep[][version 42\_65\_1709,\url{ftp://ftp.pmodwrc.ch/pub/data}]{Froehlich2006}, which provides TSI measurements over several decades. We also use the TSI output from the SATIRE-S \citep{Yeo2014} and SATIRE-T2 \citep{DasiEspuig2016} solar irradiance variability reconstruction models. In SATIRE-S the distribution of magnetic features on the solar surface is derived from full disk images and magnetograms of the Sun, whereas in SATIRE-T2 it is derived from a SFTM but with a different source term than employed in this study.

In cycle 21, both the PMOD composite and SATIRE-S contain a significant amount of data gaps that would affect the power spectra. We therefore restrict ourselves to use cycles 22 and 23 for the determination of the best parameter set.
We show the power spectra of the solar brightness variations as presented by PMOD, SATIRE-S and SATIRE-T2 in Fig.~\ref{fig:Comparison_model_data}.
One striking difference between the datasets is that SATIRE-S and SATIRE-T2 show higher power values compared to the PMOD-composite at periods below five days for both considered cycles. We attribute this to aliasing effects being present in the two SATIRE-models.
Both, SATIRE-S and SATIRE-T2, give one instantaneous value of the TSI per day, whereas the PMOD-composite gives daily averages. Consequently, the difference between the power spectra appears  because of the comparison between instantaneous values (affected by aliasing) and daily averages. To avoid aliasing issues in our model output, we calculate solar brightness with a six-hour cadence. We found that this leads to similar values of spectral power starting from timescales of about two days as the PMOD-composite.

We found our best set of parameters (see Table \ref{paper_1_best_params}) by comparing the power spectra obtained with the output of our model to those obtained with the PMOD composite. Namely, we calculated the $\chi^2$ values using the parts of the power spectra below the solar rotation period (i.e. we only considered periods shorter than 27.3 days). Despite having used only low-period parts of the power spectra  for the fit, we find that we are still able to maintain a reasonable agreement on longer timescales as well. Our calculations seem to slightly overestimate the variability on the activity timescale, which can be attributed to the absence of ephemeral regions in our model \citep[see discussion in][]{DasiEspuig2016}. 

\begin{figure*}[ht!]
\centering
\includegraphics[width=\textwidth]{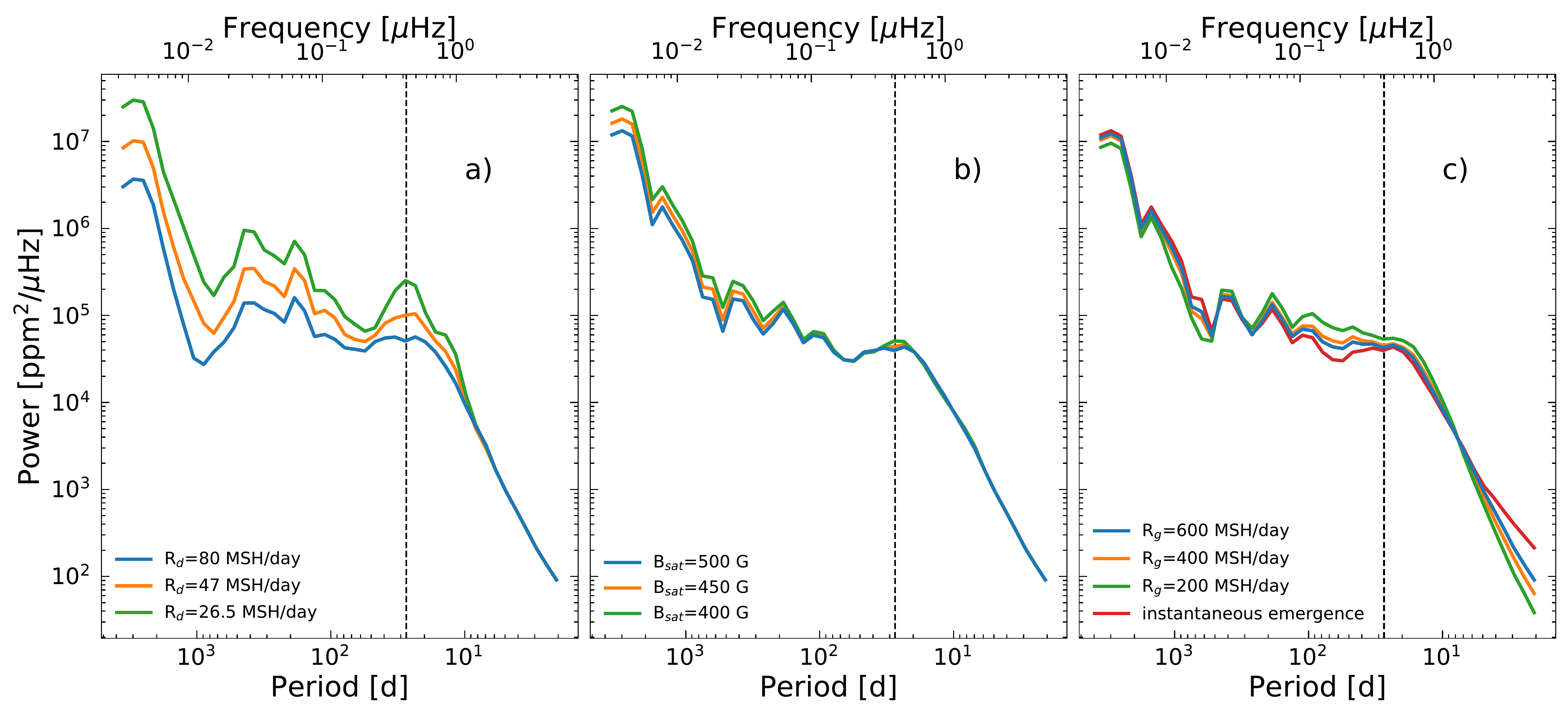}
\caption{Effect of the different parameters of the model on the brightness variations. Panel a) shows the effect of the decay rate R$_d$ on the spot component only, panel b) the effect of B$_{sat}$ on the total power spectrum, and panel c) the effect of different growth rates R$_g$ on the total power spectrum compared to not having the spot growth included as depicted by the red curve. R$_d$ and R$_g$ are in units of MSH day$^{-1}$. The vertical dashed black line indicates the synodic solar rotation period at 27.3 days.} 
\label{fig:PS_params}
\end{figure*}

Let us also check how the different free parameters of our model affect the power spectrum of solar brightness variations returned by the model. The effects of the spot decay rate $R_d$  (panel a), B$_{sat}$ value (panel b), and spot growth rate $R_g$ (panel c) are illustrated in Fig.~\ref{fig:PS_params}. With decreasing spot decay rate, $R_d$, the overall area coverage of the spots is increasing, which affects  timescales longer than about 10 days \citep{Shapiro2019}. The prominent peak at the rotation period for the $R_d=26.5$ MSH day$^{-1}$ is a result of the long lifetime of the spots. The longer the spot lives, the higher the probability it reoccurs at the next rotation which leads to the formation of the rotation harmonic in the power spectrum.

The effect of the saturation threshold, $B_{sat}$, is shown in Fig.~\ref{fig:PS_params} b. We note that the facular filling factors are primarily regulated via this parameter. On the activity cycle timescale, faculae are the dominant source of variability, whereas on timescales, below 100 days, the spot component is the main driver of the variability. A value of 500 G for $B_{\rm sat}$ leads to the best fit compared to the PMOD-composite.
In contrast to the effect of the decay rate, $R_d$, the growth rate , $R_g$, shows the highest impact on timescales below 10 days (see right panel of  Fig.~\ref{fig:PS_params}). The value of 600 MSH day$^{-1}$ gives the best agreement with the PMOD composite on those timescales.

\section{Solar brightness variations as seen by an ecliptic bound observer}\label{ecliptic}

\subsection{TSI variability during activity cycles of different strengths}

\begin{figure}
\centering
\includegraphics[width=0.5\textwidth]{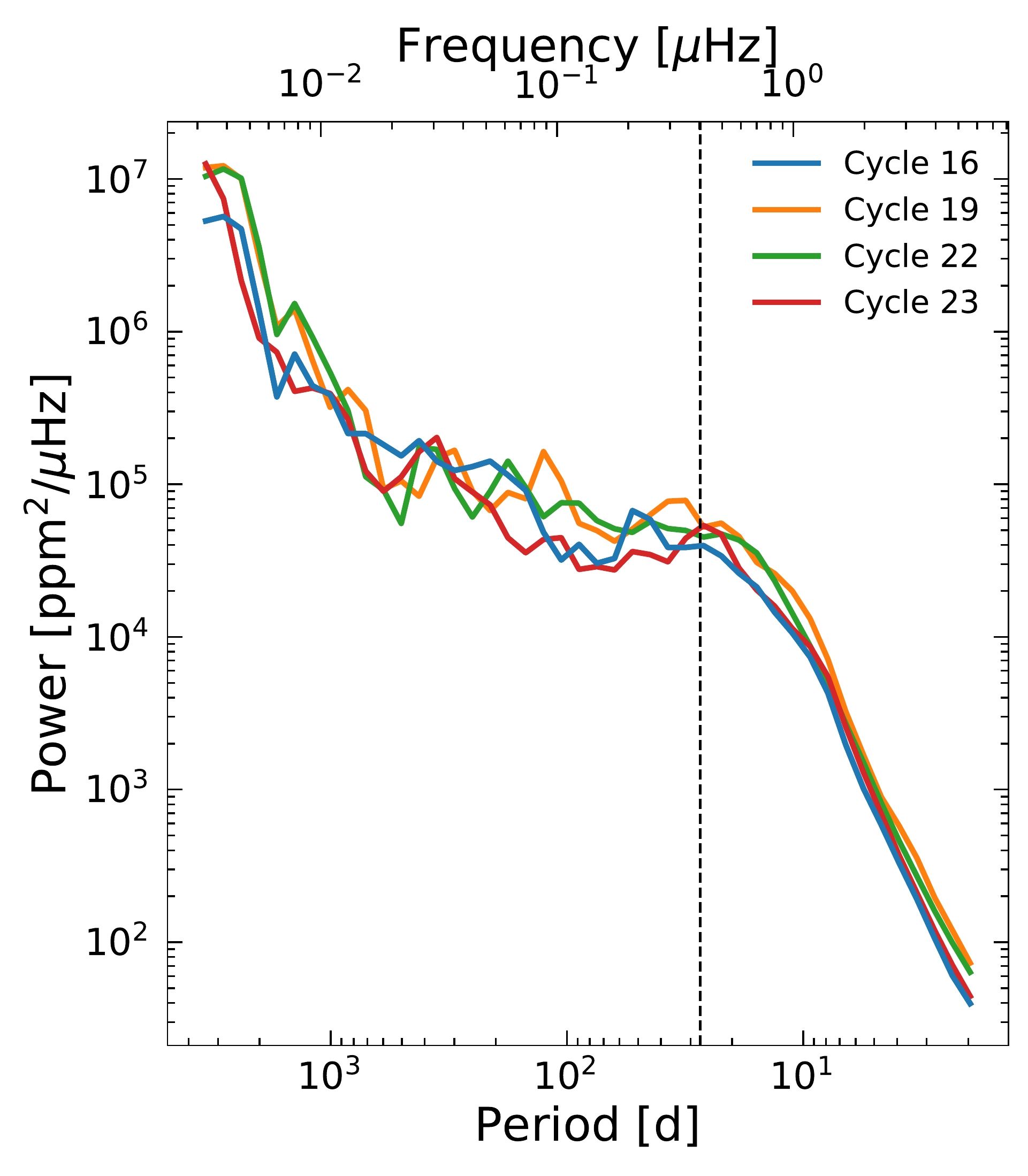}
\caption{Power spectrum of the TSI for different cycles as seen by an ecliptic bound observer. The vertical dashed black line indicates the synodic solar rotation period at 27.3 days.}
\label{fig:PS_TSI_cycles}
\end{figure}

Until now, we considered the TSI variability during cycles 22 and 23. To understand the solar brightness variations in the context of stellar variability, it is important to explore different activity levels. %As the sunspot group number is used as the input into the model, 
With our source term we can calculate solar brightness variations back to 1700. %/, i.e. over the whole period over which such data are available.
In Fig.~\ref{fig:PS_TSI_cycles}, we compare power spectra of the TSI variability as returned by our model for cycles 16 (one of the weakest cycle over the last 300 years), 19 (the strongest cycle observed so far), 22, and 23.
For cycle 16 and 23, a small peak at the rotation period of about 27 days can be seen.
 The profile of the power spectrum for cycle 19 is rather surprising, with two peaks on periods slightly below (25 days) 
  and above (32 days) the rotation period (see also Fig. \ref{fig:PS_filter_cycles} where the double peak structure is more easily visible). \cite{Shapiro2019} explained such a 
  double-peak structure by the cancellation of spot and facular contribution to the rotation signal. \cite{Veronika_rot} further analysed the connection between the power spectrum profile and detectability of the rotation period.

Recently a lot of effort has been put into determining stellar rotation periods from photometric observations by the \textit{Kepler} telescope \citep[see e.g.][]{Timo2013,McQuillan2014,Angus2018}.
In what appears to be an intriguing result, the detection of the rotation period of old stars with near-solar level of magnetic activity seems to be challenging due to the low amplitude of the irradiance variability,  short lifetime of spots, and the cancellation of the rotational signal from spots and faculae \citep[][]{Aigrain2015,Shapiro2017,Reinhold2019}. In agreement with previous studies \citep[e.g.][]{Lanza_Shkolnik2014,Aigrain2015}, our analysis indicates that the same star can be deemed as periodic or non-periodic  \citep[according to the definition of][]{McQuillan2014}, depending on whether it is observed at high or low activity.

\subsection{Solar variability in different passbands}

\begin{figure}
\centering
\includegraphics[width=0.75\textwidth]{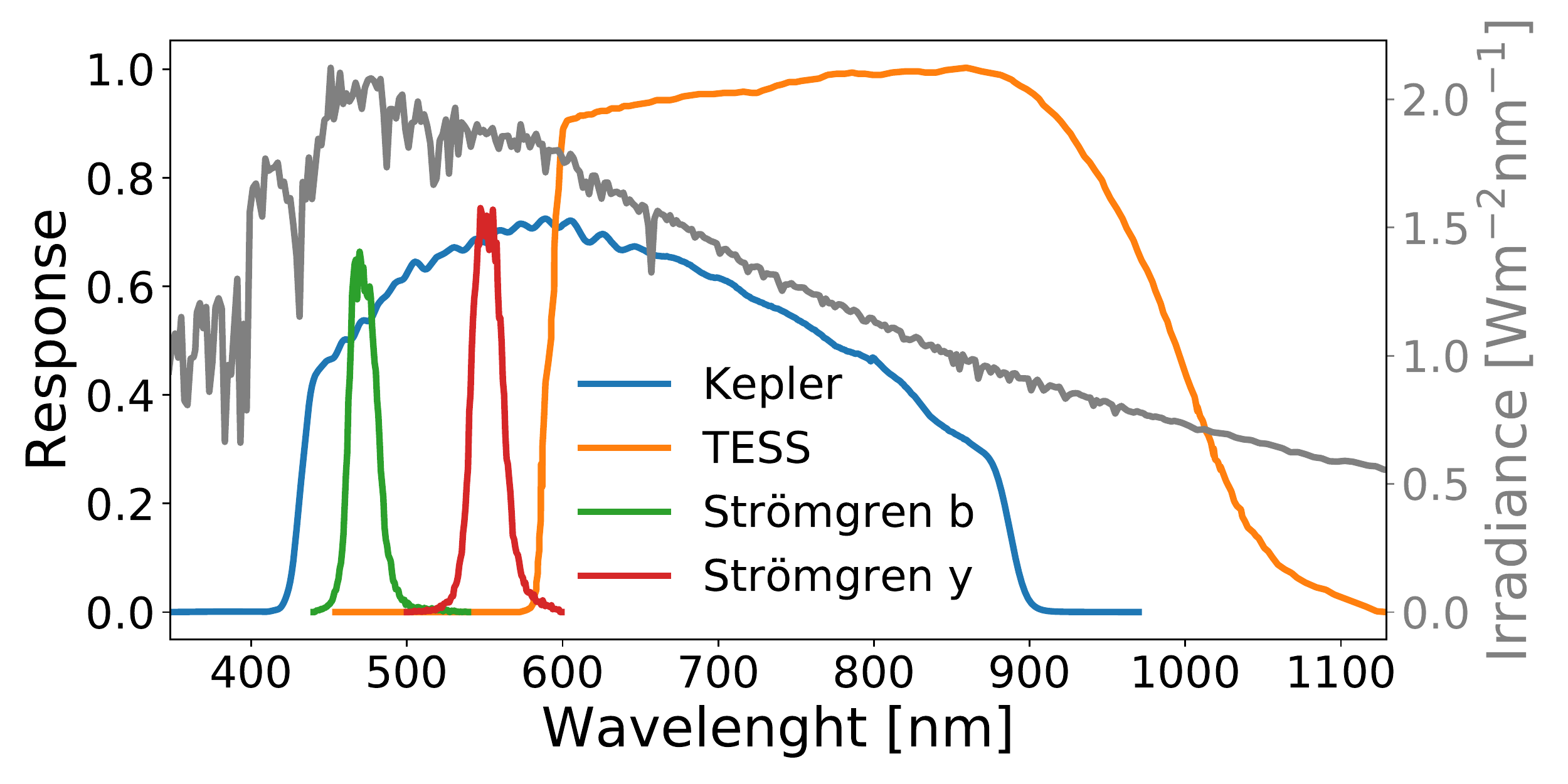}
\caption{Response functions of the different filter systems used in this work. The quiet-Sun irradiance as used by SATIRE is shown in grey.}
\label{fig:response_functions}
\end{figure}

In this section, we explore solar brightness variations as they would be observed in different passbands. 
We multiply the computed spectral irradiance given by Eq.~(\ref{SSI}) with the response functions of different filter systems and then integrate over the corresponding wavelength ranges.
We consider the Strömgren filters \textit{b} and \textit{y} which have been widely used in ground-based observations to study long-term stellar photometric variability \citep{Radick2018}, as well as the \textit{Kepler} and TESS passbands. The transmission curves and the quiet-Sun spectrum (according to the SATIRE model) are shown in Fig.~\ref{fig:response_functions}. The Strömgren \textit{b} and \textit{y} filters are centred at 476 and 547 nm, respectively, so that  Strömgren \textit{b} is located around the maximum of the solar spectrum, while Strömgren  \textit{y} is shifted to the red. The primary goal of \textit{Kepler} was to find planets around solar-type  stars and its filter profile covers almost the  whole visual wavelength range. TESS is aimed at observing a large number of M dwarfs and is, consequently, more sensitive to the red part of the spectrum.

\begin{figure*}
\centering
\includegraphics[width=\textwidth]{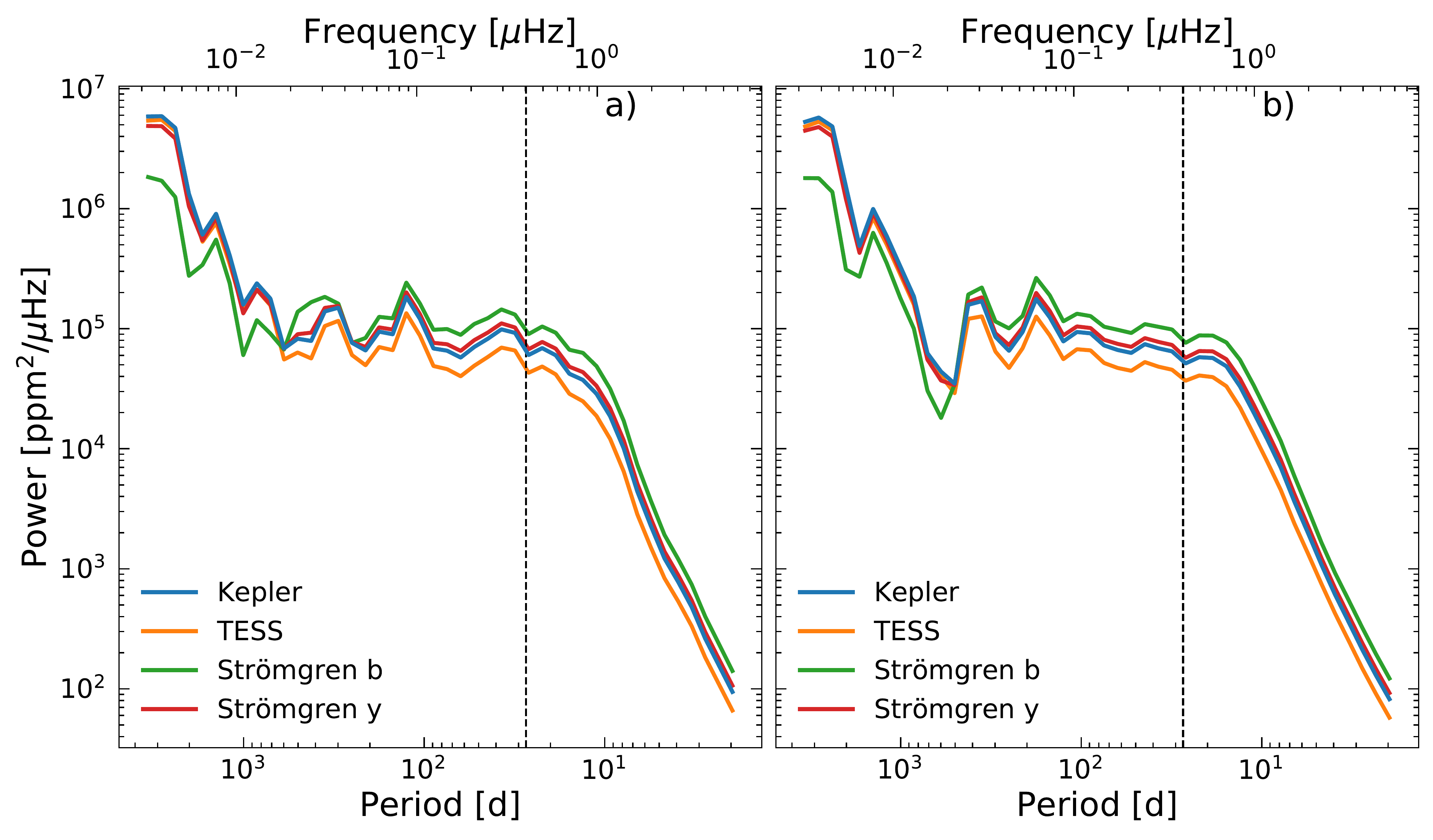}
\caption{Power spectra of solar brightness variations in different filter systems for different cycles as observed from the ecliptic.
Panel a) shows cycle 19, panel b) cycle 22. The vertical dashed black line indicates the synodic solar rotation period at 27.3 days.} 
\label{fig:PS_filter_cycles}
\end{figure*} 

We compare the different filter systems and their effect on the measured variability for different cycles as observed by a solar equator-bound observer in Fig.~\ref{fig:PS_filter_cycles}. 
Interestingly, the shapes of the power spectra are very similar on timescales below about a year.
On timescales below 1 year, the variability in the two narrow-band Strömgren filters shows the highest power,  followed by \textit{Kepler}, whereas the brightness variations as they would be observed by TESS show the lowest amplitude.

On timescales above one year the variability in the \textit{Kepler}, TESS and Strömgren y passband have similar strength, whereas the signal in Strömgren b is considerably lower.
 For the Strömgren b filter, \cite{Shapiro2016} have found that the facular and spot contributions to the variability almost cancel each other out, hence, the variability is low. The compensation is less pronounced in the other passbands.

\section{Solar brightness variations as they would be seen from out of ecliptic}\label{inclination}

In the following, we refer to the inclination as the viewing angle of the observer with respect to the solar rotation axis. An inclination of 90$^\circ$ corresponds to an observer in the solar equatorial plane, while inclinations of \textless 90$^\circ$ refer to a displacement of the observer from the equatorial plane towards the North pole.

\subsection{Effect of inclination on brightness variability}\label{North}

\begin{figure*}
\centering
\includegraphics[width=\textwidth]{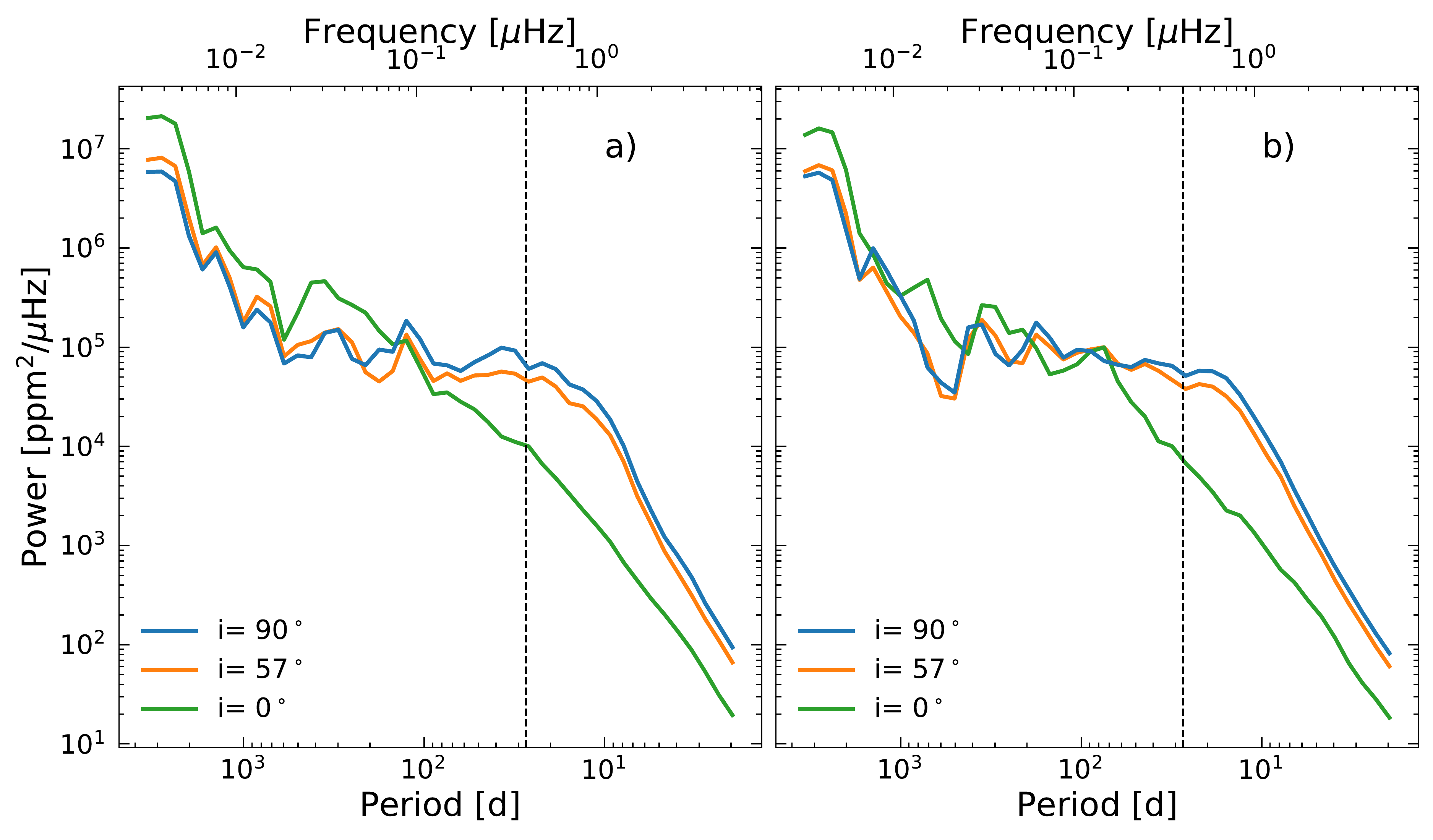}
\caption{Power spectra of solar brightness variations in the \textit{Kepler} passband with at different inclinations and two different cycles. Panel a) shows cycle 19 and b) cycle 22. The vertical dashed black lines indicate the synodic solar rotation period at 27.3 days.} 
\label{fig:PS_Kepler_TESS}
\end{figure*}

We now consider the variability during cycles 19 and 22 as it would be observed by \textit{Kepler}. The power spectra of brightness variations as they would be seen at  90$^\circ$ (i.e. from the equatorial plane), at 57$^\circ$ (which is the mean value of the inclination  for a random distribution of orientations of rotation axes), and at 0$^\circ$ (i.e. the view at the solar North pole) are plotted in Fig.~\ref{fig:PS_Kepler_TESS}.

\begin{figure*}
\centering
\includegraphics[width=\textwidth]{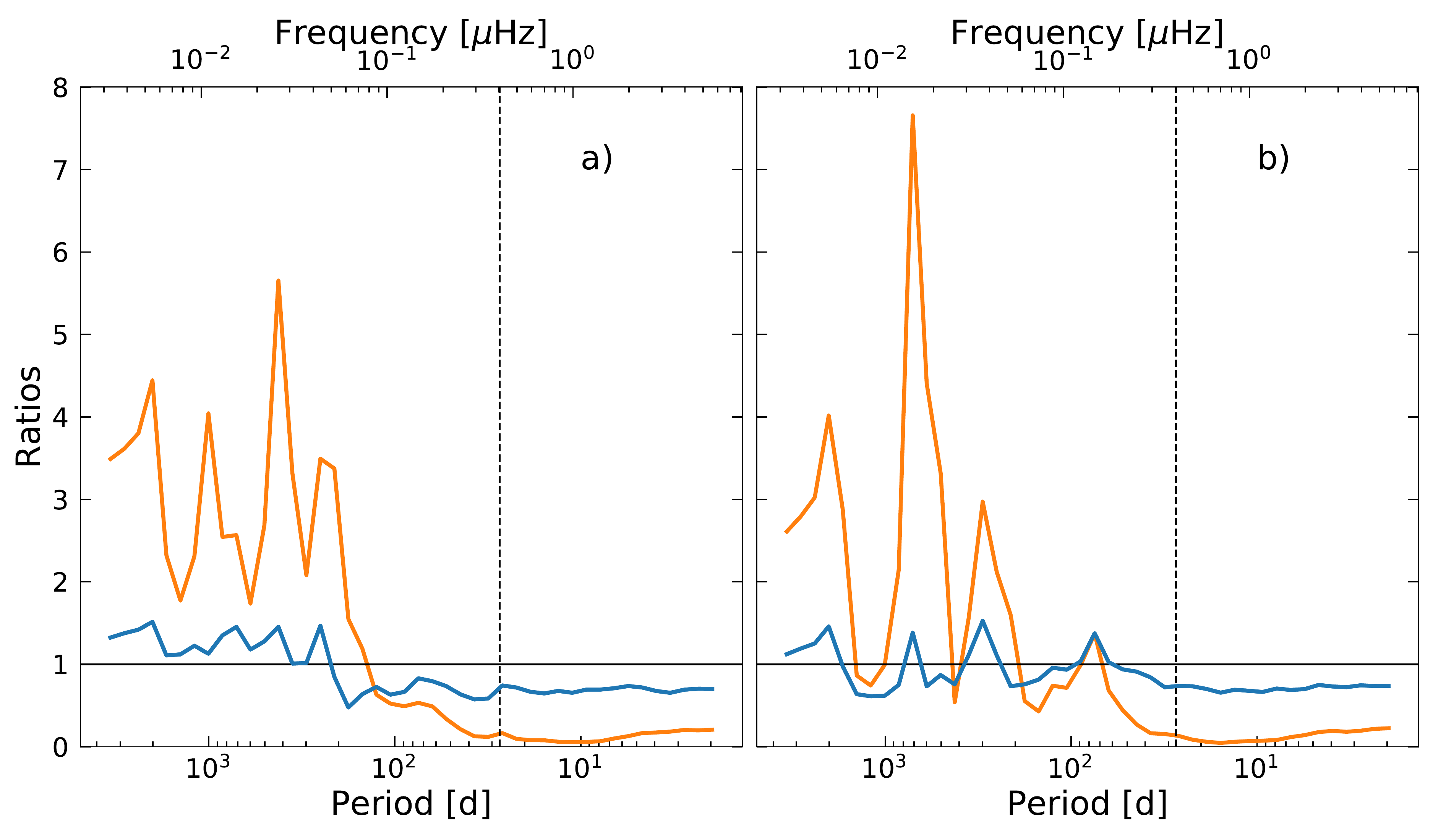}
\caption{Ratios of the power spectra of solar brightness variations for \textit{Kepler} as shown in Fig.~\ref{fig:PS_Kepler_TESS}. Blue lines represent the ratios of 57$^\circ$ to 90$^\circ$ and orange lines between 0$^\circ$ and 90$^\circ$. Panel a) shows cycle 19 and b) cycle 22. The vertical dashed black line indicates the synodic solar rotation period at 27.3 days. The horizontal solid black line indicates a ratio of 1.} 
\label{fig:PS_incl_ratios}
\end{figure*}

The power at the rotational timescale drops with decreasing inclination, but the variability on the activity timescale increases. This effect is not strong between 90 and 57$^\circ$ inclination, but significant between 90 and 0$^\circ$.
 Interestingly, the double-peak structure of cycle 19 that has been described before for the ecliptic-bound observer, is also present for the inclination of 57$^\circ$, although the peaks are less pronounced. 
 For the observer at 0$^\circ$, the power in the signal below 100 days is significantly lower than for the 90 and 57$^\circ$ vantage point. However, on timescales longer than 100 days, the power becomes higher compared to the other vantage points. We discuss this result in more detail in Sect.~\ref{disentangle}. 
 We also show the power spectra of brightness variations as observed by TESS and in the two Strömgren filters in the Appendix (Fig.~\ref{fig:PS_TESS}--\ref{fig:PS_Str_y}) for cycle 19 only. 

The impact of the inclination on the power spectrum becomes more evident in Fig.~\ref{fig:PS_incl_ratios}, where we show the ratios between the power as it would be measured at inclinations of 57$^\circ$ and 0$^\circ$ relative to that obtained by an ecliptic-bound observer.
In agreement with Fig.~\ref{fig:PS_Kepler_TESS} the power on timescales below 200 days decreases with decreasing inclinations, whereas longward of 200 days the power increases with decreasing inclination.
The reason for the increase of the variability is due to several effects. Most noteworthy are the effects of foreshortening and centre-to-limb variations (CLV). In the wavelength regime where \textit{Kepler} operates, the facular contrast (compared to the quiet Sun) is higher at the limb due to limb-darkening, whereas the spot contrast is the strongest at disc centre, as seen by an ecliptic bound observer. With decreasing inclination, the effect of CLV on the facular component is less pronounced and the facular contribution to the brightness variations is increasing (conversely, the effect of the spots is decreasing). While the effect of foreshortening is decreasing with decreasing inclination, it is not enough to compensate for the stronger contrast of the faculae.
For a more detailed discussion see \cite{Shapiro2016}. The distribution of the magnetic features (in particular the spot distribution) is also important, as we discuss in the next section.

\subsection{Disentangling evolution and rotation of magnetic features}\label{disentangle}

\begin{figure*}
\centering
\includegraphics[width=\textwidth]{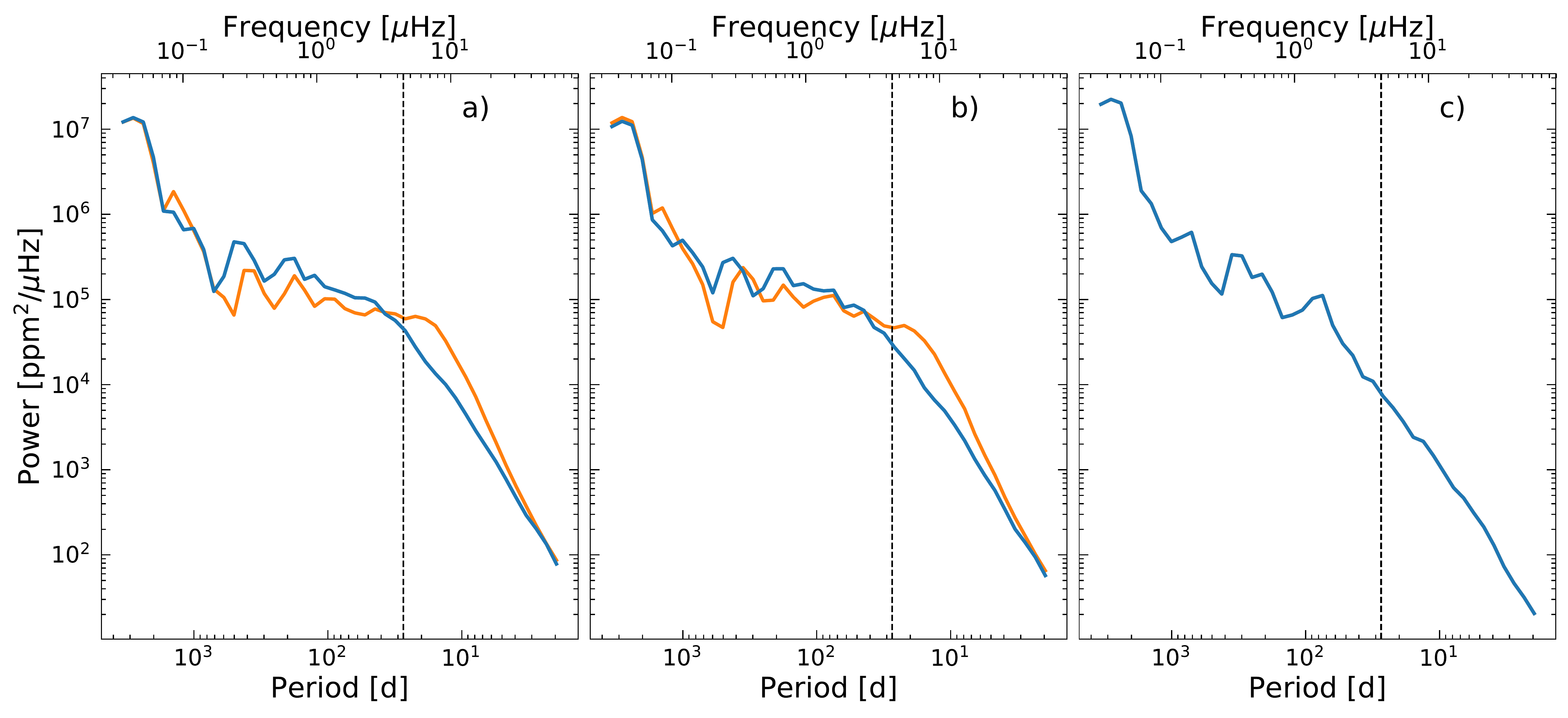}
\caption{Comparison of the power spectra of the solar brightness variations in the  \textit{Kepler} passband, with and without taking the solar rotation into account (orange and blue, respectively). Panel a) shows 90$^\circ$ inclination, b) 57$^\circ$ and c) 0$^\circ$. The vertical dashed black lines indicates the synodic solar rotation period at 27.3 days.} 
\label{fig:PS_effect_nonrot}
\end{figure*}
The solar brightness variability is caused by changes in the solar disc coverage by magnetic features. These changes are in turn due to 
(1) emergence and evolution of magnetic features and (2) the solar rotation,  which causes transits of individual magnetic features across the visible solar disk \citep[see, e.g.][and references therein]{Solanki1984}.  
Our model allows us to pinpoint the contribution of the solar rotation to the solar brightness variability. This can be done by disregarding the free term in Eq.~(\ref{diff_rot}), i.e. by looking at the non-rotating Sun from a fixed direction. We note that by doing this we still preserve the differential rotation term.

In Fig.~\ref{fig:PS_effect_nonrot} we compare the power spectra of solar brightness variations over cycle 22 calculated with and without taking solar rotation into account (orange and blue lines, respectively).
Figure~\ref{fig:PS_effect_nonrot} a shows power spectra as recorded by an ecliptic-bound observer. The solar rotation does not play a substantial role at timescales below about four to five days (the orange and blue lines in Fig.~\ref{fig:PS_effect_nonrot} a are very close to each other). The variability at such timescales is apparently due to the evolution of individual magnetic features. The variability at timescales between five days and the solar rotation period is mainly due to the solar rotation. Interestingly, while the rotation itself becomes unimportant at timescales above 
the rotation period the two power spectra are still different up to the timescale of about four to five years. This is because the variability of the rotating Sun is determined by the longitudinal-averaged distribution of magnetic features. The variability of 
the non-rotating Sun is given by the distribution seen from a fixed vantage point.
Since the emergence of magnetic features is random over longitude, the two described distributions are the same if averaged over a sufficiently long time interval (so that blue and orange lines almost coincide 
at timescales larger than four to five years). At the same time, at timescales shorter than four to five years, the distributions might still be different since they depend on the specific realisation of emergences of magnetic features. Consequently, this part of the power spectrum depends on the 
specific longitudinal location of the vantage point.

Figure~\ref{fig:PS_effect_nonrot} b illustrates the case of a 57$^\circ$ inclination, which looks very similar to the case of the ecliptic-bound observer. Fig.~\ref{fig:PS_effect_nonrot} c  represents the view from the observer located over the solar North pole.
Naturally, the solar rotation does not contribute to the brightness variability as it is determined solely by the evolution of the magnetic features and the modulation of their emergence rate over the solar activity cycle. Therefore, the blue and orange curves 
in Fig.~\ref{fig:PS_effect_nonrot} c coincide at all timescales.

Fig.~\ref{fig:PS_effect_nonrot} allows us to better understand the 
origin of the decrease of short-timescale variability  with decreasing inclination as seen in Figs.~\ref{fig:PS_Kepler_TESS}--\ref{fig:PS_incl_ratios}. The emergence of active regions is confined to about $\pm$ 30--40$^\circ$ centred around the equator. Consequently, even though the variability at timescales shorter than four to five days is not affected by the solar rotation, it is strongly decreased due to the effect of foreshortening.

\subsection{The full time series}

\begin{figure}
\centering
\includegraphics[width=0.75\textwidth]{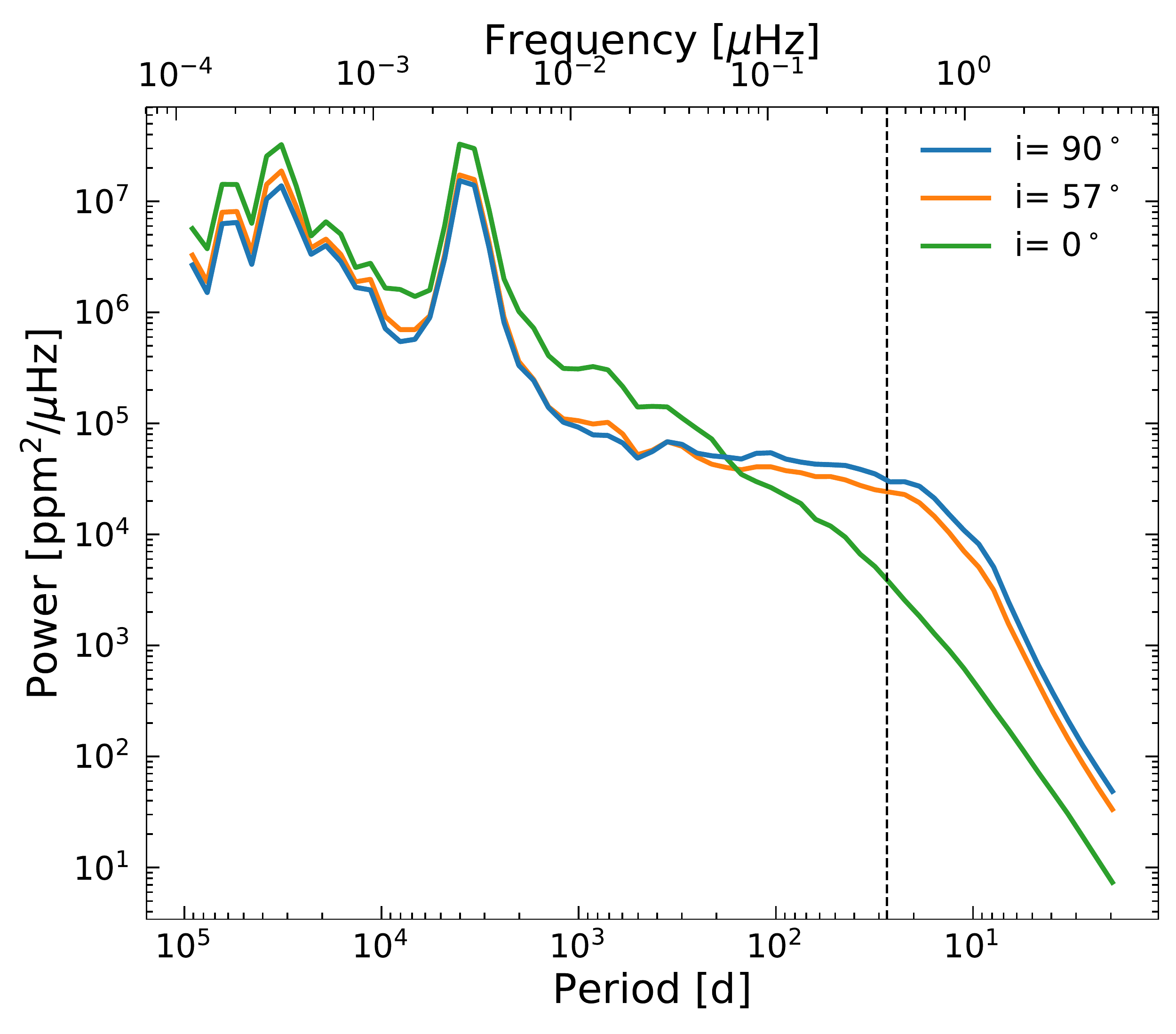}
\caption{Power spectra of the solar brightness variations for the full time series of over 300 years as it would be observed by \textit{Kepler} for different inclinations. The vertical dashed black lines indicates the synodic solar rotation period at 27.3 days.} 
\label{fig:PS_full_series}
\end{figure}

In the previous sections, we limit our analysis to selected individual solar activity cycles. The source term used in the SFTM provides information from 1700 to 2009. We now consider the solar brightness variations for this whole interval, with respect to different inclinations, limiting ourselves to calculating solar brightness variation in the \textit{Kepler} passband, which we present in Fig.~\ref{fig:PS_full_series}. 
The differences in the power spectra between the 90$^\circ$ and 57$^\circ$ degree vantage point are small (shown earlier in the paper), whereas the difference between 90 and 0$^\circ$ is pronounced.
On the timescale above one year, the variability as observed from an inclination of of 0$^\circ$ becomes stronger, due to the stronger facular contribution to the solar brightness variations.

For all inclinations, a pronounced peak at around 10.8 years is visible, which corresponds to the average length of a cycle in our considered sample.
On the rotational timescale, however, no peak is seen and no peaks above or below the rotation period appear. 

\section{Conclusions and outlook}\label{conclusion}
We employed the Surface Flux Transport model \citep[SFTM, in the form of][]{Cameron2010} with the source term from \cite{Jiang2011_1} to obtain the distribution of magnetic flux on the entire solar surface. This distribution was then  converted into surface area coverages of solar magnetic  features and the SATIRE approach was utilised for calculating brightness variations. This allowed us to model the brightness variability of the Sun at different activity levels as it would be seen from any arbitrary vantage point and in  different filter systems.
 
 We analysed the dependence of the power spectrum of solar brightness variations on the inclination. While the decrease of the inclination leads to an increase of the variability on the timescale of the solar activity cycle, the variability decreases at shorter timescales. In particular, it decreases on the timescale of solar rotation. Since the Sun is always seen equator-on, its variability is higher than of another star with the same activity level, but seen from a higher latitude. Consequently the higher variability of solar-like stars cannot be due to the inclinations of their rotation axis alone. The effect of the inclination strengthen the conclusions of \cite{Timo2020} that stars with near-solar fundamental parameters and rotation periods have on average significantly higher variability on the solar rotation timescale than the Sun. 

 Our calculations also indicate that the power spectrum of solar brightness variations does not have a clear peak at the rotation period, not only for the ecliptic-bound observer \citep[see][]{Shapiro2017, Veronika_rot}, but also for the out-of-ecliptic observer. This factor might play an important role in explaining the deficiency of stars with detected near-solar rotation periods \citep[see][]{VanSaders2019, Veronika_rot}.

 Our model also allowed us to decompose the contributions of solar rotation and evolution of magnetic features into solar brightness variability.  In particular, we have shown that the variability on timescales below five days is mainly due to the evolution of magnetic features and not due to the solar rotation.

The SFTM model is also capable of simulating stars more active than the Sun \citep{Isik2018}, so  we plan to extend the present study to model brightness variations of stars that are more active than the Sun. In combining it with the results of \cite{witzkeetal2018, Veronika_rot}, we also plan to extend the model to stars with different fundamental parameters.

%\begin{acknowledgements}
%The research leading to this paper has received funding from the European Research Council under the European Union’s Horizon 2020 research and innovation program (grant agreement No. 715947). It also got financial support  from the BK21 plus program through the National Research Foundation (NRF) funded by the Ministry of Education of Korea. We would like to thank the International Space Science Institute, Bern, for their support of science team 446 and the resulting helpful discussions.
%\end{acknowledgements}

%-------------------------------------------------------------------
%\bibliographystyle{aa}
%\bibliography{bib}

%\begin{appendix}

\chapter{Connecting measurements of solar and stellar brightness variations}\label{sec:paper_2}

The contents of this section are identical to the printed version of  N\`{e}mec, N.-E., et al.,  A\& A, 638, A56,2020, reproduced with permission \textcopyright ESO. DOI:10.1051/0004-6361/202038054 \\
\textbf{Contributions to the paper:} I produced the results and provided the main scientific interpretation.
%wrote the first draft of the manuscript, performed the calculations and the data analysis, partially utilising scripts that were provided by E. I. The manuscript has been read and approved by all authors.

\section{Introduction}
Dedicated planet-hunting photometric missions such as CoRoT \citep[Convection, Rotation and planetary Transit, see][]{COROT,COROT2}, \textit{Kepler} \citep{KEPLER}, and TESS \citep[Transiting Exoplanet Survey Satellite, see][]{TESS}, and also the \textit{Gaia} space observatory \citep{Gaia2016} have made it possible to measure stellar brightness variability with unprecedented precision. In particular, they allow studying stellar brightness variations caused by transits (as the star rotates) and the evolution of magnetic features, that is, bright faculae and dark spots. Such variations are often referred to as rotational stellar variability. 
The plethora of stellar observational data rekindled an interest in the questions of how typical our Sun is as an active star, and more specifically, how the solar rotational variability compares to that of solar-like stars. Furthermore, these data allow
probing whether the solar activity paradigm is also valid for other stars. This requires comparing the stellar properties and behaviour with those of the Sun.  
While the solar variability has been measured for more than four decades now by various dedicated space missions  \citep[see e.g.][for reviews]{Froehlich2012,Ermolli2013,Solanki2013, Greg2016},
a comparison between solar and stellar brightness measurements is far from straightforward \citep[see e.g.][]{Basri2010,Timo2020,Witzke2020}. Firstly, solar and stellar brightness variations  have been measured in different spectral passbands. Because the amplitude of the solar rotational variability strongly depends on the wavelength \citep{Solanki2013,Ermolli2013}, the solar and stellar brightness records can be reliably compared only after conversion from one passband to another. Secondly, the solar brightness variations have (so far) only been measured from the ecliptic plane,  which is very close to the solar equatorial plane (the angle between the solar equator and ecliptic plane is about 7.25$^\circ$). The values of the angle between the line of sight of the observer and the rotation axes of the observed stars (hereinafter referred to as the inclination) are mostly unknown. 

Studies comparing solar and stellar rotational brightness variations have used different types of solar brightness measurements. \cite{Timo2020}, for instance, used the total solar irradiance (TSI), that is, the solar radiative flux at 1~AU integrated over all wavelengths. More commonly, however, the solar variability was characterised \citep[see e.g.][]{Basri2010,Gilliland2011,Harrison2012} using measurements  by the Variability of solar IRradiance and Gravity Oscillations / Sun PhotoMeters (VIRGO/SPM) \citep[][]{Froehlich1995,Froehlich1997} instrument on board the Solar and Heliospheric Observatory (SoHO). VIRGO/SPM measures solar brightness in three filters with a bandwidth of 5 nm each. Neither VIRGO/SPM nor TSI measurements can be directly compared to records of stellar brightness variability, which typically cover wavelength ranges broader than the VIRGO/SPM filters, but much narrower than the TSI. Accurate estimations of solar variability in passbands used for stellar measurements have therefore so far been missing.
Some effort has previously been made to model the solar rotational variability as it would be observed out of ecliptic \citep[e.g.][]{Vieira2012,Shapiro2016,Nina1}. In particular, 
\cite{Shapiro2016} and \cite{Nina1} (hereinafter N20) have shown that the amplitude of the solar brightness variations on the rotational timescale decreases with decreasing inclination. Because of its almost equator-on view, the Sun would therefore appear on average more variable than stars with the same activity level that are observed at random inclinations. At the same time, an easy-to-use receipt for correcting the variability for the inclination effect  is lacking so far, and consequently,  the inclination has not yet been quantitatively accounted for in solar-stellar comparison studies.

In this paper we seek to overcome these two hurdles and quantify solar variability in passbands that are used by different stellar space missions and at different inclinations. In Sect.~\ref{Equator} we employ the spectral and total irradiance reconstruction \citep[SATIRE;][]{Fligge2000,Krivova2003} model of solar brightness variations to show how the actual solar brightness variations are related to solar brightness variations as they would be observed in spectral passbands used by stellar missions.  We also establish the connection between the TSI and VIRGO/SPM measurements. In Sect. \ref{Inclination} we follow the approach developed by N20 to quantify the effect of the inclination on the brightness variations. We discuss how the Sun as observed by \textit{Kepler} can be modelled using light curves obtained by VIRGO/SPM in Sect. \ref{Kepler-VIRGO} 
before we summarise our results and draw conclusions in Sect.~\ref{Conclusion}.

\section{Conversion from solar to stellar passbands}\label{Equator}

\subsection{SATIRE-S}
The SATIRE model \cite[][]{Fligge2000,Krivova2003} attributes the brightness variations of the Sun on timescales longer than a day to the presence of magnetic features on its surface, such as bright faculae and dark spots. The two main building blocks of SATIRE are the areas and the positions of the magnetic features on the solar disc as well as contrasts of these features relative to the quiet Sun (i.e. regions on the solar surface free from any apparent manifestations of magnetic activity).
The contrasts of the magnetic features as a function of disc position and wavelength  were computed
by \cite{Unruh1999} with the spectral synthesis block of the  ATLAS9 code \citep{Kurucz1992, Castelli1994}. The 1D atmospheric structure of the two spot components (umbra and penumbra) and of the quiet Sun were calculated using radiative equilibrium models produced with the ATLAS9 code, while the facular model is a modified version of FAL-P by \cite{Fontenla1993}.

Various versions of the SATIRE model exist. In this section we employ the most precise version, which is SATIRE-S, where the suffix ``S'' stands for the satellite era \cite[][]{Ball2014,Yeo2014}.  SATIRE-S uses the distribution of magnetic features on the solar disc obtained from observed magnetograms and continuum disc images and spans from 1974 to today, covering four solar cycles. As especially the early ground-based  observations contain gaps in the data, we used the SATIRE-S model as presented by \cite{Yeo2014} (version 20190621), where the gaps in spectral solar irradiance (SSI) and TSI have been filled using the information provided by solar activity indices. SATIRE-S was shown to reproduce the apparent variability of the Sun as observed, in both the SSI and in the TSI \cite[see][and references therein]{Ball2012,Ball2014,Yeo2014,Danilovic2016}.
The spectral resolution of the SATIRE output is 1 nm below 290 nm, 2 nm between 290 nm and 999 nm, and 5 nm above 1000 nm. This is fully sufficient for the calculations presented in this study.

\subsection{Filter systems}\label{Filters}
In this section we multiply the SATIRE-S SSI output with  the response function of a given filter and integrate it over the entire filter passband to obtain the solar light curve in the corresponding filter. It is important to take the nature of the detectors used in different instruments into account \citep[see e.g.][]{Maxted2018}.
In particular, while solar instruments (e.g. VIRGO/SPM and all TSI instruments) measure the energy of the incoming radiation, charge-coupled devices (CCDs) used in \textit{Kepler},  \textit{Gaia}, and TESS count the number of photons  and not their energy. In order to obtain the solar light curve, $LC$, as it would be measured  by the instrument counting photons, we therefore follow
\begin{equation}
LC =  \int\limits_{\lambda_1}^{\lambda_2} R(\lambda) \cdot I(\lambda)  \frac{\lambda}{h\cdot c} \, d\lambda,
\label{eq:filter}
\end{equation}
\noindent where $\lambda_1$ and $\lambda_2$ are the blue and red threshold wavelengths of the filter passband, $R(\lambda)$ is the response function of the filter, and $I(\lambda)$ is the spectral irradiance at a given wavelength, $h$ is the Planck constant, and $c$ the speed of light.

\begin{figure}
\centering
\includegraphics[width=0.6\columnwidth]{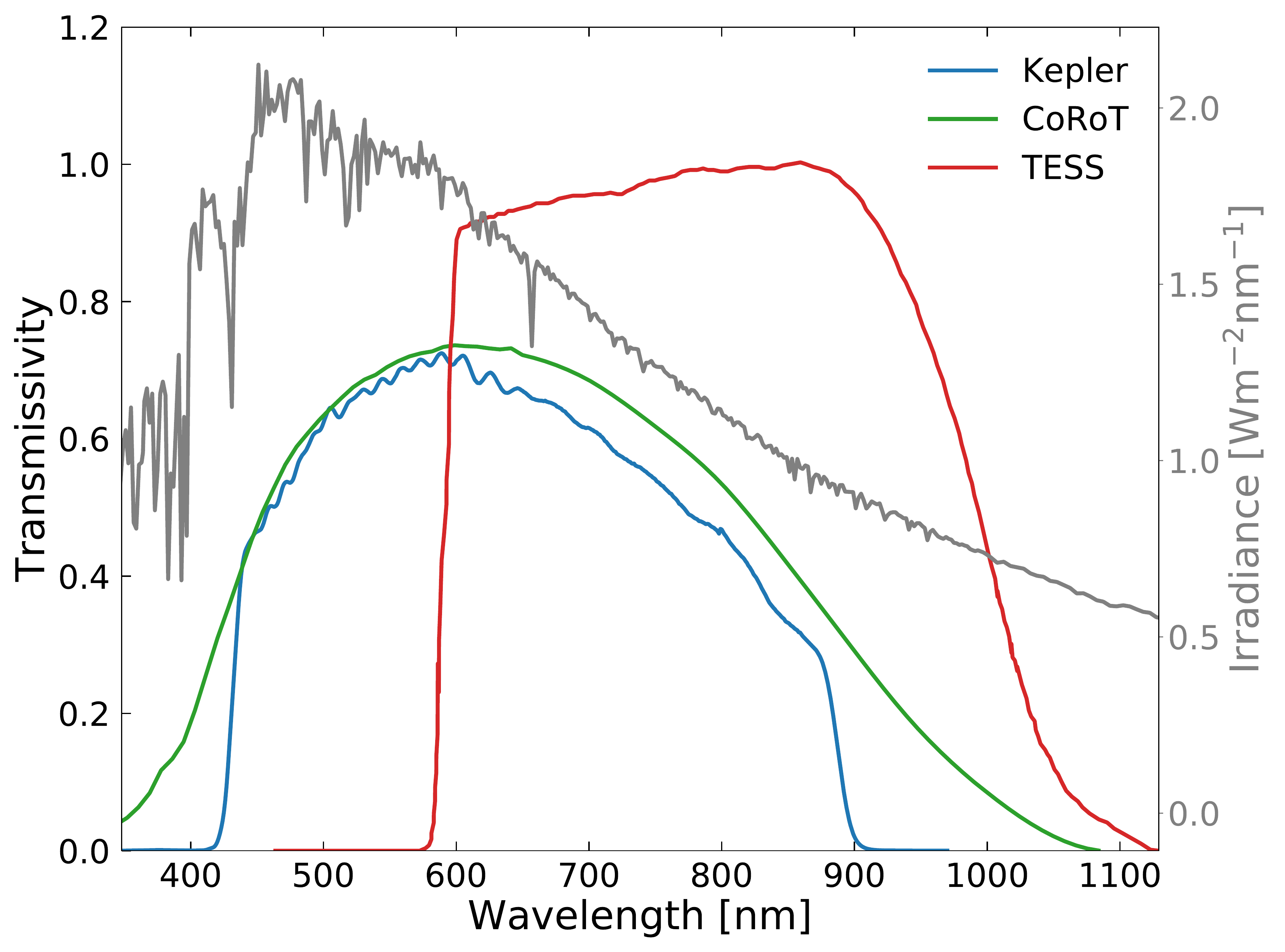}
\includegraphics[width=0.6\columnwidth]{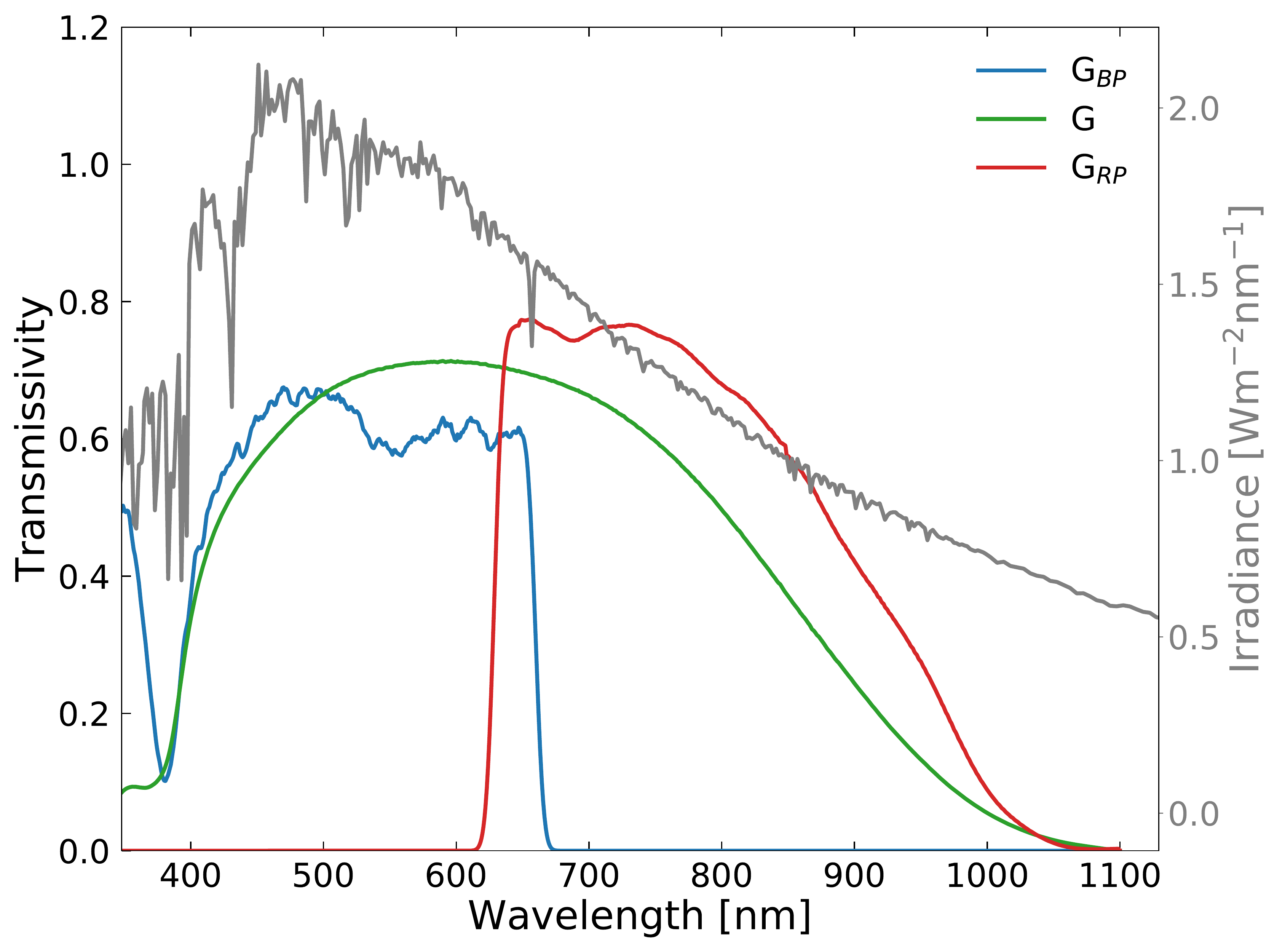}
\includegraphics[width=0.6\columnwidth]{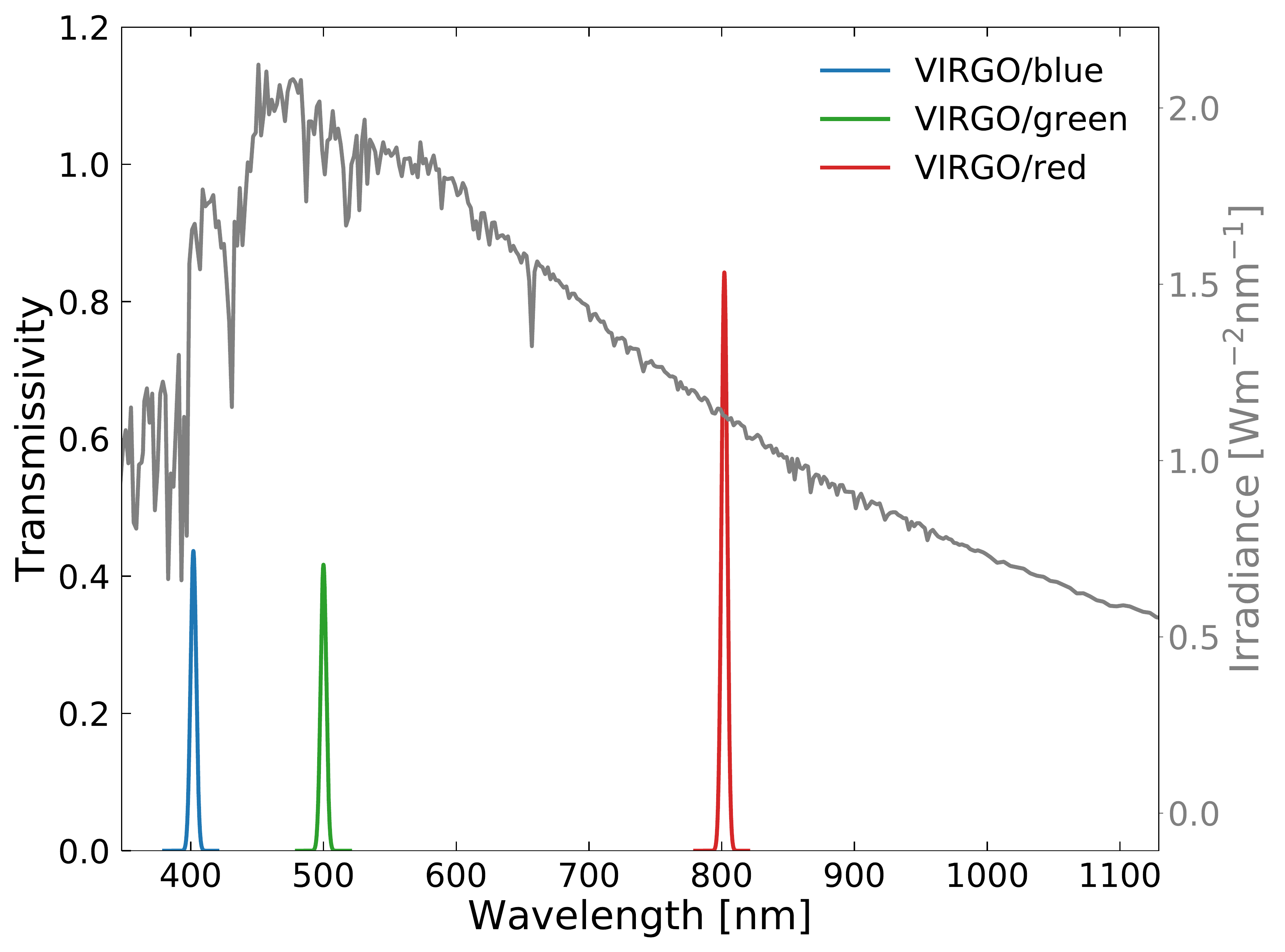}
\caption{Response functions for the various filter systems used in this study. For comparison, the quiet-Sun spectrum used by SATIRE-S is plotted in grey in each panel. Top panel: \textit{Kepler}, TESS, and CoRoT. Middle panel: Three Gaia passbands. Bottom panel: Three VIRGO/SPM channels.}
\label{fig:filters}
\end{figure}

First we consider several broad-band filters used by the planet-hunting missions:  CoRoT, \textit{Kepler}, and TESS. The spectral passbands employed in these missions are shown in the top panel of Fig.~\ref{fig:filters}, along with the quiet-Sun spectrum calculated by \cite{Unruh1999} and used in SATIRE-S. Clearly, the  CoRoT and \textit{Kepler} response functions are very similar to each other because both missions focused on G stars. TESS is designed to observe cooler stars than \textit{Kepler}, hence the response function is shifted towards the red part of the spectrum.

\textit{Gaia} measures stellar brightness in three different channels \citep{Gaia2016}. \textit{Gaia G}  is sensitive to photons between 350 and 1000 nm. Additionally, two prisms disperse 
 the incoming light between 330 and 680 nm for the Blue Photometer (hereafter referred to as \textit{Gaia G$_{BP}$}) and between 640--1050 nm for the Red Photometer (hereafter, referred to as \textit{Gaia G$_{RP}$}). The response functions are shown in the middle panel in Fig.~\ref{fig:filters}. We employ the revised passbands used for the second data release of Gaia \citep[Gaia DR2,][]{Evans2018} for the calculations.

Solar-stellar comparison studies have often used the solar variability as measured by the VIRGO/SPM instrument. SPM comprises three photometers, with a bandwidth of 5 nm operating at 402 nm (blue), 500 nm (green), and 862 nm (red). The response functions are shown in Fig.~\ref{fig:filters} in the bottom panel. We refer to these filters from now on as VIRGO-blue, -green, and -red.

\subsection{Results} \label{results_eq}

\begin{figure*}
\centering
\includegraphics[width=\textwidth]{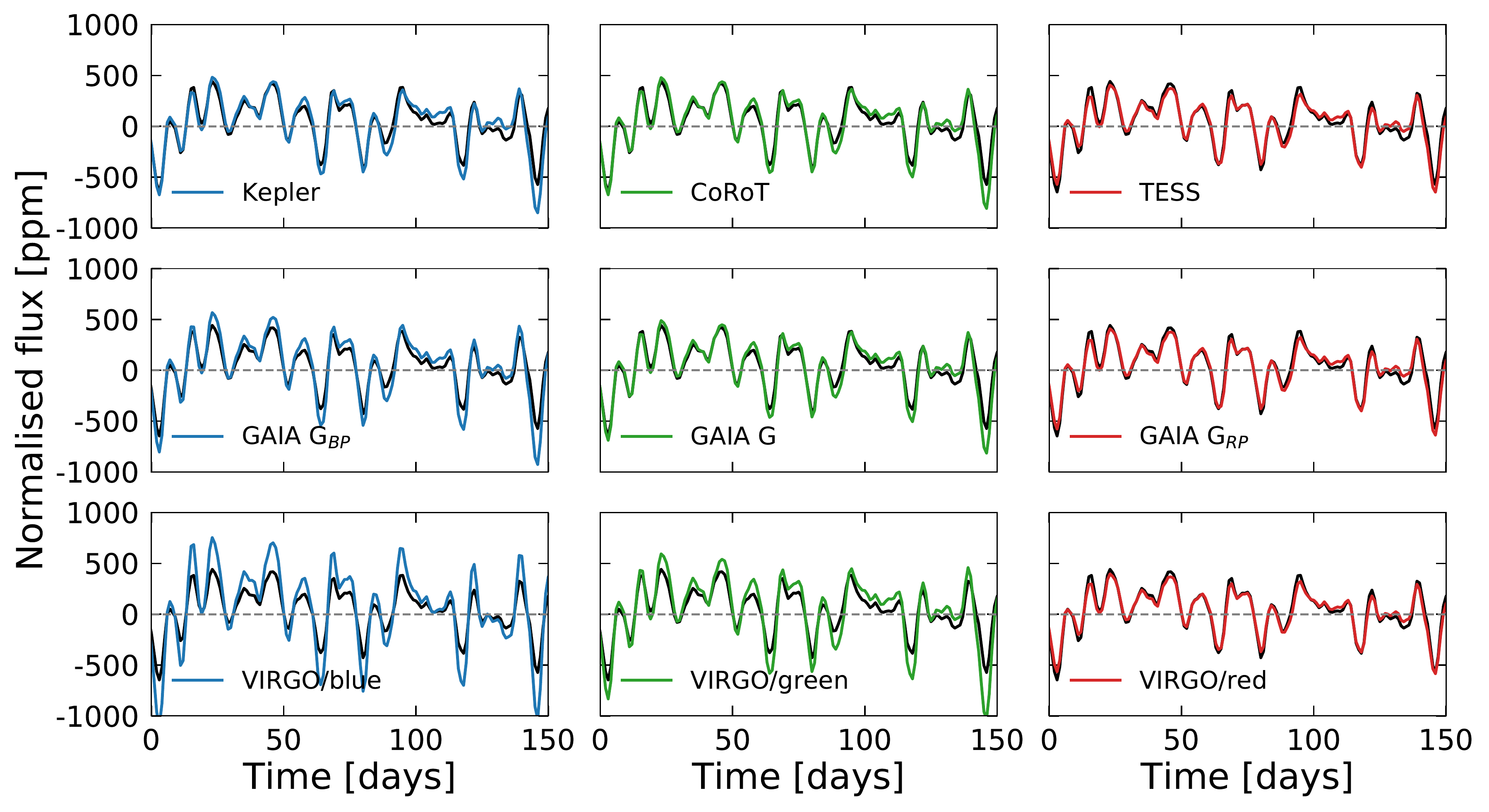}
\caption{Normalised fluxes for the different filter systems compared to the TSI (black solid line). Top panels: \textit{Kepler}, CoRoT, and TESS. Middle panels: Three \textit{Gaia} passbands. Bottom panels: Three VIRGO/SPM channels.}
\label{fig:filters_LC}
\end{figure*}

\begin{figure*}
\centering
\includegraphics[width=\textwidth]{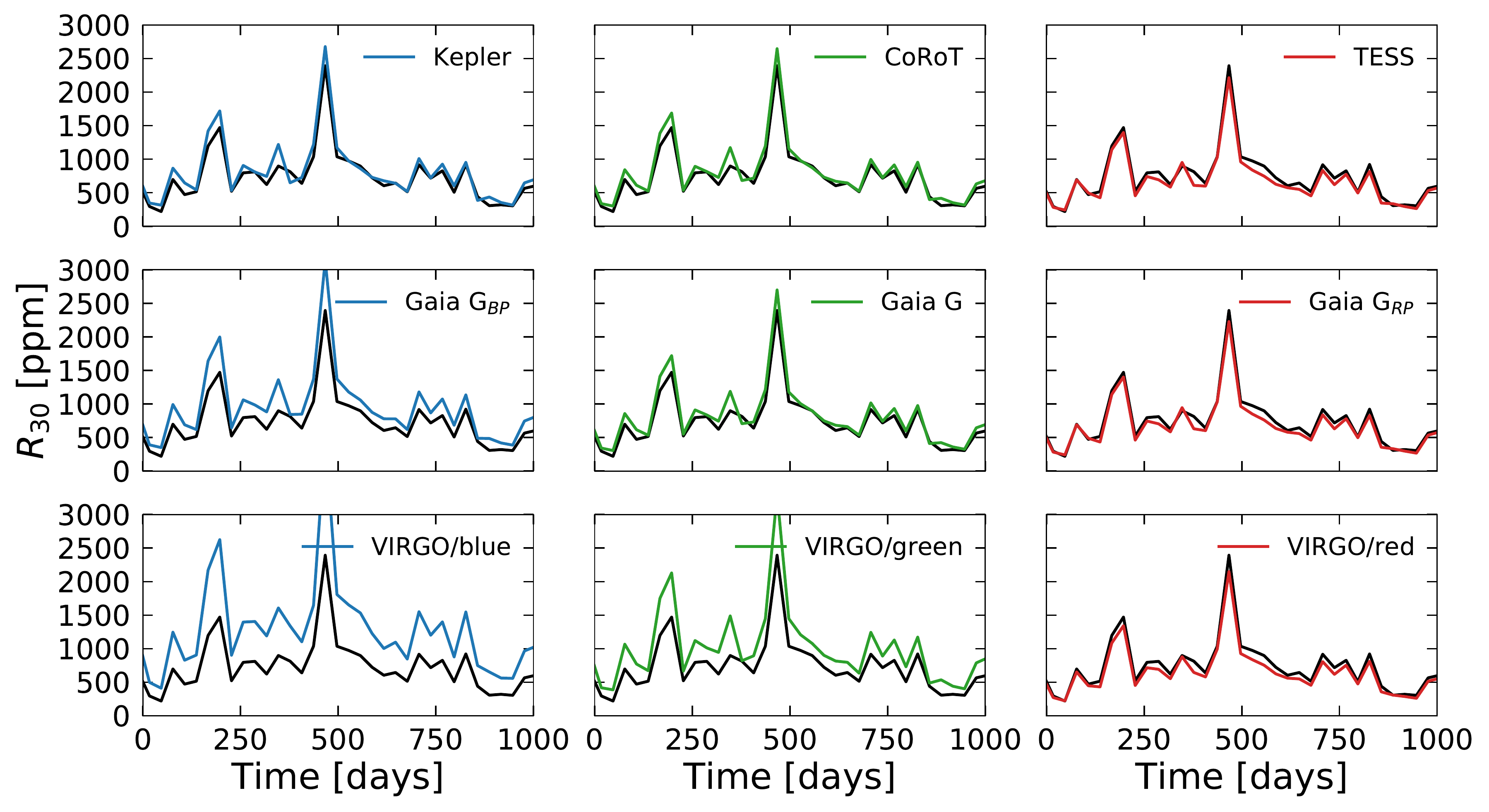}
\caption{$R_{30}$ in different filter systems compared to the TSI (black solid line) for a time span of 1000 days over solar cycle 22. Top panels: \textit{Kepler}, CoRoT, and TESS. Middle panels: Three \textit{Gaia} passbands. Bottom panels: Three VIRGO/SPM channels. See the main text for the definition of $R_{30}$.} 
\label{fig:filters_R_30}
\end{figure*}

\begin{figure*}
\centering
\includegraphics[width=\textwidth]{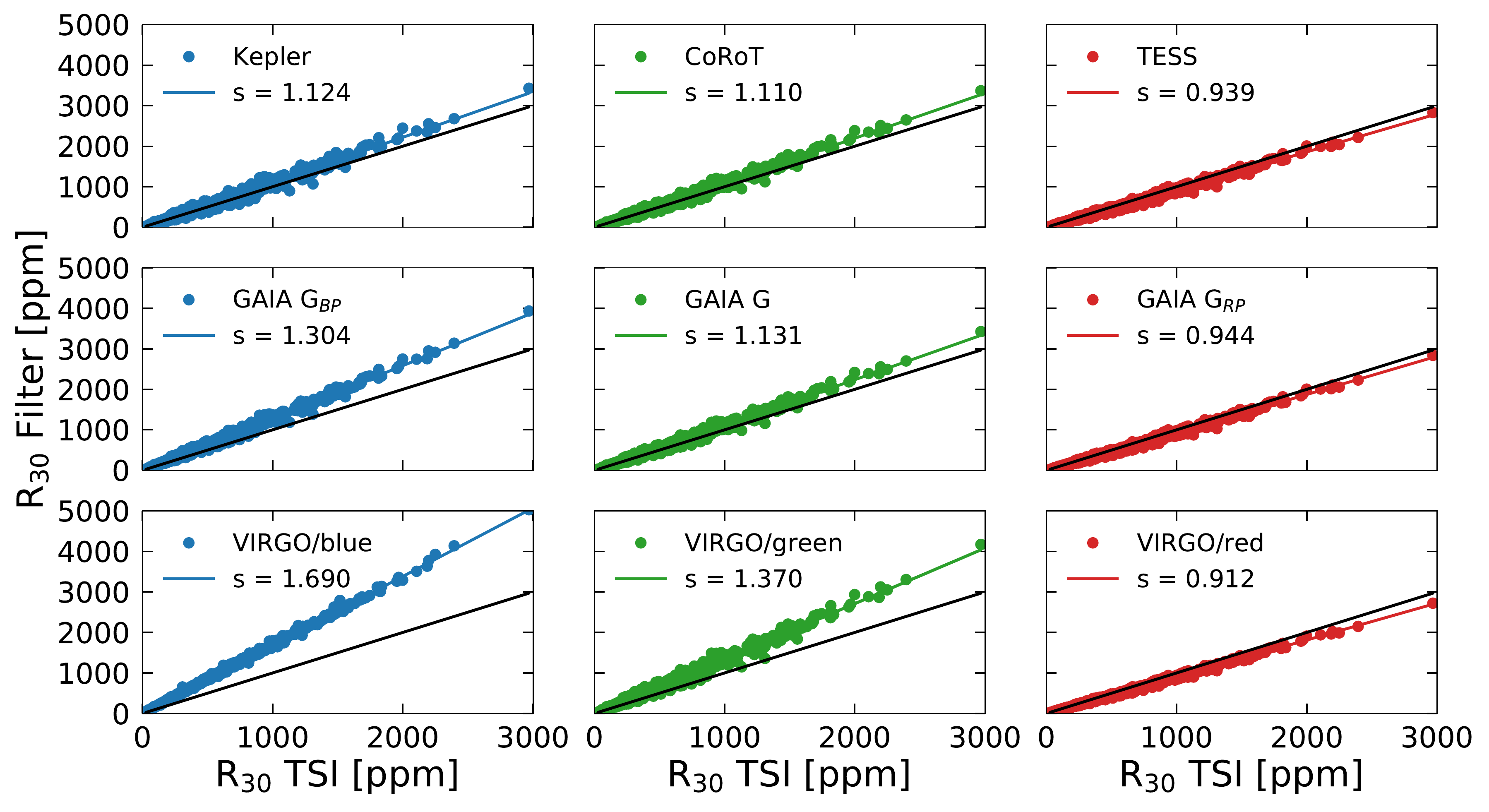}
\caption{Linear regression between the $R_{30}$ values calculated with the TSI and with the solar light curves as they would be recorded in different filter systems. The values of the slope, $s$, are given in the legend. The black lines have a slope equal to 1.}
\label{fig:R_30_regressions}
\end{figure*}

Figure~\ref{fig:filters_LC} shows the solar light curve for the period of 2456700 -- 2456850 JD (24 February 2014 -- 11 July 2014) as it would be observed in different passbands. This corresponds to a 150-day interval during the maximum of cycle 24. This interval was chosen arbitrarily to display the effect of the filter systems on the solar variability. For this, we first divided each light curve in 90-day segments. This time span corresponds to  \textit{Kepler} quarters. This is motivated by the way \textit{Kepler} observations are gathered and reduced. We note that the detrending by the \textit{Kepler} operational mode is applied here for purely illustrative purposes and was not used for the calculations presented below. Within each segment, we subtracted the mean value from the fluxes before dividing the corresponding values by the mean flux in each segment. In all stellar broad-band filters, the light curve is remarkably similar in shape to the TSI (solid black curve), although the amplitude can differ. As might be expected, the difference in the amplitude of the variability is somewhat more conspicuous in the blue filters. The $Gaia~G_{RP}$ light curve is basically identical to the TSI light curve, whereas the variability in $Gaia~G_{BP}$ and in the narrow VIRGO-blue filter show far stronger variability than the TSI.

To quantify the rotational variability, we computed the $R_{30}$ values \citep[see e.g.][]{Basri2013}. To do this, the obtained light curves were split into 30-day segments, and within each segment, we calculated the difference between the extrema and divided this value  by the mean flux in the segment to derive the relative variability. For the SATIRE-S time series, we directly considered the difference between the extrema instead of the differences between the 95th and 5th percentiles of sorted flux values, as is usually done in the literature with the more noisy \textit{Kepler} measurements. We calculated $R_{30}$ values for the period  1974--2019 (i.e. cycles 22--24). This allowed us to quantify the mean level of solar variability in $R_{30}$ that represents the full four decades of TSI measurements.

 \begin{table}
\begin{center}
\begin{tabular}{ll}
\hline \hline
                &       slope   \\
\hline
\textit{Kepler}             &   1.123 ($\pm$0.007) \\
CoRoT                           &       1.110 ($\pm$0.006) \\
TESS                            &       0.939 ($\pm$0.004) \\
\textit{Gaia G$_{BP}$}          &       1.304 ($\pm$0.005) \\
\textit{Gaia G}                 &       1.131 ($\pm$0.005) \\
\textit{Gaia G$_{RP}$}          &       0.944 ($\pm$0.004) \\
VIRGO/blue                      &       1.689 ($\pm$0.003) \\
VIRGO/green                     &       1.370 ($\pm$0.007) \\
VIRGO/red                           &   0.912 ($\pm$0.003) \\
\hline \hline
\end{tabular}
\end{center}
\caption{Slopes of the linear regressions in Fig. \ref{fig:R_30_regressions}.}
\label{tab:regressions}
\end{table}

 \begin{table}
\begin{center}
\begin{tabular}{llllll}
\hline \hline
        &       21      &       22      &       23      &       24      &       mean    \\
\hline
TSI     &       0.743   &       0.806   &       0.682   &       0.492   &       0.681   \\
\textit{Kepler} &       0.808   &       0.872   &       0.731   &       0.530   &       0.735   \\
CoRoT   &       0.801   &       0.866   &       0.726   &       0.526   &       0.730   \\
TESS    &       0.680   &       0.739   &       0.615   &       0.445   &       0.620   \\
\textit{Gaia G$_{BP}$}  &       0.944   &       1.020   &       0.861   &       0.625   &       0.862   \\
\textit{Gaia G} &       0.817   &       0.883   &       0.741   &       0.537   &       0.744   \\
\textit{Gaia G$_{RP}$}  &       0.684   &       0.742   &       0.617   &       0.447   &       0.623   \\
VIRGO/blue      &       1.252   &       1.352   &       1.167   &       0.846   &       1.154   \\
VIRGO/green     &       0.983   &       1.056   &       0.894   &       0.653   &       0.897   \\
VIRGO/red       &       0.665   &       0.722   &       0.600   &       0.435   &       0.606   \\
\hline \hline   
%\hline \hline
\end{tabular}
\end{center}
\caption{Cycle-averaged $R_{30}$ in mmag.}
\label{tab:mean_R_30}
\end{table}

In Fig.~\ref{fig:filters_R_30} we compare the $R_{30}$ values for all the filter systems introduced in Sect.~\ref{Filters} to $R_{30}$ of the TSI for a 1000-day interval starting 26 July 2013. This interval therefore includes the maximum of solar 24 as well.
To better quantify the dependence of $R_{30}$  on the passband, we show linear regressions between the variability $R_{30}$ in each filter system and the TSI in Fig.~\ref{fig:R_30_regressions}.
The slopes of the linear regressions are listed in Table \ref{tab:regressions}. The Pearson correlation coefficient is above 0.98 for all of the filter systems.
 The slope of the linear regressions depends on the filter system that is considered. For example, TESS and \textit{Gaia~G} regressions have a slope close to 1, but $G_{BP}$ displays a slope \textgreater 1, whereas VIRGO-red exhibits  a slope \textless 1. As expected, the slope is highest for the blue VIRGO filter, where the amplitude of the variability is highest.   We note that the good agreement of the TSI with the red filters is expected to be valid only for the rotational variability, which is dominated by spots. In contrast, the solar irradiance variability on the activity cycle timescale is given by the delicate balance between facular and spot components, and consequently has a very sophisticated spectral profile \citep{Shapiro2016,witzkeetal2018}. Thus, values of slopes from Table~\ref{tab:regressions} cannot be extrapolated from rotational to activity cycle timescales \citep[see][for the detailed discussion]{Shapiro2016}.

Table \ref{tab:mean_R_30} lists the cycle-averaged values of $R_{30}$ for all passbands in mmag. Together, Fig.~\ref{fig:R_30_regressions}, and Tables \ref{tab:regressions} and \ref{tab:mean_R_30} show that the TSI is a passable representative for the variability on the solar rotation timescale as it would be observed in the TESS, \textit{Kepler}, CoRoT, \textit{Gaia G}, \textit{Gaia G$_{RP}$} , and VIRGO-red filters, but it noticeably underestimates the variability in $G_{BP}$, VIRGO-green, and VIRGO-blue.

\begin{figure}
\centering
\includegraphics[width=0.75\textwidth]{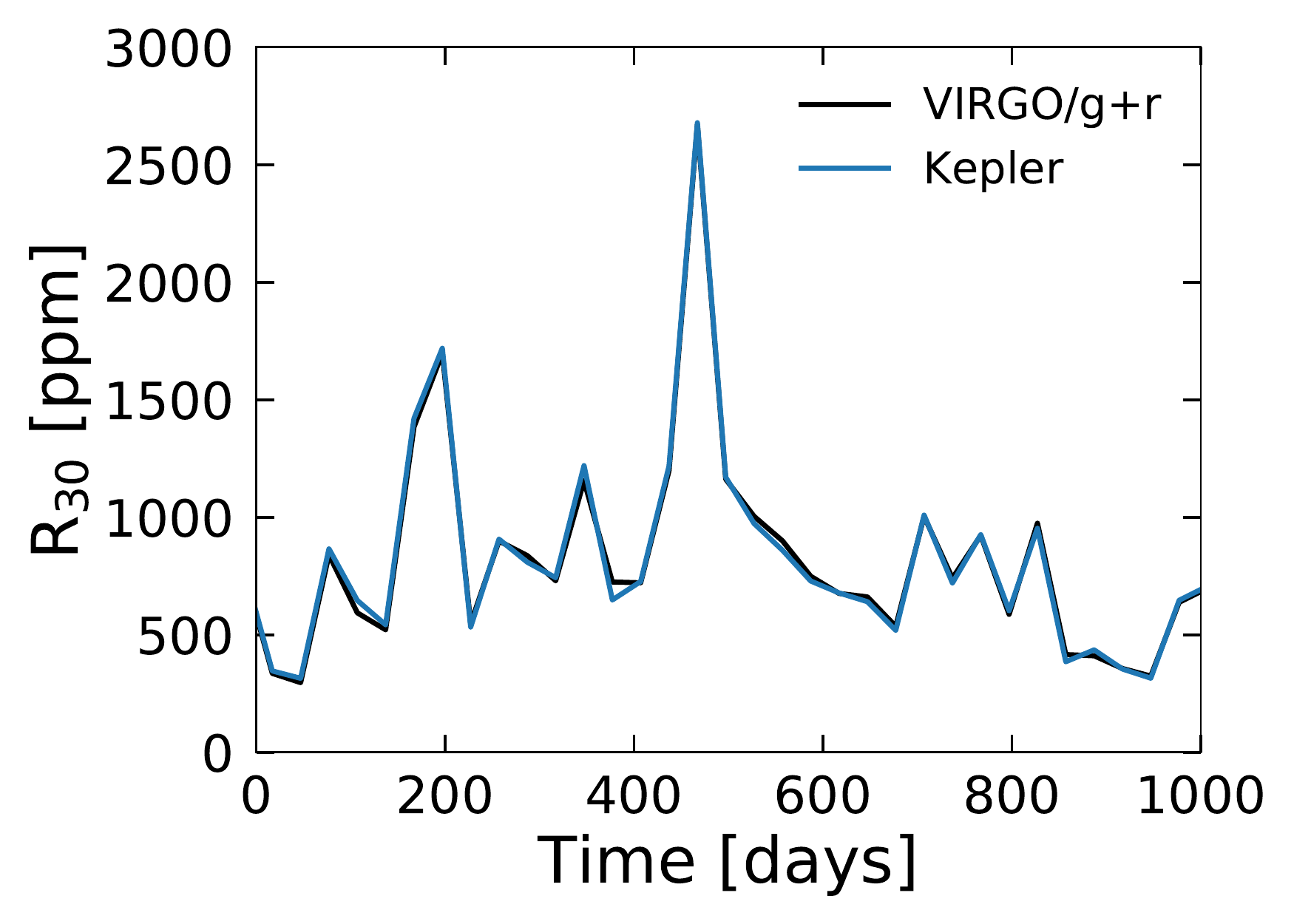}
\caption{$R_{30}$ over cycle 22 for VIRGO/g+r and \textit{Kepler}.} 
\label{fig:K_V_LC}
\end{figure}

\begin{figure}
\centering
\includegraphics[width=0.75\textwidth]{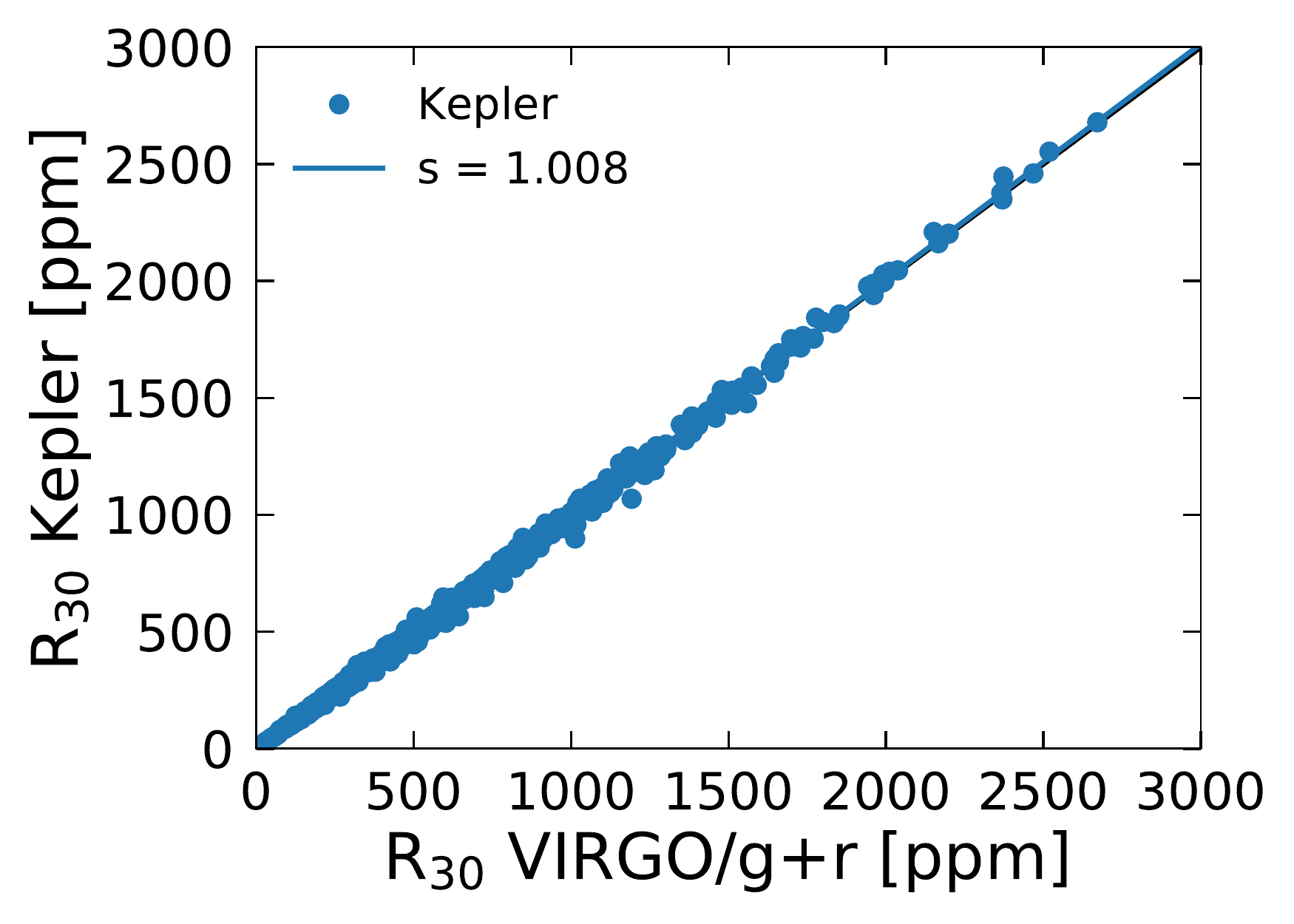}
\caption{Linear regression between the $R_{30}$ calculated for  VIRGO/g+r and for \textit{Kepler} light curves. The black solid line represents a linear regression with slope $=$ 1.}
\label{fig:K_V_regression}
\end{figure}

Several studies \citep[see e.g.][]{Basri2010,Harrison2012} have assumed that the amplitude of the rotational solar variability as it would be measured by Kepler is very close to the amplitude calculated for the combined green and red VIRGO/SPM light curves (in the following VIRGO/g+r). %  as a representative of the solar variability in the \textit{Kepler} passband.
Here we test this hypothesis.  %investigate, how the variability in the composite VIRGO/g+r band compares to the Sun as it would be observed by \textit{Kepler}.
The variability $R_{30}$ for \textit{Kepler} compared to  VIRGO/g+r is shown in Fig.~\ref{fig:K_V_LC}, which is limited to the same time interval as Fig.~\ref{fig:filters_R_30}. The two curves are remarkably similar to one another.  To test the similarity quantitatively, we show the linear regression of $R_{30}$ between {\it Kepler} and VIRGO/g+r for four solar cycles (21-24) in Fig.~\ref{fig:K_V_regression}. The  Pearson correlation coefficient is very high (0.999) and the slope deviates by only  +0.8\% of unity averaged over four solar cycles.  While these calculations are related to the amplitude of the rotational variability, $R_{30}$,  we additionally calculate regressions between {\it Kepler} and Virgo light curves in Sect.~\ref{Kepler-VIRGO}. We also directly connect the TSI and VIRGO/g+r rotational variability. The linear regression between $R_{30}$ in TSI and VIRGO/g+r  results in a slope of 0.88 ($\pm$ 0.002) and a Pearson correlation coefficient of 0.995.

\section{Correction for the inclination}\label{Inclination}
\subsection{Approach}

The results presented in Sect. \ref{results_eq} are for the Sun viewed from the ecliptic plane and apply to stars that are viewed approximately equator-on. However, this is not always the case, and the inclination of a star is often unknown.
Calculations of the solar variability as it would be measured by an  out-of-ecliptic observer  demand information about the distribution of magnetic features on the far side (for the Earth-bound observer) of the Sun. N20 have used a surface flux transport model (SFTM) to obtain the distribution of magnetic features of the solar surface, which was then fed into the  SATIRE model to calculate solar brightness variations as they would be seen at different inclinations. 

The SFTM is an advective-diffusive model for the passive transport of the radial magnetic field on the surface of a star, under the effects of large-scale surface flows. In this model, magnetic flux emerges on the stellar surface in the form of bipolar magnetic regions (BMRs). We employed the SFTM in the form given by \cite{Cameron2010} and followed the approach of N20 to simulate light curves of the Sun at different inclinations and with various filter systems. 
The emergence times, positions, and sizes of active regions in our calculations were determined using the semi-empirical sunspot-group record produced by \cite{Jiang2011_1}. This synthetic record was constructed to represent statistical properties of the  Royal Greenwich Observatory sunspot record. We additionally randomised the longitudes of the active-region emergences in the \cite{Jiang2011_1} records. Such a randomisation is needed to ensure that the near and far side of the Sun have on average equal activity, which is a necessary condition for reliable calculations of the inclination effect. As a result, our calculations reproduce the statistical properties of a given solar cycle, but they do not represent  the actual observed BMR emergences for that specific cycle. We stress that in N20 we developed the model outlined above to study the effect of the inclination on the power spectra of solar brightness variations. Here we use this model to explicitly study the dependence of the variability amplitude on the rotational timescale and its dependence on the inclination in different filters.

\subsection{Results}

In the following, we place the observer out of the solar equator towards the solar north pole. This corresponds to inclinations below 90$^{\circ}$ . We quantify the rotational variability using the $R_{30}$ metric introduced in the previous section. To represent an average level of solar activity, we limit the analysis to cycle 23, which was a cycle of moderate strength.
 
  \begin{figure*}%[hbt!]
\centering
\includegraphics[width=\textwidth]{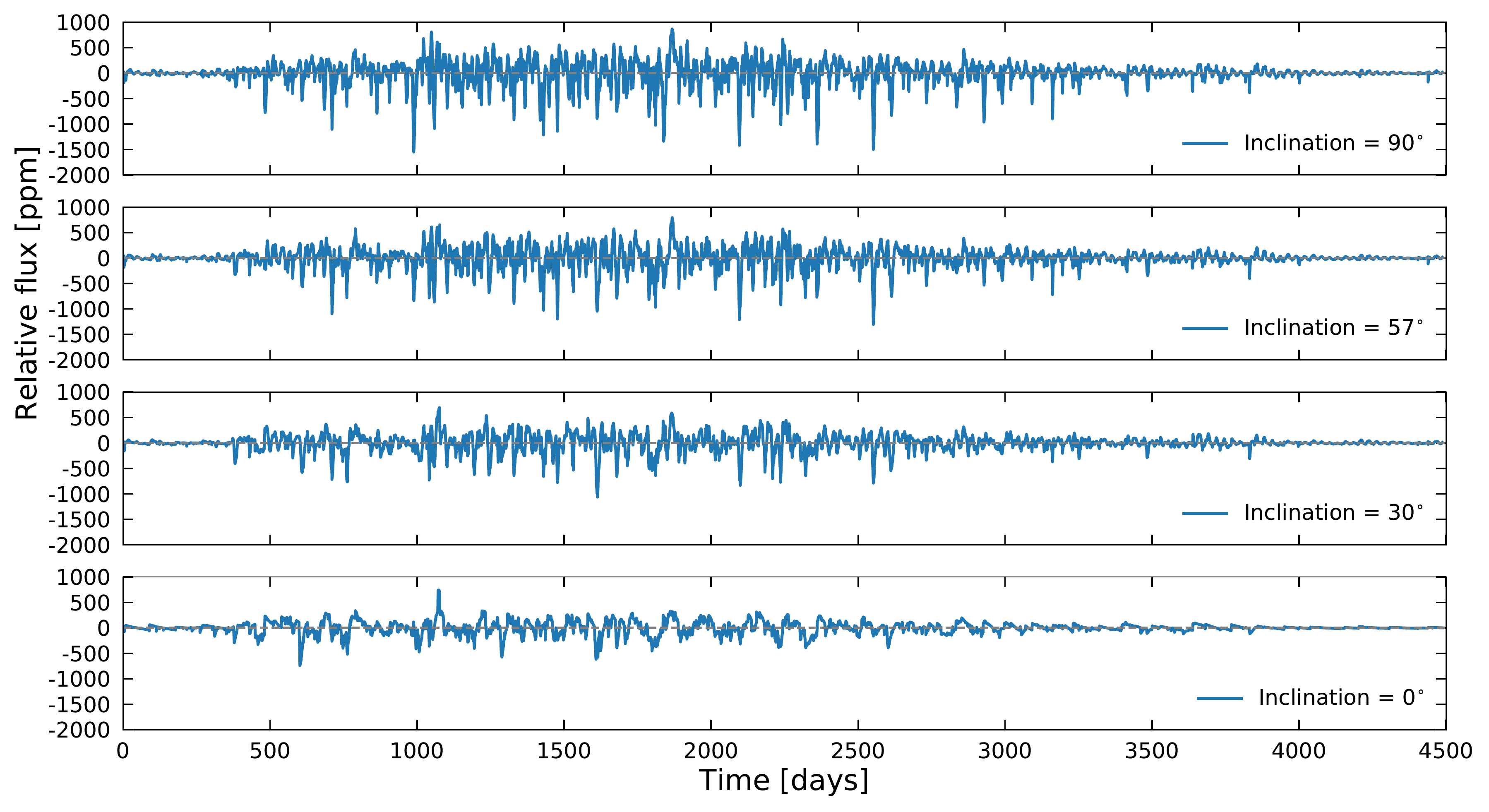}
\caption{Synthetic solar light curves covering solar cycle 23 in the \textit{Kepler} passband as it would appear at different inclinations.}
\label{fig:incl_LC}
\end{figure*}

  \begin{figure*}%[hbt!]
\centering
\includegraphics[width=\textwidth]{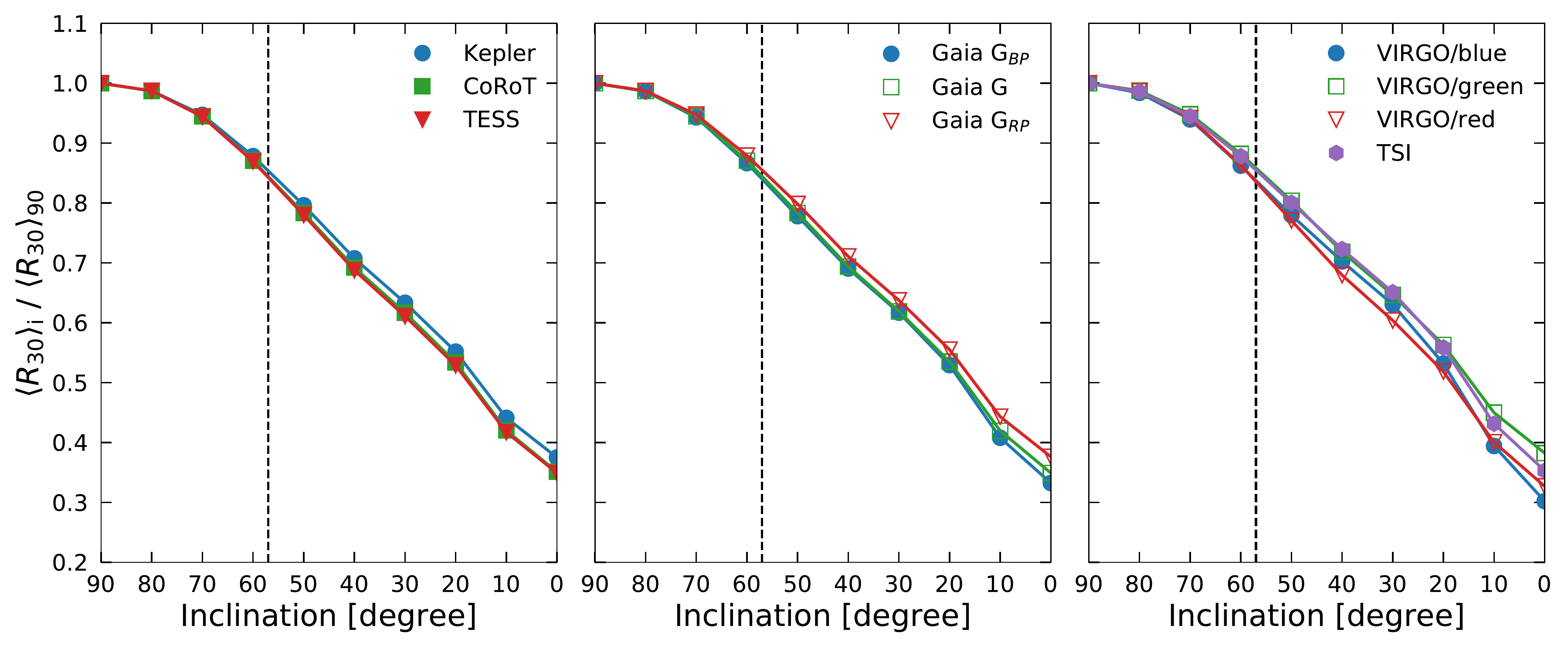}
\caption{Dependence of the mean variability in $R_{30}$ on the inclination (termed $\langle R_{30}\rangle_{\rm i}$, where $i$ stands for the inclination). All values have been normalised to the respective equator-on ($i=90^{\circ}$) value, here called $\langle R_{30}\rangle_{\rm 90}$. Individual curves represent different filter systems. Left panel: \textit{Kepler}, TESS, and CoRoT. Middle panel: Three \textit{Gaia} filters. Right panel: Three VIRGO filters and the TSI. The vertical dashed black line indicates an inclination of 57$^{\circ}$.}
\label{fig:ratios_inclination}
\end{figure*}

We show the calculated solar light curve as it would be observed by \textit{Kepler} at various inclinations in Fig.~\ref{fig:incl_LC}. For this, we divided the time series for cycle 23 into 90-day segments, which correspond to \textit{Kepler} quarters. Within each quarter, we de-trended the light curves. 90$^{\circ}$ corresponds to an ecliptic-bound observer, 57$^{\circ}$  represents a weighted mean value of the inclination with weights equal to the probability of observing a given inclination ($sin(\mathrm{i})$) for the inclination i), and 0$^{\circ}$ corresponds to an observer facing the north pole. We additionally show 30$^{\circ}$ as an intermediate point between 57$^{\circ}$ and 0$^{\circ}$. For the inclination values of 30$^{\circ}$, 57$^{\circ}$, and 90$^{\circ}$ , the variability is brought about by the solar rotation, as well as the emergence and evolution of magnetic features. 
A polar-bound observer does not observe the rotational modulation because there is no transit of magnetic features and the variability is merely generated by their emergence and evolution \citep[see][for further details]{Nina1}. We emphasise just the reduction in the amplitude of the variability with decreasing inclination, but also the change in the shape of the light curve. This is particularly visible in a comparison of the top and bottom panels of Fig.~\ref{fig:incl_LC}. 
In Fig.~\ref{fig:ratios_inclination} we show the change in  $R_{30}$ averaged over cycle 23 when we place the observer out of the ecliptic plane. To facilitate comparison, we normalised each value of the cycle-averaged variability, $\langle R_{30}\rangle_{\rm i}$, to the corresponding value for the ecliptic view, $\langle R_{30}\rangle_{\rm 90}$. Figure~\ref{fig:ratios_inclination} shows that the rotational variability decreases monotonically with decreasing inclination.
This trend is seen across all considered filter systems.
The differences in the inclination effect among the filter systems are due to the different dependencies of the facular and spot contrasts (as well as to their centre-to-limb variations) on the wavelengths.

To evaluate the averaged effect of the inclination, we introduce a new measure that we call $\langle R_{30}\rangle$  and define as
\begin{equation}
\langle R_{30}\rangle =\frac{\sum_{i}\langle R_{30}\rangle_{i}\cdot \mathrm{sin}(i)}{\sum_{i}\mathrm{sin}(i)},
\end{equation}
\noindent where $i$ is the inclination. The factor $\mathrm{sin}(i)$ ensures that the corresponding values of $\langle R_{30}\rangle_{i}$ are weighted according to the probability that a star is observed at inclination $i$. The $\langle R_{30} \rangle$ value represents the variability of the Sun averaged over all possible inclinations. In other words, if we observed many stars analogous to the Sun with  random orientations of the rotation axes, their mean variability would be given by the $\langle R_{30}\rangle$ value. Therefore $\langle R_{30}\rangle$ should be used for the solar-stellar comparison rather than the $\langle R_{30}\rangle_{90}$ value.

We present $\langle R_{30}\rangle$ normalised to  $\langle R_{30}\rangle_{90}$ for all considered filter systems as well as the $\langle R_{30}\rangle$ values themselves in Table~\ref{tab:inclination}. In the second column of Table~\ref{tab:inclination} we give the inclination-corrected value of the mean rotational variability from Table~\ref{tab:mean_R_30} for easier application of our results. On average, all filter systems show a 15\% lower variability than the equatorial case.
This implies that the slopes of the linear regressions between the $R_{30}$ values in different passbands  and the TSI have to be corrected for the inclination. When stellar measurements in the \textit{Kepler} passband are compared with the TSI records, this means the following: when stars are observed from their equatorial planes, the variability in \textit{Kepler} is about 12\% higher than in the TSI (see the slope given in Table \ref{tab:regressions}). However, when the Sun is compared to a group of stars with random orientations of rotation axes, the inclination effect must be taken into account as well. It will reduce the stellar variability observed in the \textit{Kepler} passband by approximately 15\%. Coincidentally, these two effects almost exactly cancel each other, and the observed TSI variability appears to be a very good metric for the solar-stellar comparison of \textit{Kepler} stars in a statistical sense. 
\begin{table}%[!htbp] 
\begin{center}
\begin{tabular}{lcc} 
\hline \hline%
        &$\langle R_{30}\rangle$/$\langle R_{30}\rangle_{90}$ [\%]      &$\langle R_{30}\rangle$  [mmag]  \\
\hline
TSI                     &       85.1    &       0.58    \\ %showing the TSI here probably does not make much sense
\textit{Kepler}         &       84.8    &       0.62    \\
CoRoT                   &       83.8    &       0.61    \\
TESS                    &       84.0    &       0.56    \\
\textit{Gaia G$_{BP}$}  &       83.7    &       0.72    \\
\textit{Gaia G}         &       84.1    &       0.62    \\
\textit{Gaia G$_{RP}$}  &       84.9    &       0.52    \\
VIRGO/blue                  &   83.9    &       0.98    \\
VIRGO/green                 &   85.3    &       0.76    \\
VIRGO/red                   &   83.4    &       0.50    \\
\hline \hline
\end{tabular}
\end{center}
\caption{$\langle R_{30}\rangle$/$\langle R_{30}\rangle_{90}$    and      $\langle R_{30}\rangle$ values for different filter systems. For $\langle R_{30}\rangle$ we multiplied $\langle R_{30}\rangle$/$\langle R_{30}\rangle_{90}$ by the corresponding value for $\langle R_{30}\rangle_{90}$ from Table \ref{tab:mean_R_30}.  Time-averaging is performed over solar cycle 23.}
\label{tab:inclination}
\end{table}

We have shown in Sect.~\ref{Equator} that the amplitude of solar rotational variability as it would be measured by \textit{Kepler} can be very accurately approximated by calculating the amplitude of the VIRGO/g+r light curve. However, when brightness variations of the Sun are compared to those of a large group of stars with unknown inclinations, the use solar variability averaged over all possible inclinations should be used rather than the solar variability observed from the ecliptic plane.
We have established  that the effect of a random inclination decreases the variability in the \textit{Kepler} passband by 15\%. Taking this into account, the relative difference between the variability in VIRGO/g+r and the solar variability in \textit{Kepler}  averaged over inclinations is $-14$\%. Unlike for the TSI, corrections for the passband and the inclination only partly compensate for each other. We therefore suggest that the TSI is a better representative of the Sun as it would be observed by \textit{Kepler} than VIRGO/g+r if the inclination of a star is unknown.

\section{Modelling \textit{Kepler} light curves using VIRGO/SPM}\label{Kepler-VIRGO}

The the previous sections, we have quantitatively validated the argument of \cite{Basri2010} that the VIRGO/g+r light curve corresponds to the same variability as the \textit{Kepler}  light curve if both light curves are recorded from the solar equatorial plane. 
In this section we perform complementary calculations: we test if the  \textit{Kepler} light curve can be modelled as a linear combination of solar light curves in the different VIRGO/SPM channels. We restrict our calculations to solar cycle 23. All light curves are computed with the N20 model.

We divided all light curves into 90-day segments and calculated the relative flux within these segments (i.e. we considered the same normalisation of light curves as shown in Figs. \ref{fig:filters_LC} and \ref{fig:incl_LC}). Next, we applied multiple linear regression to fit the \textit{Kepler} light curve with the VIRGO/SPM green+red light curves to determine the best set of coefficients for the linear fit.
 unity averaged over four solar cycles. 
 
 \begin{figure*}[hbt!]
\centering
\includegraphics[width=\textwidth]{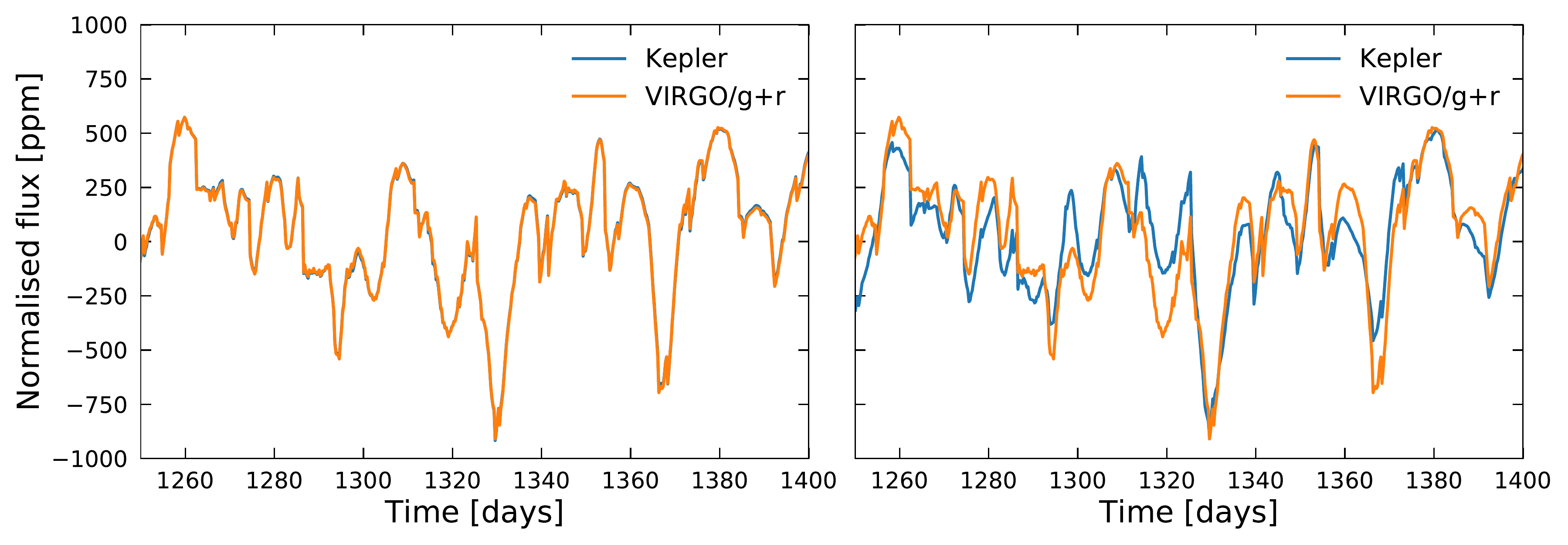}
\caption{Comparison between the \textit{Kepler} and the regressed VIRGO/green+red light curve. Left:i = 90$^{\circ}$,  and right:i=57$^{\circ}$. For the coefficients of the fit, see the main text.} 
\label{fig:fit}
\end{figure*}

For an ecliptic-bound observer, we write the multiple linear regression in the form
\begin{equation}
    K_{i,\,SPM} =   a\cdot V_g+ b\cdot V_r ,
\label{eq:fit_2_90}
\end{equation}
where $i$ is the inclination, $V_g$ and $V_r$ are solar light curves  in VIRGO/SPM blue, green, and red filters (corresponding to the equatorial plane). The best fit for $i=90^{\circ}$ yields  a$=$0.275 ($\pm$ 0.001) and b$=$0.619 ($\pm$ 0.002). The $r^2$ value is 0.999. Such a high correlation is expected because the VIRGO/SPM and \textit{Kepler} rotational variability are similar, as discussed in Sect.~\ref{Equator}.
Next, we applied the multiple-regression model to simulate the out-of-ecliptic \textit{Kepler}-like light curve using the light curves in the SPM channels as input.
Because an inclination of 57$^{\circ}$ is often used to represent the statistical mean of all possible inclination values, we fit the \textit{Kepler} light curve observed at $i=57^{\circ}$ with a linear combination of VIRGO light curves (observed at $i=90^{\circ}$). The best fit results in the following coefficients: a$=-$0.421 ($\pm$ 0.018) and b$=$1.418 ($\pm$ 0.027), with $r^2 = 0.83$.

Figure~\ref{fig:fit} compares the Kepler-like light curve with the regression model using light curves in all three VIRGO filters for $i=90^{\circ}$ and $i=57^{\circ}$.
For the 90$^{\circ}$ inclination case (left panel in Fig.~\ref{fig:fit}), the differences between the two light curves are basically invisible, but for the 57$^{\circ}$ case, the differences are quite pronounced. These differences have various origins. In particular, the transits of magnetic features would take different times for the ecliptic and out-of-ecliptic observer. Furthermore, with decreasing inclination, the facular contribution becomes stronger, while the spot contribution weakens \citep[see e.g.][]{Shapiro2016}, which changes the shape of the light curve. It is therefore necessary to take the actual distribution of magnetic features into account when light curves are modelled that were observed at different inclination angles.

\section{Conclusions}\label{Conclusion}
We presented recipes for treating two problems that hamper  the comparison between solar and stellar rotational brightness variations: the difference between the spectral passbands that are used for solar and stellar observations, and the effect of inclination.
To quantify the effect of different spectral passbands on the rotational variability represented through the $R_{30}$ metric, we employed the SATIRE-S model. We found that the rotational variability observed through the filter systems used by the \textit{Kepler} and CoRoT missions is about 12\% higher than the TSI, whereas the variability in the TESS passband is about 7\% lower. For \textit{Gaia G,} we find +15\% and for \textit{Gaia G$_{BP}$} +30\% difference in the amplitude of the rotational variability compared to the TSI, whereas \textit{Gaia G$_{RP}$} shows a difference of -7\%. These numbers are valid for equator-on observations on rotational timescales.

 Previous studies have used combinations of the red and green VIRGO light curves for solar-stellar comparisons \citep[see e.g.][]{Basri2010,Gilliland2011,Harrison2012}. 
 We used linear regressions of the rotational variability between the two combined VIRGO/SPM passbands and the solar variability as \textit{Kepler} would observe it, to test the goodness of this comparison. We find that the variability in \textit{Kepler} is 7\% higher than that of VIRGO/green+red. Moreover, we found that  the sum of the VIRGO green and red light curves very accurately represents the solar light curve in the \textit{Kepler} passband. This is only valid for the Sun observed from the ecliptic, however. We showed that a linear combination of VIRGO/SPM passbands cannot accurately reproduce the solar \textit{Kepler} light curve observed out of ecliptic.
 
 We have calculated the dependence of the rotational variability on inclination by following the approach in \cite{Nina1}. In this approach, an SFTM was used to simulate the distribution of magnetic features on the surface of the Sun, which was then used to compute the brightness variations with SATIRE. We find that across all filter systems discussed in this study, the rotational variability drops by about 15\% when it is averaged over all possible directions of the rotation axis. Because the \textit{Kepler} rotational variability as observed from the ecliptic plane is 12\% higher than the TSI rotational variability, we conclude that the TSI is the best proxy for the solar rotational brightness variations if they were observed by \textit{Kepler} when the inclination effect is considered.

%\begin{acknowledgements}
%We thank Chi-Ju Wu, whose master thesis has sparked the idea for the manuscript.
%The research leading to this paper has received funding from the European Research Council under the European Union’s Horizon 2020 research and innovation program (grant agreement No. 715947). 
%SKS acknowledges financial support  from the BK21 plus program through the National Research Foundation (NRF) funded by the Ministry of Education of Korea. YCU acknowledges support through STFC consolidated grants ST/N000838/1 and ST/S000372/1.
%\end{acknowledgements}

\chapter{Where have all the faculae gone? Explaining the transition from faculae to spot domination}\label{sec:paper_3}

This section is the draft of a paper by N.-E. N\`{e}mec, A. I. Shapiro, E. I\c{s}{\i}k, S. K. Solanki, and N. A. Krivova \\
\textbf{Contributions to the paper:} I produced the results and provided the main scientific interpretation.

\section{Introduction}
The magnetic fields emerging on solar and stellar surfaces lead to the formation of surface magnetic features, such as dark spots and bright faculae \citep[see, e.g.,][for the review of the solar case]{Sami_rev2006}. One of the exciting manifestations of these magnetic features are the brightness variations of the Sun and solar-like stars \citep{Solanki2013,Ermolli2013}.

Solar  brightness variations have been routinely monitored since the 1970s by dedicated spaceborne instrumentation \citep[for review see][]{Kopp2014}. Using radiometers, these missions measured the total solar irradiance (TSI), which is defined as the wavelength integrated time--depended solar radiative flux at a distance of 1 AU. The TSI varies with an 11-year periodicity. Compared to the minimum, the TSI is 0.1\% higher at solar maximum. This implies that the TSI variability on the cycle timescale is faculae dominated \citep[see, e.g. detailed discussion and references in][]{Shapiro2016}. This means that the decrease of the TSI due to the spots is overcompensated by the increase due to the faculae. However, there is observational evidence, that with increasing spot coverage, the fractional coverage of faculae to spots decreases, i.e. the relative importance of the faculae decreases \citep{Foukal1993, Chapman1997}.

Long-term monitoring programs of stellar variability started in the 1960's, as Olin Wilson founded the Mount-Wilson observations of the Ca II H and K lines to search for activity cycles in cool main sequence stars \citep{Wilson1978}. The emission in the cores of the Ca II H\&K lines is formed in stellar chromospheres and is a good proxy of the magnetic heating in this atmospheric layer. Thus, it is routinely used to characterise stellar magnetic activity. 

The next big step in observations of stellar variability became the establishment of the Lowell observatory monitoring program \citep{Lockwood1992}. The program observed stars in the visual spectral domain in the \textit{Strömgren b} and \textit{y} filters, which are centered at wavelengths of 467 and 547 nm, respectively. Simultaneous measurements of stellar photometric brightness and  chromospheric activity revealed a distinction between active younger and less active older stars \citep[e.g.][]{Soderblom1991}. 
The Ca II emission and brightness appear to be correlated for less active stars (i.e. the increase of the Ca II emission is accompanied by an increase in the photometric brightness) and anti-correlated in active stars. Interestingly, this transition from faculae to the spot-domination happens at an  activity level just a bit higher than the Sun has demonstrated over the last 400 years. \cite{Shapiro2014} (hereinafter \citetalias{Shapiro2014}) showed that this transition can be explained by extrapolating the solar dependencies of spot and facular disk coverages on  the chromospheric activity. This indicates that the spot disk coverages increase faster with the activity level than the facular coverages, so that at certain activity, the stellar variability becomes spot-dominated. The \citetalias{Shapiro2014} results imply that the dependencies of solar spot and facular disk coverages on the activity level are valid not only for the Sun, but also for other solar-like stars, and, in particular, for stars more active than the Sun. 

We seek to explain the physical mechanism behind this dependence of facular and spot area coverages with increasing activity level by following the approach presented in \citealt{Nina1} (in the following \citetalias{Nina1}). \citetalias{Nina1} have used a surface flux transport model \citep[SFTM,][]{Cameron2010} together with an irradiance model to calculate the brightness variations of the Sun as represented via power spectra. In the present work, we do not calculate the brightness per se, but only the area coverages of faculae and spots, following the  \citetalias{Nina1} approach. We summarize this approach in Sect. \ref{paper3_model}, before we extend the model to more active stars utilizing the SFTM in the form presented by \cite{Isik2018}, hereafter \citetalias{Isik2018} in Sect. \ref{active}. We discuss our findings and the underlying mechanism in Sect. \ref{cancellation}, before concluding our work in Sect. \ref{paper3_conclusio}.

\section{Model}\label{paper3_model}

\subsection{Surface flux transport and filling factors}

We obtain the surface distribution and coverage of magnetic features by following the approach presented in  \citetalias{Nina1}. This method relies on two steps. Firstly, the surface distribution of magnetic flux is simulated by employing a surface flux transport model (SFTM), which produces full surface magnetograms. Secondly, the flux from these magnetograms is attributed to spots and faculae. We describe the steps needed briefly in the following, but refer to the full discussion in \citetalias{Nina1} for more details.

In the SFTM the magnetic flux on the surface of stars is passively transported due to advection, diffusion and large scale flows, in this instance the differential rotation and the meridional flow \citep[see][and references therein]{Cameron2010}. The magnetic flux in the model is injected as bipolar magnetic regions (BMRs, e.g. one BMR consists of two flux patches with opposite polarities). The emergence of the BMRs is introduced via the source term that determines the position and time of the emerging flux, as well as the flux of the polarity patches, namely
\begin{equation}
B^{\pm}(\lambda,\phi) = B_{\mathrm{max}}\left(\frac{0.4 \Delta \beta}{\delta}\right)^{2} e^{-2[1-\cos(\beta_{\pm}(\lambda,\phi))]/\delta^{2}} ,
\label{source2}
\end{equation}
\noindent where $\beta_{\pm}$($\lambda$,$\phi$) is the heliocentric angle between a given point ($\lambda$, $\phi$) and the centres of the polarity patches, $\Delta \beta$ is the separation between the two polarities and $\delta$ is the size of the individual polarity patch, taken to be 4$^\circ$. $B_{\mathrm{max}}$ is a scaling factor set to 374 G \citep{Cameron2010,Jiang2011_2}.

The source term was compiled following the approach of \cite{Jiang2011_1} by constructing a semi-empirical input record. This input record contains statistical properties of the latitude, the size distribution and the tilt-angles of emergences, reflecting the properties of sunspot observations from the Royal Greenwich Observatory for the time period 1986--1996, corresponding to solar cycle 22. Additionally, the longitudes of the emergences are randomised. This allowed \citetalias{Nina1} to study the impact of different vantage points with respect to the ecliptic plane on the solar brightness variations represented through power spectra. We follow the approach of randomising the longitudes in  this work as well. We remark that this is fully sufficient for our purposes in the present work, as we are concerned with modelling general trends of stellar variability. An advantage of employing the SFTM is that the source term  can be changed depending on the objective of research.

To obtain the area coverages of the spots, \citetalias{Nina1} have used the following approach: (i) calculate the spot area at the day of emergence from the input record following the relation between the separation $\Delta \beta$ and the spot area from \cite{Cameron2010}, (ii) follow a linear decay law with a single decay rate $R_d$ for all spots independent of size and (iii) add a linear growth law with a single growth rate $R_g$. Next, the facular coverages were obtained by first, subtracting the magnetic flux of the spots from the magnetograms and then using the remaining flux to calculate the faculae areas following the approach of \cite{Krivova2003} of setting a saturation threshold $B_{sat}$.
While we used the SFTM simulations from \cite{Cameron2010} in \citetalias{Nina1} and \cite{Nina2}, the results in this study are based on the model in \citetalias{Isik2018} in the present work. We discuss the differences in these two models in more detail in the next section.

\subsection{Fixing the parameters}\label{parameters}

In \cite{Nina1} we have fixed the free parameters ($R_g$, $R_d$, and $B_{sat}$) by comparing the brightness variations returned by our model to observations and the SATIRE-S model \citep{Yeo2014}. Since in this study we are interested in explaining general trends in the behaviour of facular and spot coverages, we fix the parameters by comparing the coverages returned by our model to those returned by the SATIRE-S model. The area coverage of the magnetic features in SATIRE-S are obtained from observed magnetograms and full disk images and are available starting from 1979. The output of \citetalias{Isik2018} SFTM runs is available for solar cycle 22. 

The statistical source term employed in this work prohibits direct day-to-day comparison between the area coverages from SATIRE-S and our model. We therefore sort the area coverages according to ascending facular areas, before we split the resulting monotonous area coverages into bins containing 58 days. Within each bin we then calculate the mean value.

First, we fit the output of SATIRE-S following, 
\begin{equation}
A_f = a\cdot \sqrt{A_{s}-c} + d.
\label{eq:fit}
\end{equation}
Then we find a combination of the free parameters  of our model ($R_d$, $R_g$ and $B_{sat}$) which leads to values for $a$, $c$, and $d$ as close to those obtained with fitting SATIRE-S as possible.

\begin{figure}%[hbt!]
\centering
\includegraphics[width=0.75\textwidth]{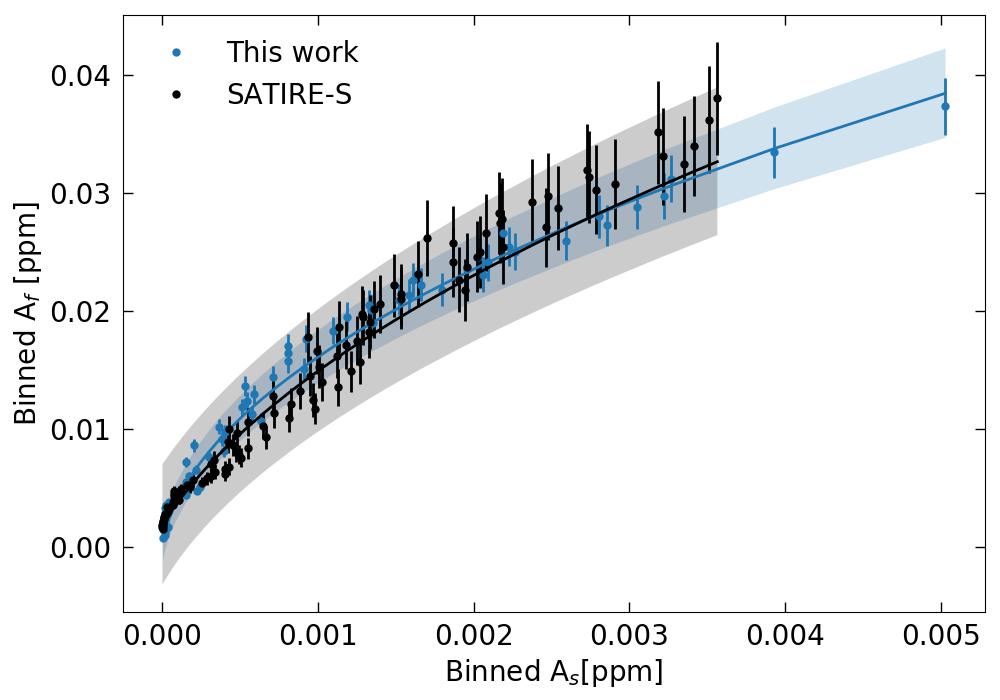}
\caption{Comparison between the binned filling factors returned by the SATIRE-S model (black) and the model presented in this work (blue). Solid lines represent corresponding fits (see text for details) and the shaded areas are the 1-$\sigma$ intervals. Vertical bars display the standard deviation within each bin. Each bin contains 58 data points and is represented by the mean value.}
\label{fig:params_model}
\end{figure}

We display the data-sets and their corresponding fits in Fig. \ref{fig:params_model} and give the coefficients of the fits according to Eq. \ref{eq:fit} in Table \ref{tab:fits}. Additionally, the cycle averaged spot coverages (0.00108 for SATIRE-S and 0.00104 for our model) and the cycle averaged facular coverages (0.0144 for SATIRE-S and 0.0139 for this work) are consistent.
We compare the parameters obtained in this work to the ones obtained by \citetalias{Nina1} in Table \ref{tab:params}. The values of $R_d$ and $B_{sat}$ are slightly different in the two models. The reason for this is that for the compilation of the input record, the sunspot group number is used, which needs to be converted into individual spot pairs. In \citetalias{Nina1} we followed the approach by \cite{Jiang2011_1}, who used a conversion factor of 2.1 from sunspot group number to sunspot number, whereas the \citetalias{Isik2018} approach used a conversion factor of 2.75. The larger factor in \citetalias{Isik2018} resulted in fewer region emergences. To compensate this, the spot lifetime must be longer and the faculae threshold has to be lowered to attribute more flux to the faculae in order to preserve the area coverages.

\begin{table}[]
\centering
\begin{tabular}{cccc}
\hline
\hline
         & a            & c                 & d               \\ \hline
SATIRE-S & 0.67  & -0.0003  & -0.009   \\ 
          & ($\pm$ 0.057) & ($\pm$ 0.00008) & ($\pm$ 0.003)\\ \hline
Model    & 0.57   & -0.0002  & -0.002 \\ 
        & ($\pm$ 0.05)  & ($\pm$ 0.00002) & ($\pm$ 0.0008) \\ \hline
\hline
\end{tabular}
\caption{Coefficients of the fit according to Eq. \ref{eq:fit} of the curves displayed in Fig. \ref{fig:params_model}.}
\label{tab:fits}
\end{table}

\begin{table}[]
\centering
\begin{tabular}{llll}
\hline
\hline
Parameter  & This work  & \citetalias{Nina1}  & Description\\ \hline
$R_d$ [MSH/day]  & 47 & 80 & decay rate of spots\\
$R_g$ [MSH/day]  & 600 & 600 &growth rate of spots \\
$B_{sat}$ [G]   &470  & 500  &saturation threshold for faculae\\ \hline
\hline
\end{tabular}
\caption{List of the parameters of our model compared to the \citetalias{Nina1} values.}
\label{tab:params}
\end{table}

\section{Extending the model to stars more active than the Sun}\label{active}

\begin{figure*}
\centering
\includegraphics[width=\textwidth]{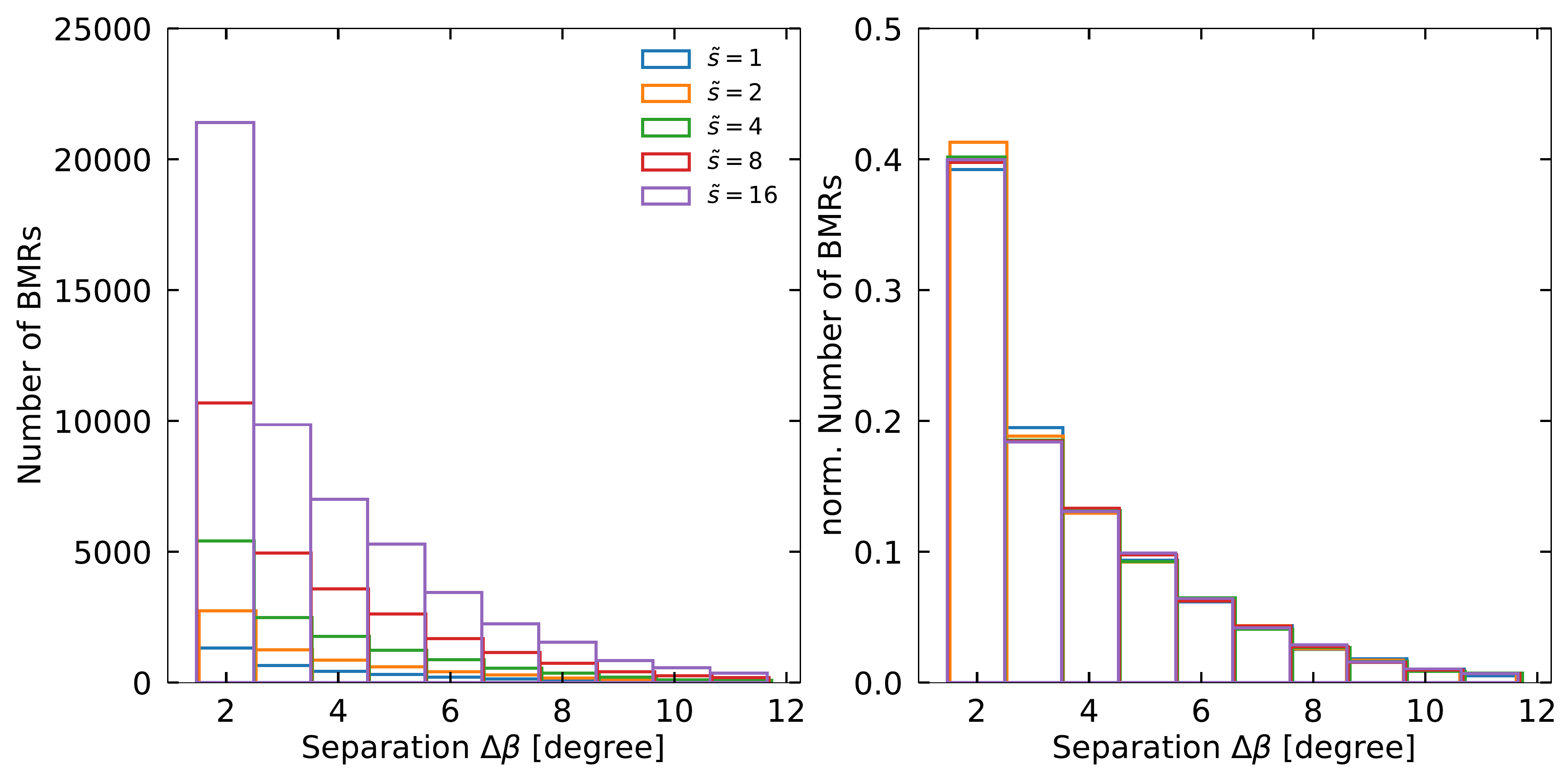}
\caption{Distribution of the separation of the bipolar magnetic regions (BMRs) of stars with different activity levels in terms of their number of emergences. Left: absolute numbers of BMRs per $\Delta \beta$ bin, right: normalised numbers of BMRs per $\Delta \beta$. The integral for each distribution in this panel is 1.}
\label{fig:histograms_scaling}
\end{figure*}

In Sect. \ref{parameters} we have shown that our model is able to reproduce the observed drop in solar facular area with increasing spot area. In the following, we investigate how $A_f$ as a function of $A_s$ changes for  stars more active than the Sun. In particular, we address the physical mechanism behind this behaviour. % For this, we extend the model to stars that are more active than the Sun. 
We can define the time-dependent emergence rate of BMRs on a star $S_{\star}$ as 
$S_{\star}(t) = S_{\odot}(t)\cdot \tilde{s}$, where  $S_{\odot}(t)$ is the-time dependent solar BMR emergence rate and $\tilde{s}$ is a scaling factor. This means, that a star with $\tilde{s}=2$ exhibits two times more BMR emergences compared to the solar case, $\tilde{s}=1$. We note that in contrast to \cite{Isik2018} we consider stars with solar rotation rate so that while we change the emergence rate of BMRs, we preserve the solar distribution of latitudes of emergences \citep[see,][]{Jiang2011_1}. We also preserve the solar distribution of areas of active regions  \citep[see][Appendix A.3]{Isik2018} and the histograms in Fig. \ref{fig:histograms_scaling}. We extend the model in this work to $\tilde{s}=2,4,8$ and $16$.

\begin{figure}%[hbt!]
\centering
\includegraphics[width=\textwidth]{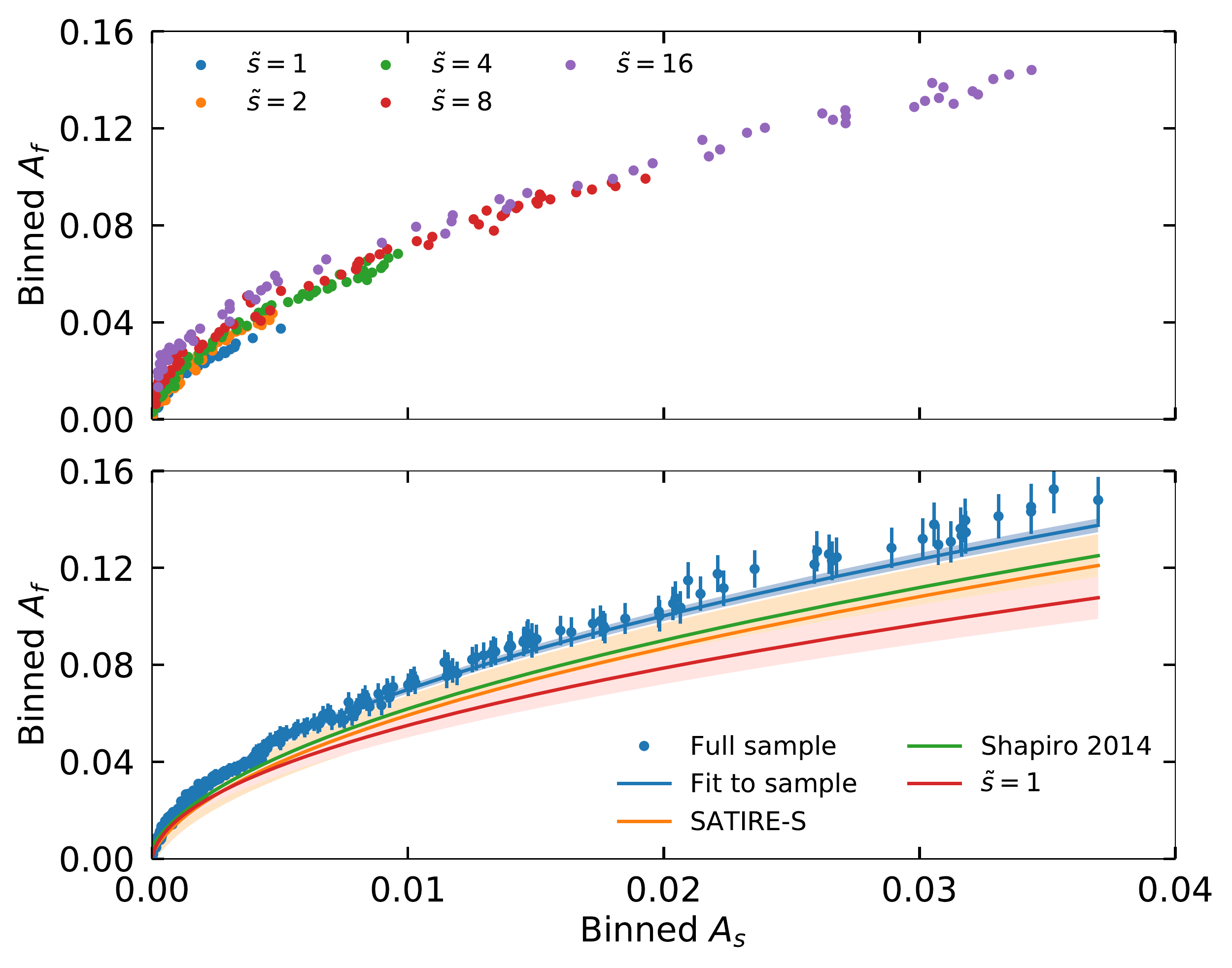}
\caption{Binned facular disk areas coverages $A_f$ as a function of the spot disk area coverages $A_s$ for different activity levels $\tilde{s}$. Top: binned values for the different simulated stellar activity levels. Bottom: comparison between the area coverages of the full set of simulations in this work (blue) to the fit to the observed area coverages returned by SATIRE-S (orange), empirical relationships \citep[][green]{Shapiro2014}, and the fit to the solar activity level $\tilde{s}=1$ (red). }
\label{fig:bins_emre}
\end{figure}

The calculated dependencies of $A_f$ on $A_s$ are shown in Fig.~\ref{fig:bins_emre} top panel. The binning was done in the same way as described in Sect. 4.2.2. While the dependencies calculated for different $\tilde{s}$ generally follow the form of Eq. \ref{eq:fit},  Fig.~\ref{fig:bins_emre} shows that the cases $\tilde{s}>=1$ are not simply a continuation of the solar case ($\tilde{s}=1$). In general, for a given spot coverage, calculations with larger 
$\tilde{s}$ return higher facular coverage. The exact reason for such a tendency is out of the scope of the present study but we speculate that it can be attributed to our binning procedure and the phase shift between maximum contribution of a given BMR to spot and facular coverages \citep[see discussion in][]{Foukal1998}.  

In Fig. \ref{fig:bins_emre} bottom panel we compare the fit to the data (blue) with various other results: the fit to the SATIRE-S values (orange), the relationship between $A_f$ and $A_s$ published by \citetalias{Shapiro2014} and the fit to the $s=1$ case of our model. While the results agree with each other within error-bars and the shown 1$\sigma$ intervals (shaded areas in Fig. \ref{fig:bins_emre} bottom panel), this figure also indicates, that there might not be an universal dependence of $A_f$ on $A_s$. As discussed above the activity level used for calculating the dependence of  $A_f$ on $A_s$ is an  important factor. %SATIRE-S in particular uses solar cycles 21--23, which were of moderate strength. 

%This findings can be backed by various studies \citep[see e.g.][]{Foukal1998} and have crucial implications for solar--stellar comparisons. As \cite{Radick1998} tried to put the solar chromospheric variability in the stellar context, they noted that depending on the strength of the chosen solar cycle, the solar chromospheric variability might be over- or underestimated by the general trends determined from stellar observations.

\begin{figure*}
\centering
\includegraphics[width=\textwidth]{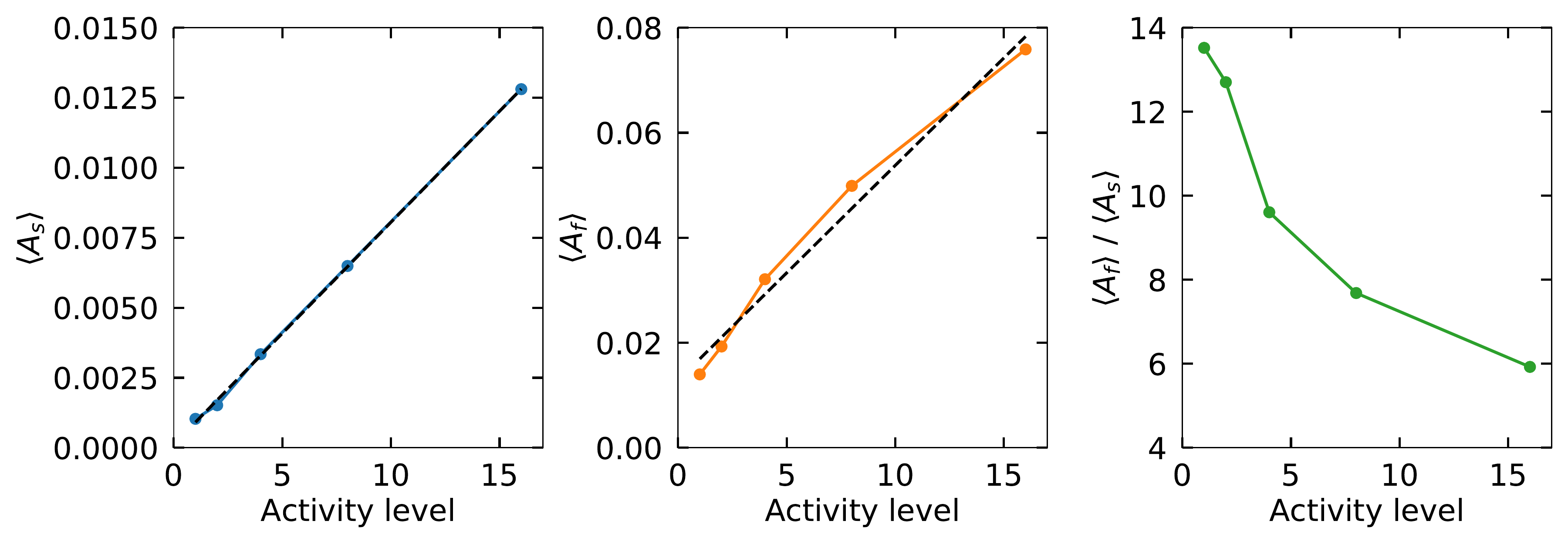}
\caption{Mean cycle averaged facular disk areas coverages $\langle A_f \rangle $, mean cycle averaged spot disk areas coverages $\langle A_s \rangle $ and their ratios as a function of the activity scaling factor, $\tilde{s}$, (see text for details).. The dashed black lines are linear fits to the data, the colored solid lines connect the points to guide the eye.}
\label{fig:mean_emre}
\end{figure*}

Fig. \ref{fig:mean_emre} we additionally show the mean values of the disc area coverages by spots ($\langle A_s \rangle $, left panel), faculae ($\langle A_f \rangle $ ,middle panel) and the ratio between these two values (right panel) as a function of activity scaling factor, $\tilde{s}$. Clearly, $\langle A_s \rangle $ increases linearly, while  $\langle A_f \rangle $ deviates from a linear increase. With higher activity level, the ratio $\langle A_f \rangle$/$\langle A_s \rangle$ is dropping, which is inline with the above discussed behavior of $\langle A_f \rangle$ with increasing $\langle A_s \rangle$.

\section{What happens to the flux?}\label{cancellation}

It is not surprising that $\langle A_s \rangle $ is increasing linearly with increasing activity level. The spot area coverages are determined by the separation $\Delta \beta$ of the BMRs given by the source term. As Fig. \ref{fig:histograms_scaling} shows, with increasing activity level $\tilde s$, the distribution of the regions with respect to $\Delta \beta$ stays the same, therefore the spot area coverages are linearly increasing as well.

The same should in principle also apply to the faculae. With the increasing BMR emergence rate for more active stars, more flux (that is attributed to the faculae) is emerged on the surface of a star. However, as seen in Fig. \ref{fig:mean_emre} middle panel, something is happening to the flux once it is emerged, that decreases $\langle A_f \rangle $ with higher activity levels.
In the SFTM, flux is present in both positive and negative polarity. As the model does not directly include convection, flux is only distributed across the surface by the surface flows and the surface diffusivities. All in all, there is only one mechanism to remove flux from the surface and that is through the cancellation of fluxes with opposite polarities.
As activity increases the opposite polarities come closer to each other so that the cancellation gets more effective and the ratio between facular and spot coverages drops. We note that our model also allows to take into account the cancellation of spots with different polarities.  However, since spot areas are much smaller than those of faculae and also spot lifetimes are much shorter, such cancellations are barely occurring in our model.

%This happens, if there is an encounter of opposite polarities. 
%The field attributed to faculae is diffuse and the areas attributed to faculae is larger than the spots, hence the chances of opposite polarity encounters is increased. 

%We note, that our masking process also takes this cancellation effect into account for the spot masking, however, as their areas are small and their lifetimes are short, events of opposite polarity encounters are barely occurring.

%some observational evidence
This explanation in reasonably, given that cancellation events have been observed on the solar surface in active regions \citep[e.g.][]{Green2011,Tiwari2019}, in active region plage \citep[][]{Chitta2019} and even in quiet Sun regions \citep{WANGJINGXIU198879}. Cancellation events and the consequent release of magnetic energy have also propose as possible explanations of both chromospheric and coronal heating through nanoflares \citep[see e.g.][]{Priest2018} and an possible trigger mechanism of coronal jets \citep{McGlasson2019}.

\section{Conclusion}\label{paper3_conclusio}

We have used a surface flux transport model \citep[SFTM, as described in][]{Isik2018} to simulate the emergence and evolution of magnetic flux on the surfaces of stars, together with the algorithm of obtaining the area coverages of magnetic features from \cite{Nina1} to investigate the curious case of the decrease of the ratio between facular and spot coverages  with increasing stellar activity.

This allowed us to explain the empirical dependence between facular and spot coverages reported by \cite{Shapiro2014} as well as to confirm the extrapolation of this dependence beyond the level of solar activity. At higher activity levels, the facular area coverage gradually saturates, whereas the spot area coverage is increasing. This results in a transition from faculae to spot domination, as the ratio between facular to spot area drops.
Our results strongly point towards the cancellation of magnetic flux in the form of opposite polarity encounters as the reason for such a behaviour of the faculae towards higher activity levels.

\chapter{Explaining the dependence of stellar brightness variability on the rotation period}\label{sec:paper_4}

This section is the draft of a paper by N.-E. N\`{e}mec, A. I. Shapiro, E. I\c{s}{\i}k, T. Reinhold, S. K. Solanki. \\
\textbf{Contributions to the paper:} I produced the results and provided the main scientific interpretation. 

\section{Introduction}

 Planetary hunting missions such as the Convection, Rotation and planetary Transits \citep[CoRoT][]{COROT,COROT2}, \textit{Kepler} \citep{KEPLER} and the Transiting Exoplanet Survey Satellite \citep[TESS,][]{TESS} allowed studying the day-to-day brightness variations of stars with high precision and cadence. Such brightness variations are well understood for the Sun and are brought about by the transit of dark spots and bright faculae across the visible stellar disc \citep[see][for review]{Solanki2013, Ermolli2013}.
 
 Models based on the solar paradigm are a useful tool-set in providing explanations for the plethora of stellar photometric data. For example, \cite{Nina1} have combined the Spectral And Total Irradiance Reconstruction \citep[SATIRE,][]{Fligge2000,Krivova2003} model together with a surface flux transport model \citep[SFTM,][]{Cameron2010} to model the power spectra of solar brightness variations at various inclinations down to several days. This approach allowed to remove a number of important observational biases when comparing the solar variability to that of other stars \citep{Nina2,Timo2020}. By employing the \cite{Nina1} model, \cite{Timo2020_2} have found that rotation periods of most of the G-dwarfs of near-solar age go undetected. This results provide an explanation of the discrepancy between the predictions of the number of Sun-like rotators in the \textit{Kepler} field \citep{VanSaders2019} and the actual number of detected ones.

Modelling brightness variations of stars rotating faster than the Sun poses several challenges. The most prominent one being that the latitudinal distribution of magnetic features of fast rotators is vastly different from that of the Sun \citep{Schuessler1992,Schrijver2001} For instance, for stars with rotation periods below about 3 day various, studies have reported the presence of polar starspots \citep{Jeffers2002,Marsden2004,Jarvinen2006,Waite2015}. \cite{Isik2018} have developed a model, that is able to explain the formation of high latitude spots for those faster rotating stars, by combining thin-flux tube simulations with the SFTM.

In the present work, we extend the model of \citetalias{Isik2018} to calculate the brightness variations of stars with various rotation rates, degree of nesting of magnetic features on their surfaces (i.e. the tendency of magnetic features to emerge in the vicinity of previous emergences), and observed at various inclinations (i.e. angles between rotation axis and direction to the observer).
We compare our results to the observational trends deduced by \cite{McQuillan2014} from the {\it Kepler} data and propose a possible explanation of these trends.

\section{Model}

Our model consists of two building blocks: calculations of the surface distribution of magnetic features and the subsequent calculations of the stellar radiative fluxes. The first block is based on the model developed by \cite{Isik2018}, while the second block builds on the model presented in \cite{Nina1}.

\subsection{Flux emergence and transport}\label{FEAT}

%\textbf{Emre's text}

At its core, the flux emergence and transport (FEAT) model extends the pattern of emergence and evolution of the magnetic fields observed on the surface of the Sun to more active stars.
At first, an 11--year semi-synthetic active region record based on the statistics of solar cycle 22 \citep{Jiang2011_1} is generated. This record contains a number of properties of active regions. Most importantly, how many regions are emerged in total throughout the cycle (we call this the emergence rate), the latitude of emergence and the tilt angles. 

The faster rotation rate and the resulting stronger Coriolis force implies that rising flux tubes are pushed towards higher latitudes. This is taken into account by thin flux tube simulations \citep{Schuessler1996,Isik2018}. For this, the time-latitude distribution of the active regions given by the input record is mapped to the base of the convection zone, for the case of $\Omega_\odot$. Then the flux-tube simulations were run to follow the evolution of flux tubes throughout the convection zone until the surface of the star (for $\Omega_\star$), where they emerge in the form of a loop with two magnetic field patches (we will refer to those as bipolar magnetic regions, BMRs for short) of opposite polarity as footpoints. Using the lookup table from flux-tube simulations for a given stellar rotation rate, ($\Omega_\star$), the latitudes and tilt angles of emerging flux loops are then fed into the surface flux transport model (SFTM). Additionally, in agreement with the observed activity-rotation relation, the BMR emergence rate can be scaled with the rotation rate. Namely, we  define the time-dependent emergence rate of BMRs on a star as $S_{\star}(t)= S_{\odot}(t) \cdot \tilde{s}  $, where $\tilde{s}$ is a scaling factor. 
In most of the runs we put it to $\tilde{s}= \tilde{\omega} \equiv \Omega_{\star}/\Omega_{\odot}$, where $\Omega_{\star}$ is the rotation rate of a given star and $\Omega_{\odot}$ is the solar rotation rate. 

The SFTM adopted here \citep{Cameron2010,Isik2018} describes the passive transport of the radial component of the magnetic field under the effects of differential rotation, meridional flow, and supergranular diffusion. In the SFTM, the flux emerges instantaneously in the form of bipolar-magnetic regions (BMRs).
Once the flux is emerged, it is subject to differential rotation and the meridional flow, while diffusing at timescales corresponding to supergranular motions on the Sun.
Additionally, the emergence pattern of BMRs is altered by introducing BMR nesting, which is a feature observed on the Sun. In our study, we follow the approach of "free-nesting", where a probability, $p$, is set for a given BMR to be part of a nest \citep[for further discussion see][Appendix C]{Isik2018}.

 % We will call $p$ the degree of nesting in the remaining of this Chapter. 

\subsection{Defining the filling factors}\label{filling_factors}

The output of the FEAT model are full solar surface maps  (360$\degree$ by 180$\degree$) of the distribution of magnetic fields, with a resolution of 1$\degree$ x 1$\degree$ per pixel. As our brightness calculations rely on the filling factors of spots and faculae, we need to convert the magnetic fields into filling factors. 

The approach presented in \citetalias{Nina1}
follows the evolution of sunspots after their emergence to calculate the coverage of the solar disk by spots.
However, this approach neglects the spontaneous formation of spots from magnetic flux which without superposition would emerge as faculae \citep{Kitiashvili2010}.  This is not an issue for calculating solar variability since such spontaneous spot formations have not been observed on the Sun. However, it might pose problem for more active stars. Therefore we utilise another approach in this study.  %change the approach in the present work to take the effect of this possible superpositions into account. 
In contrast to \citetalias{Nina1}, we calculate the disc coverages by spots directly from the value of the magnetic field returned by the FEAT model. We only take the spot component of the variability into account, as spots become the dominant source of the variability for stars more active than the Sun \citep[see][]{Shapiro2014}.  To calculate the coverage of a given pixel of the synthetic magnetogram returned by FEAT we define two thresholds: a lower cut-off, $B_{min}$ for the fields and an upper, saturation level, $B_{max}$. The spot coverage of a given pixel is related to the field in the pixel as
\begin{equation}
    \alpha^{m,n}=
  \begin{cases}
    0 \quad \text{if} \quad |B_{mn}|< B_{min} \\
	 \frac{|B_{mn}|}{(B_{max}-B_{min})} + 1-\frac{B_{max}}{(B_{max}-B_{min})} \quad \text{if} \quad B_{min} <=|B_{mn}| < B_{max} \\
	1 \quad \text{if}  \quad |B_{mn}| >= B_{max},
	\end{cases}
	\label{eq:filling_factors}
\end{equation} 
\noindent where $\alpha^{m,n}$ is the spot filling factor in a given pixel with coordinates m and n on the map, and $|B_{mn}|$ is the absolute value of the field in said pixel. We find that with this simplified approach and parameters of $B_{min} = 60$G and $B_{max}=700$G, we are able to reproduce the spot component of the solar  variability as returned by the  the more sophisticated \citetalias{Nina1} model. We further distinguish between the umbral and penumbral spot component by assuming a ratio of 1-to-5 following \cite{Baumann2004}.
For a more detailed discussion about $B_{min}$ and $B_{max}$ we refer to the more detailed discussion in Sect. \ref{model_details}.
We note that \citetalias{Isik2018} used a single threshold approach to define the spot filling factor, however as demonstrated in Appendix \ref{model_details} Fig. \ref{fig:params} in top panel, this leads to an overestimation of the variability of the solar case, if compared to the results of \citetalias{Nina1}.
As the FEAT model produces maps with a one day cadence, we interpolate between the maps to obtain a cadence of 6h \citep[see][for further discussion]{Nina1}.

\subsection{Calculating the brightness variations}\label{bright_var}

The last step in our model is to combine the spot filling factors
with intensities of spot and quiet stellar regions (i.e. regions of a star free from spots).

The intensity spectra of the spots and the quiet Sun are the same that are used by \citetalias{Nina1}, namely they have been computed  by \cite{Unruh1999} with the spectral synthesis block of the  ATLAS9 code \citep{Kurucz1992, Castelli1994}, assuming radiative equilibrium.

We calculate the corresponding spectral irradiance of the spots, $S(t,\lambda)$ (i.e. the stellar fluxes, normalized to 1 AU), where $t$ is the time and $\lambda$ the wavelength, is calculated by summing up the intensities weighted by the corresponding spot filling factors of a pixel as given by
\begin{equation}
    S(t,\lambda) = S^{q}(\lambda_w) +\sum_{mn} \left(I_{mn}(\lambda)-I_{mn}^{q}(\lambda)\right) \, \alpha_{mn}(t)
    \Delta\Omega_{mn}.
%\label{SSI}
\end{equation}
\noindent The summation is done over the pixels of the surface maps and the $m$ and $n$ indices are the pixel coordinates (longitude and latitude, respectively), $\alpha_{mn}$ is the spot filling factor of pixel($m$,$n$), $\Delta \Omega_{mn}$ is the solid angle of the area on the stellar disc corresponding to one pixel, as seen from the distance of 1 AU, and S$^{q}$ is the quiet star irradiance, defined as
\begin{equation}
   S^{q}(\lambda_w) = \sum_{mn} I_{mn}^{q}(\lambda_w)\Delta\Omega_{mn}.
\end{equation}
\noindent 
Note, that the solid angles of the pixels, as well as the corresponding intensity values depend on the vantage point. Hence $S(t,\lambda)$  is sensitive to the stellar inclination.
As we are concerned with the range of stellar variabilities as observed by the \textit{Kepler} telescope in this work, we also need to take into account the nature of its detectors \citep[see e.g.][]{Maxted2018}, as \textit{Kepler} counts the number of photons and not their energy. In order to obtain the light curves (LCs) of the simulated stars as they would be observed by \textit{Kepler}, we follow
\begin{equation}
%LC =  \int\limits_{\lambda_1}^{\lambda_2} R(\lambda) \cdot I(\lambda)  \frac{\lambda}{h\cdot c} \, d\lambda,
LC(t) =  \int\limits_{\lambda_1}^{\lambda_2} R(\lambda) S(\lambda,t)  \frac{\lambda}{hc} \, d\lambda,
%\label{eq:filter}
\end{equation}
\noindent where $\lambda_1$ and $\lambda_2$ are the blue and red threshold wavelengths of the filter passband, $R(\lambda)$ is the response function of the filter and $S(\lambda,t)$ is the spectral irradiance at a given wavelength and time $t$, $h$ is the Planck constant, and $c$ the speed of light.

\section{Light curves}

We employ the FEAT model (see Sect. \ref{FEAT}) to compute the distribution of magnetic field for an 4--year interval (using the solar distribution for 1986--1996, corresponding to solar cycle 22, as an input), and then calculate filling factors of spots and their impact on stellar brightness (Sect. 5.2.2 and 5.2.3., respectively). We focus on the impact of different distributions of spots on stellar surfaces and the inclination on the light curves (LCs) in this section. To make our calculations comparable to the {\it Kepler} data, each calculated LC has been split into 90-day segments (e.g. similar to the \textit{Kepler} quarters). Each chunk is then normalised by its mean value. In this section, if not otherwise stated, we show a zoomed in version the LCs in Appendix \ref{full_LC} of the time-span of 600--800 days.

\begin{figure}[h!]
\centering
\includegraphics[width=\textwidth]{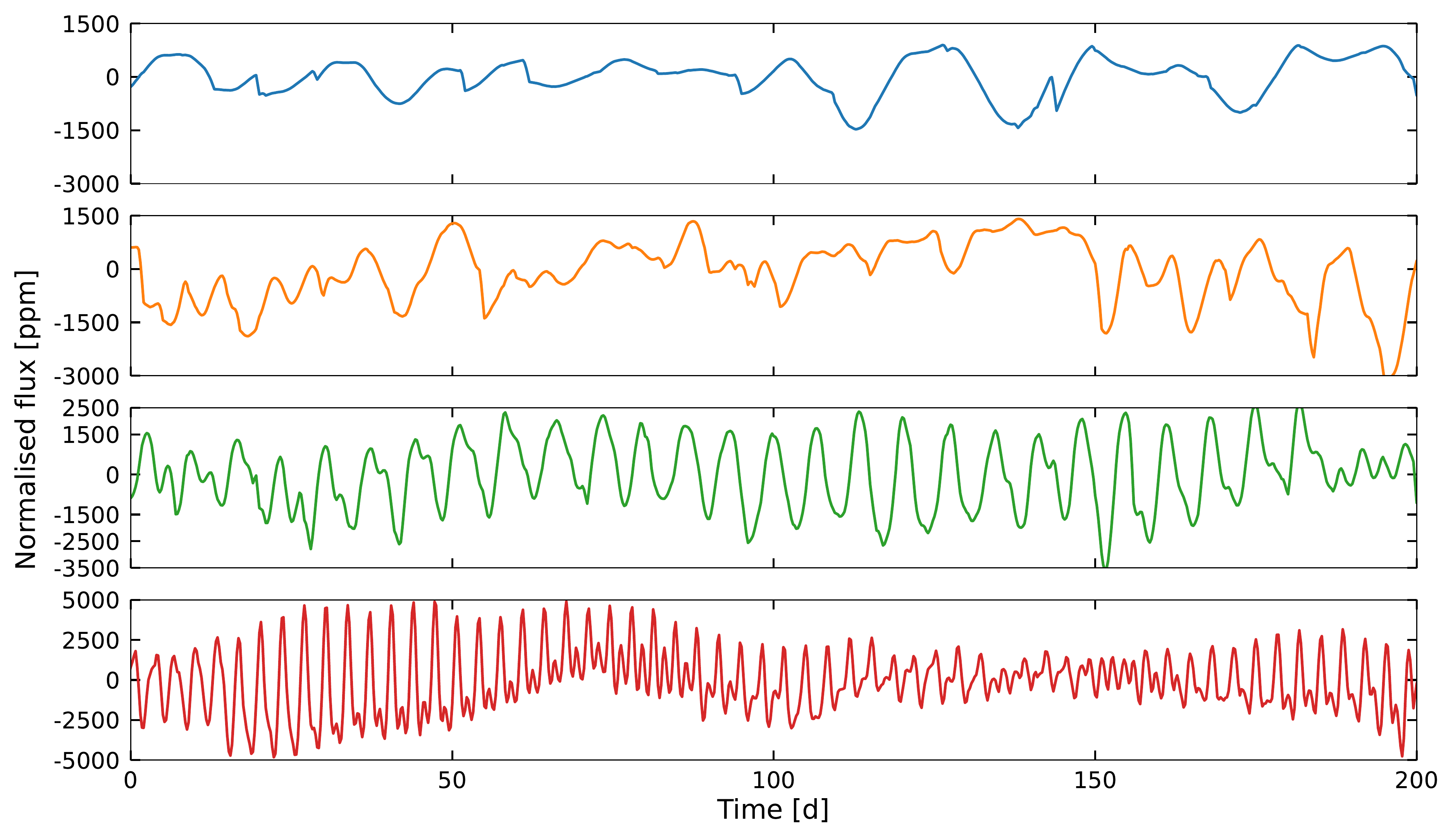}
\caption{Synthetic stellar light curves (LCs) for stars with different rotation rates as they would be observed in the \textit{Kepler} passband at an inclination of $i=90\degree$. Shown are rotation rates values of 1\rotrate{} (blue),  2\rotrate{} (orange), 4\rotrate{} (green), and  8\rotrate{} (red).  The degree of nesting is set to 0\%. Plotted is a 200-day fragment of the time-span from day 600--800 from the full 4--years of the synthetic LCs shown in Fig. \ref{fig:LCs_eq1}.}%Simulated light curves as they would be observed by \textit{Kepler} at an inclination of 90$^{\circ}$ with $p=$0. Blue solid lines represent 1\rotrate{}, orange 2\rotrate{}, green 4\rotrate{}, and red 8\rotrate{}.}
\label{fig:LCs_eq1_zoom}
\end{figure}

Firstly, we consider the LCs of stars that are observed equator-on ($i=$90$^{\circ}$).
Fig. \ref{fig:LCs_eq1_zoom} displays the de-trended LCs for the for the non-nested case ($p=0$) for different rotation rates. One can see that the faster the star is rotating, the higher its amplitude of variability. %Moreover, while individual transits of the spots are visible for the slower rotating stars, the rotation rate is too high to identify individual transits of the faster rotators.  
On the top panel (corresponding to the 1\rotrate{} case) individual dips in the LCs correspond to transits of different features, as the lifetime of the spots is generally smaller than the rotation rate. As this features appear randomly in time, the solar LCs appears quite irregular. In contrast, Figure~\ref{fig:LCs_eq1_zoom} bottom panel indicates that the LCs of the faster rotating stars appear more regular.% compared to their slower rotating counterparts.
This is because contrary to the 1\rotrate{} case, the spots on stars rotating faster than the Sun can last for several stellar rotations leading to a regular profile of the light curve. 

\begin{figure}[h!]
\centering
\includegraphics[width=\textwidth]{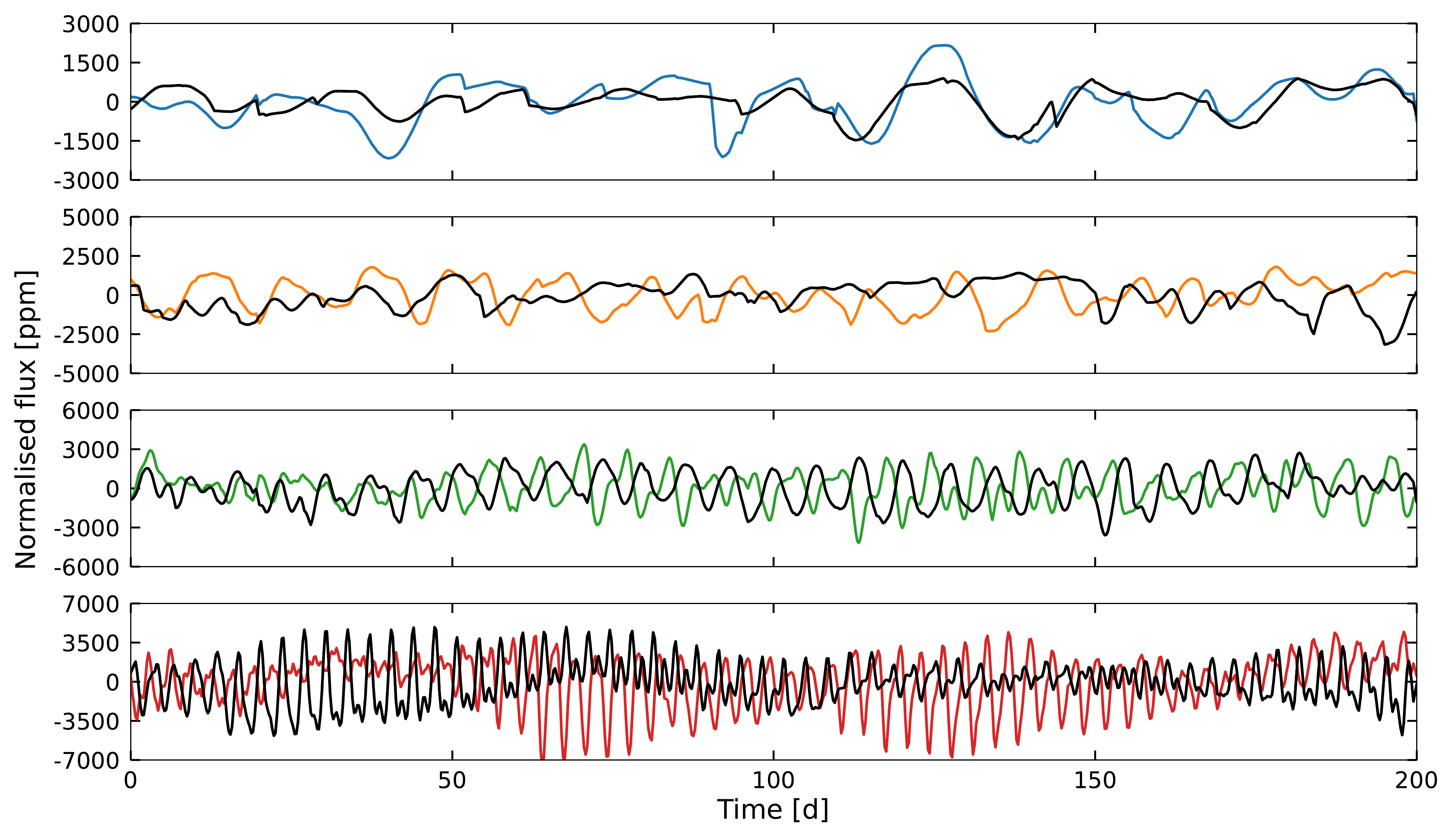}
\caption{The same as in Fig. \ref{fig:LCs_eq1_zoom} but the light curves are calculated with a nesting degree of 70\%. The light curves from Fig. \ref{fig:LCs_eq1_zoom}  are shown as black lines for comparison. Plotted is a 200-day fragment of the time-span from day 600--800 from the full 4--years of the synthetic LCs shown in Fig. \ref{fig:LCs_eq2}.}
\label{fig:LCs_eq2_zoom}
\end{figure}

\begin{figure}[h!]
\centering
\includegraphics[width=\textwidth]{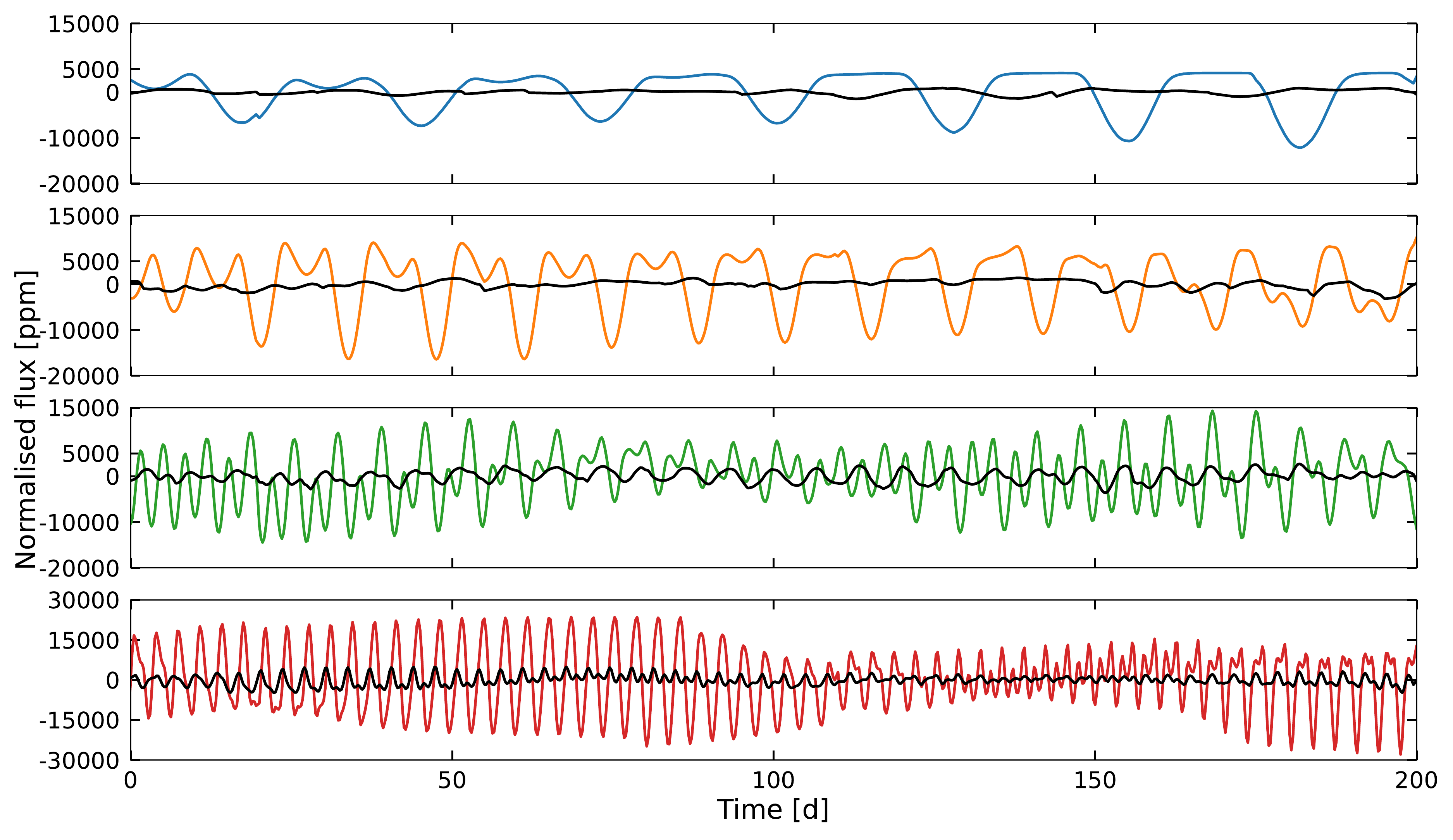}
\caption{The same as in Fig. \ref{fig:LCs_eq1_zoom} but the light curves are calculated with a nesting degree of 99\%. The light curves from Fig. \ref{fig:LCs_eq1_zoom}  are shown as black lines for comparison.Plotted is a 200-day fragment of the time-span from day 600--800 from the full 4--years of the synthetic LCs shown in Fig. \ref{fig:LCs_eq3}.}
\label{fig:LCs_eq3_zoom}
\end{figure}

The LCs shown in Fig. \ref{fig:LCs_eq2_zoom} are calculated with nesting $p=0.7$. One can see that the amplitudes of the variability increase for all four shown rotation periods.
The light curves also look more regular than those calculated with $p=0$.
We increase the nesting even further to $p=0.99$ in Fig.\ref{fig:LCs_eq3_zoom}. The change in the LCs with this high degree of nesting compared to the non-nested case is remarkable. The amplitude of the LCs is enhanced for all cases and the runs with $p=0.99$ exhibit regular patterns, even the solar case. Such a behaviour can be understood by looking at the corresponding butterfly diagrams shown in 
Sect. \ref{butterfly}. 

\begin{figure}
\centering
\includegraphics[width=\textwidth]{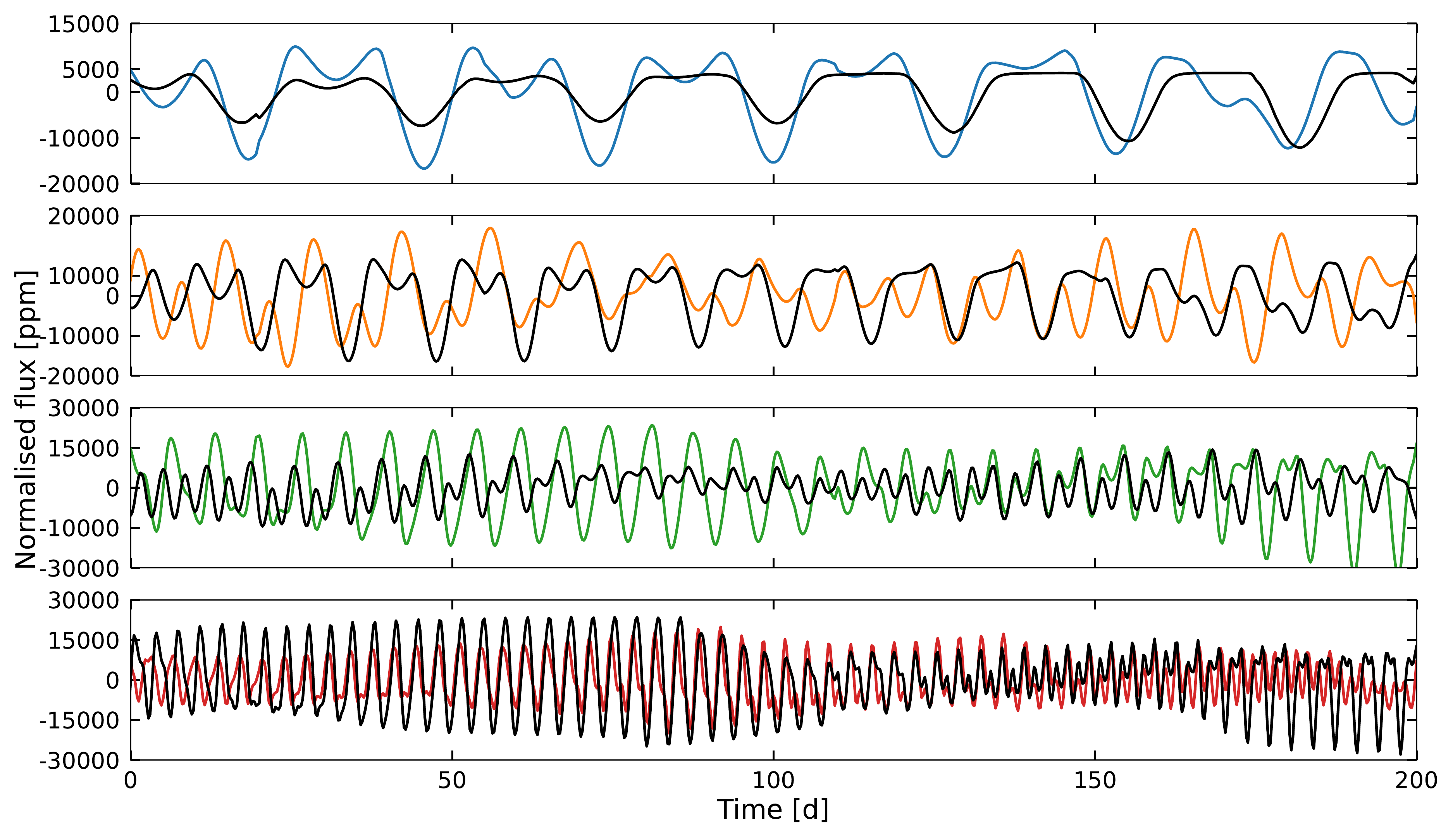}
\caption{The same as in Fig. \ref{fig:LCs_eq1_zoom} but the light curves are calculated with a nesting degree of 99\% and an activity scaling of $\tilde{s} = 2\cdot\tilde{\omega}$. The black lines in each panel are the corresponding cases $\tilde{s} = \tilde{\omega}$ from Fig. \ref{fig:LCs_eq3_zoom}.
Plotted is a 200-day fragment of the time-span from day 600--800 from the full 4--years of the synthetic LCs shown in Fig. \ref{fig:LC_eq_act}.}

%Synthetic stellar light curves (LCs) for stars with different rotation rates as they would be observed by \textit{Kepler} at an inclination of $i=90\degree$. Blue solid lines represent 1\rotrate{}, orange 2\rotrate{}, green 4\rotrate{}, and red 8\rotrate{} with a nesting degree of 99\% and an activity scaling of $\tilde{s} = 2\cdot\tilde{\omega}$. The black lines in each panel are the corresponding cases from Fig. \ref{fig:LCs_eq3_zoom}. Shown is a 200-day fragment of the time-span from day 600--800 from the full 4--years of the synthetic LCs shown in Fig. \ref{fig:LC_eq_act}.}
\label{fig:LC_eq_act_zoom}
\end{figure}

It is possible that solar cycle 22 does not represent the maximum level of activity the Sun is capable of  \citep[see, e.g.,][]{Timo2020}. Thus, we also investigate what happens when we scale our calculations to a cycle twice as active as cycle 22. While we used $\tilde{s} = \tilde{\omega}$ before, we are scaling the activity level now as $2\cdot \tilde{\omega}$, meaning that a star with solar rotation rate exhibit two times more emergences than the Sun in cycle 22, a star with twice the solar rotation rate  exhibits four times more emergences, and so forth. We show the comparison between the $p=0.99$ cases for $\tilde{s} = \tilde{\omega}$ (black solid lines) and $\tilde{s} = 2\cdot\tilde{\omega}$ (coloured lines) in Fig. \ref{fig:LC_eq_act_zoom}. Not surprisingly, the variability increases if the number of emergences is increasing for 1\rotrate{}, 2\rotrate{}, and 4\rotrate{} cases. However, the  8\rotrate{} $\tilde{s} = 2\cdot\tilde{\omega}$ displays a lower amplitude than the $\tilde{s} = \tilde{\omega}$ case.  This implies some sort of "saturation" of the variability at high activity and high nesting degree. %We will address this point later in this manuscript in more detail.  
The full 4--years of the simulated LCs presented in this section so far can be found in Appendix \ref{full_LC} Fig. \ref{fig:LCs_eq1}--\ref{fig:LC_eq_act}.

\begin{figure}
\centering
\includegraphics[width=\textwidth]{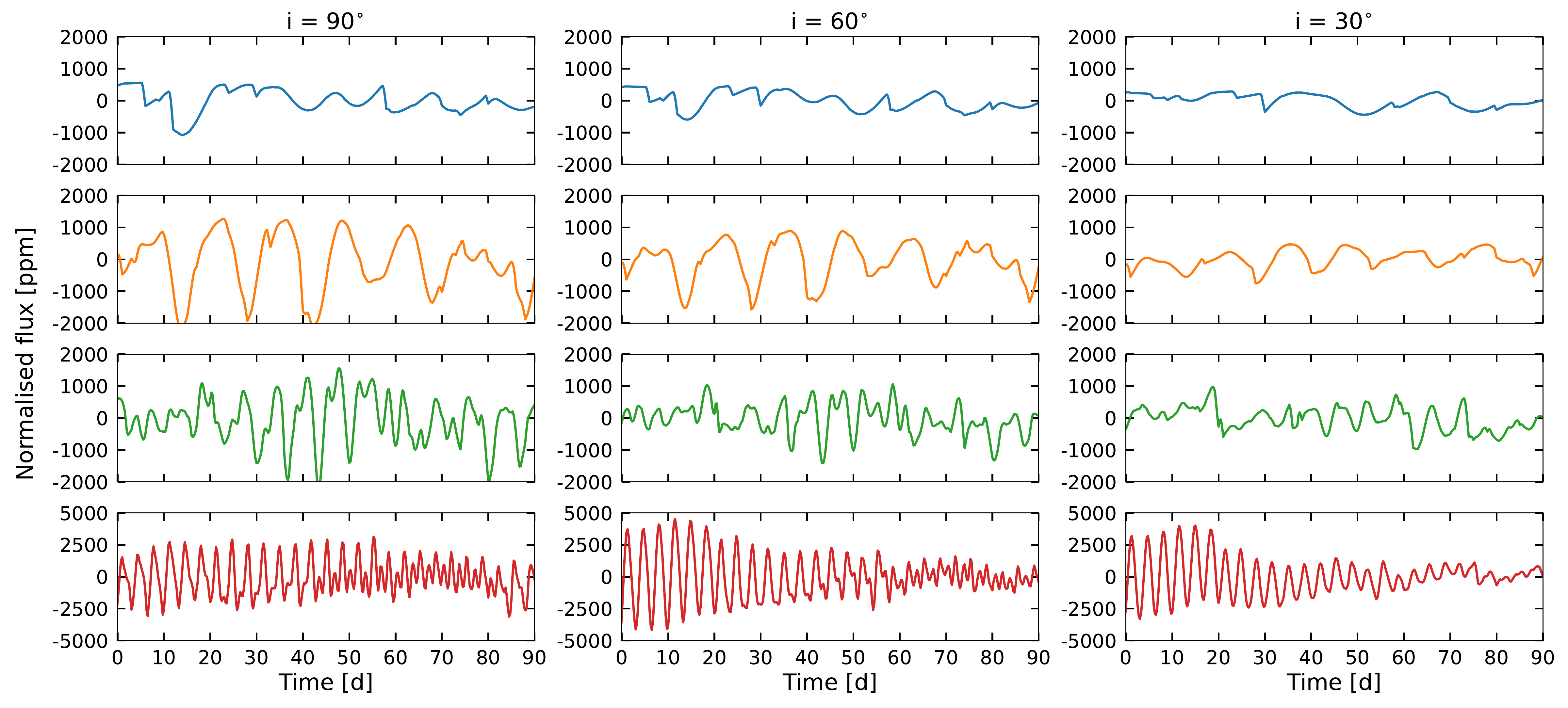}
\caption{Synthetic stellar light curves (LCs) for stars with different rotation rates as they would be observed by \textit{Kepler} at different inclinations. Blue solid lines represent 1\rotrate{}, orange 2\rotrate{}, green 4\rotrate{}, and red 8\rotrate{} with a nesting degree of 0\%. Shown is a 90-day fragment of the time-span from days 900-990 from the full 4--years of the synthetic LCs shown in Fig. \ref{fig:LCs_eq1}.} 
\label{fig:LC_0_nest_incl}
\end{figure}

Next, we consider the inclination effect. For demonstration, we limit ourselves to the non-nested cases with different rotation rates. 
We show the time-span of 0--90 days from Fig.~\ref{fig:LCs_eq1_zoom} for inclinations of 90\degree, 60\degree and 30\degree in Fig. \ref{fig:LC_0_nest_incl}.
For 1\rotrate{}, the amplitude of the variability decreases with decreasing inclination. The shape of the transits change due to projection effects (foreshortening) and centre-to-limb variations of the spot contrasts. 
For 2 and 4\rotrate{}, the amplitudes of the LCs are decreasing for the inclinations shown here as well. Interestingly, the situation changes for the  8\rotrate{} case. The amplitude increases from $i=90\degree$ to $60\degree$ and then decreases from $i=60\degree$ to $i=30\degree$. The amplitude of variability observed at  $i=30\degree$ is larger than that at $i=$90\degree. Even for lower inclinations, the 8\rotrate{} runs exhibit large amplitudes in its variability. This can be explained with the help of the butterfly diagrams in Sect. \ref{butterfly} Fig. \ref{fig:butterfly} and \ref{fig:butterfly2}. For the case of 1\rotrate{} all regions emerge within $\pm$30 \degree around the equator. The Coriolis-force is getting stronger with increasing rotation rate and for the 8\rotrate{} star, the BMRs can emerge at latitudes up  to $\pm$70 \degree around the equator. However, a "zone of avoidance" of about $\pm$10\degree opens up at the equatorial region. For $i=90\degree$ case the high-latitude spots appear close to the limb where they effect on brightness is significantly reduced by the foreshortening.  %their contrast is lower for $i=$90\degree. 
With decreasing inclination the spots, contrary to the solar case, shift towards the centre of the visual disc and their effect on brightness increases.

\section{Comparison to observations}

In the following we test the trends established in the previous section against observations. For this, we take the rotation periods from \cite{McQuillan2014} and the measurements of the variability from \cite{Timo2020}. \cite{McQuillan2014} were able to measure rotation periods of 133,100 stars. As we focus on Sun-like stars in the present work, we limit the effective temperatures between 5500--6000 K according to the effective temperatures published by \cite{Mathur2017}. This temperature cut limits the sample to 8921 stars.
We express the variability through \rvar{} \citep{Timo2020}. For this, we divide the simulated LCs into 90--day segments. Within each segment, we calculate the difference between the extrema \citep[instead of taking the difference between 95th and 5th percentiles since our calcualtions are free form the observational noise, see][]{Nina1} before dividing by the mean value. 

\begin{figure}
\begin{subfigure}[b]{.48\textwidth}
\centering
\includegraphics[width=\linewidth]{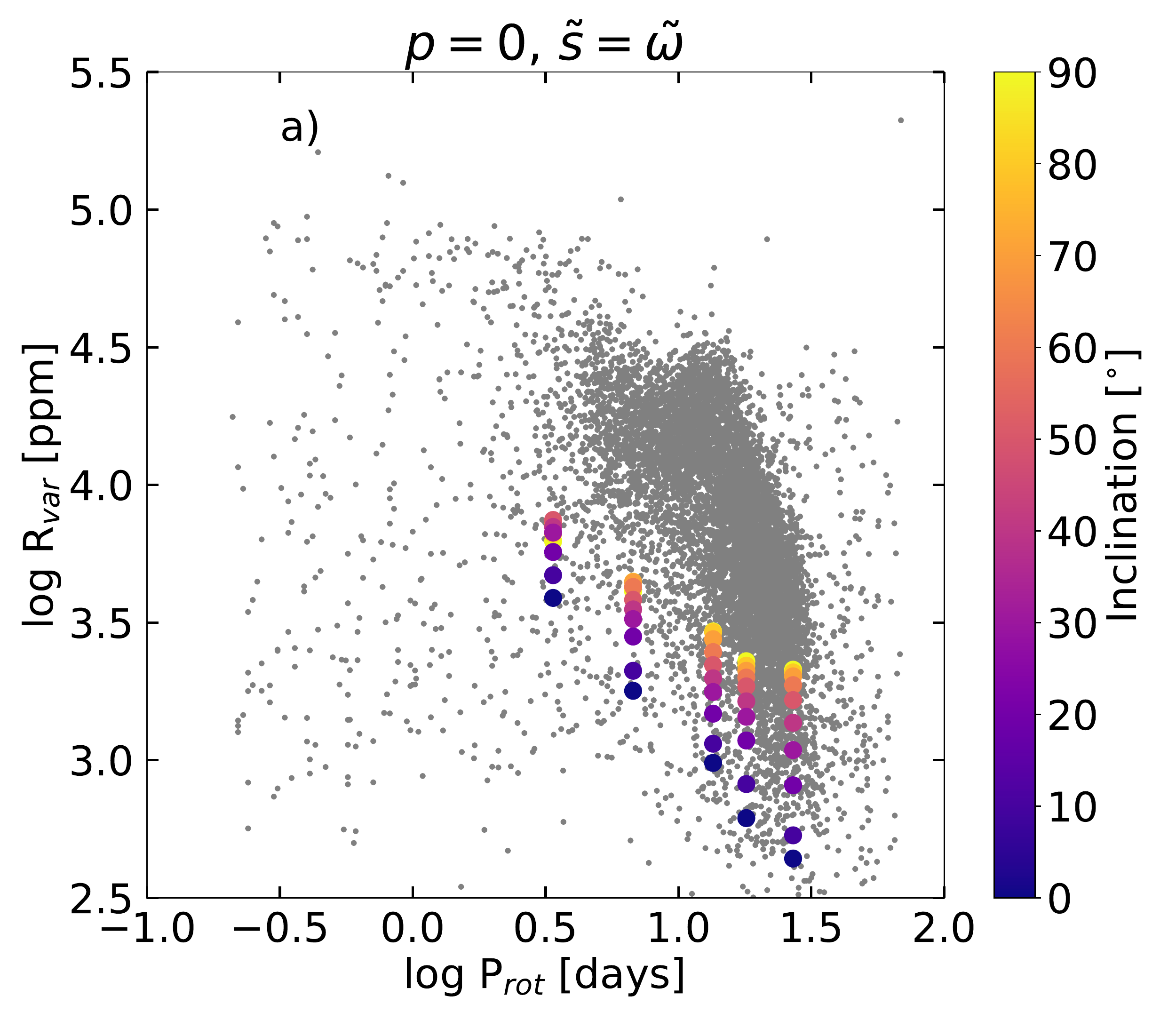}
        %\caption{}\label{fig:fig_a}
\end{subfigure}
\begin{subfigure}[b]{.48\textwidth}
\centering
\includegraphics[width=\linewidth]{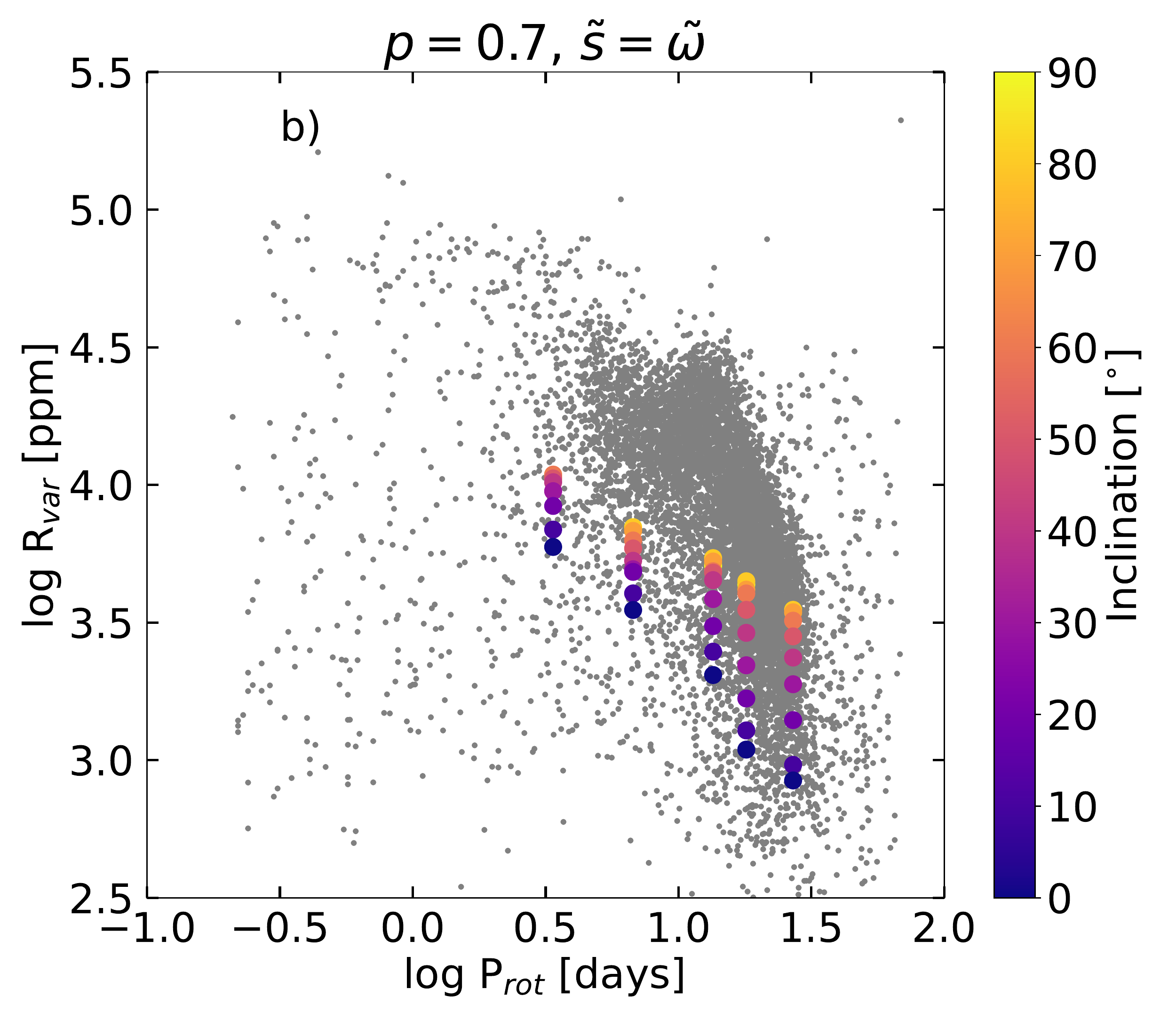}
%\caption{}\label{fig:fig_b}
\end{subfigure}

\medskip

\begin{subfigure}[b]{.48\textwidth}
\centering
\includegraphics[width=\linewidth]{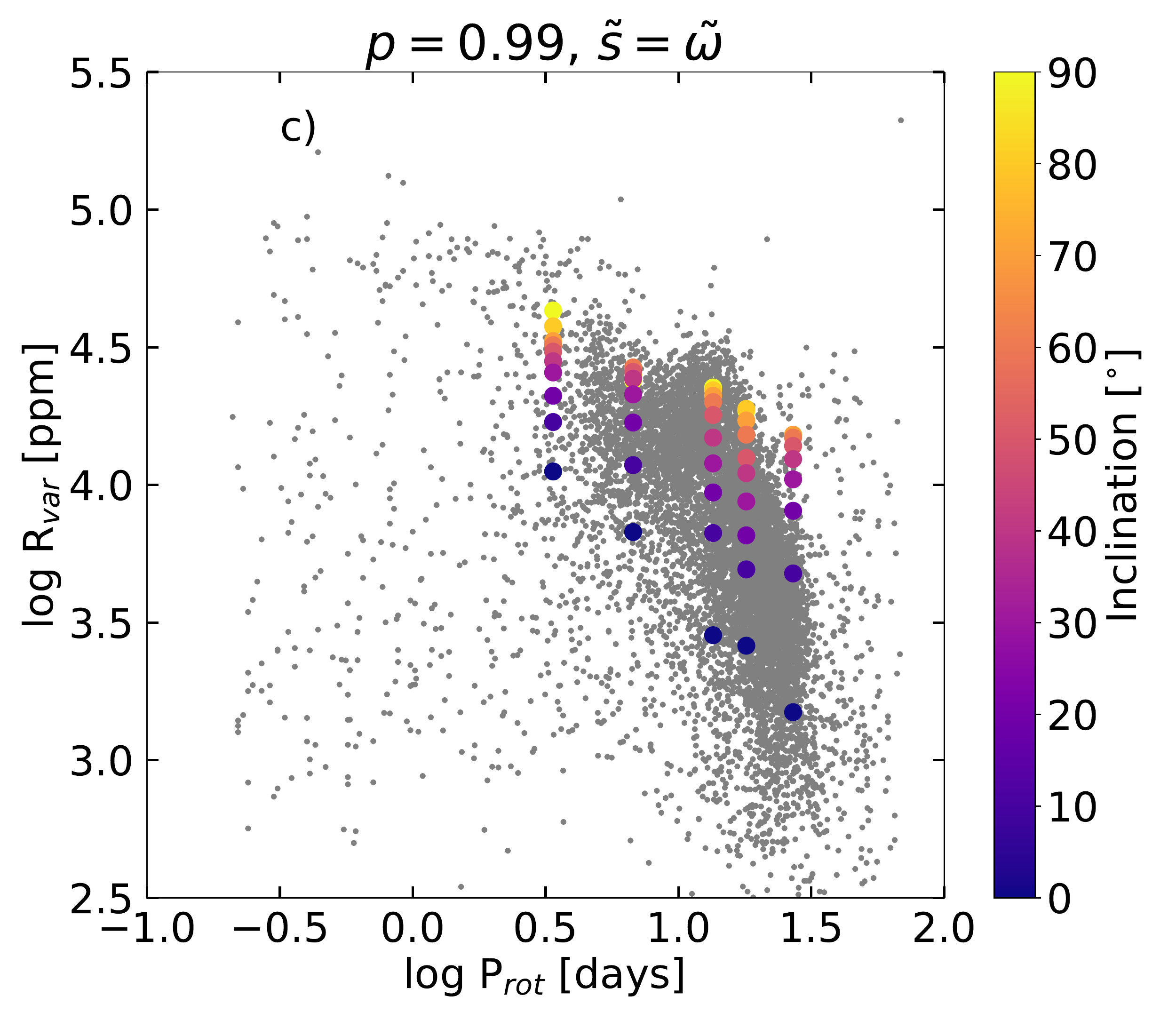}
        %\caption{}\label{fig:fig_a}
\end{subfigure}
\begin{subfigure}[b]{.48\textwidth}
\centering
\includegraphics[width=\linewidth]{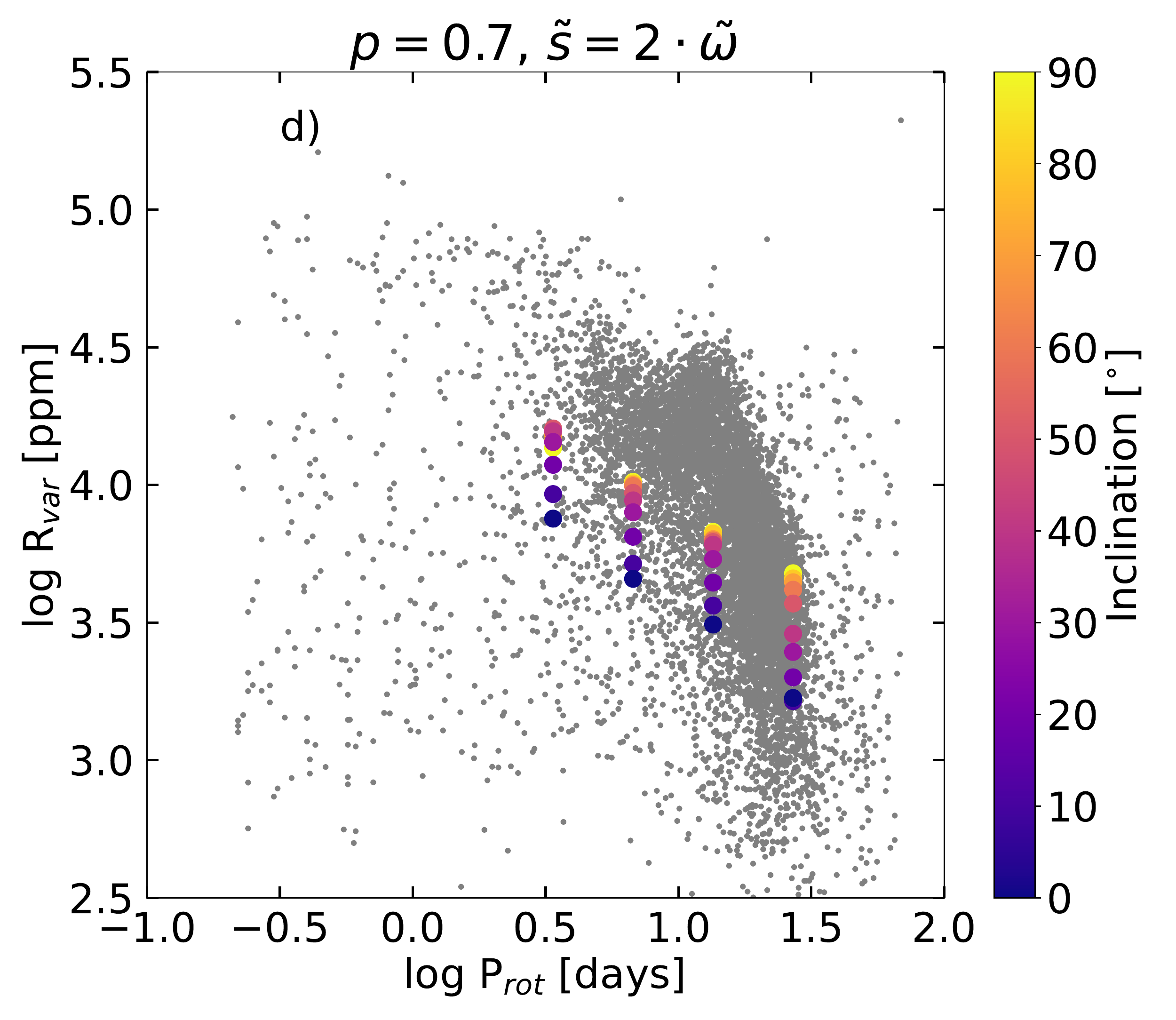}
%\caption{}\label{fig:fig_b}
\end{subfigure}

\medskip

\begin{subfigure}[b]{.48\textwidth}
\centering
\includegraphics[width=\linewidth]{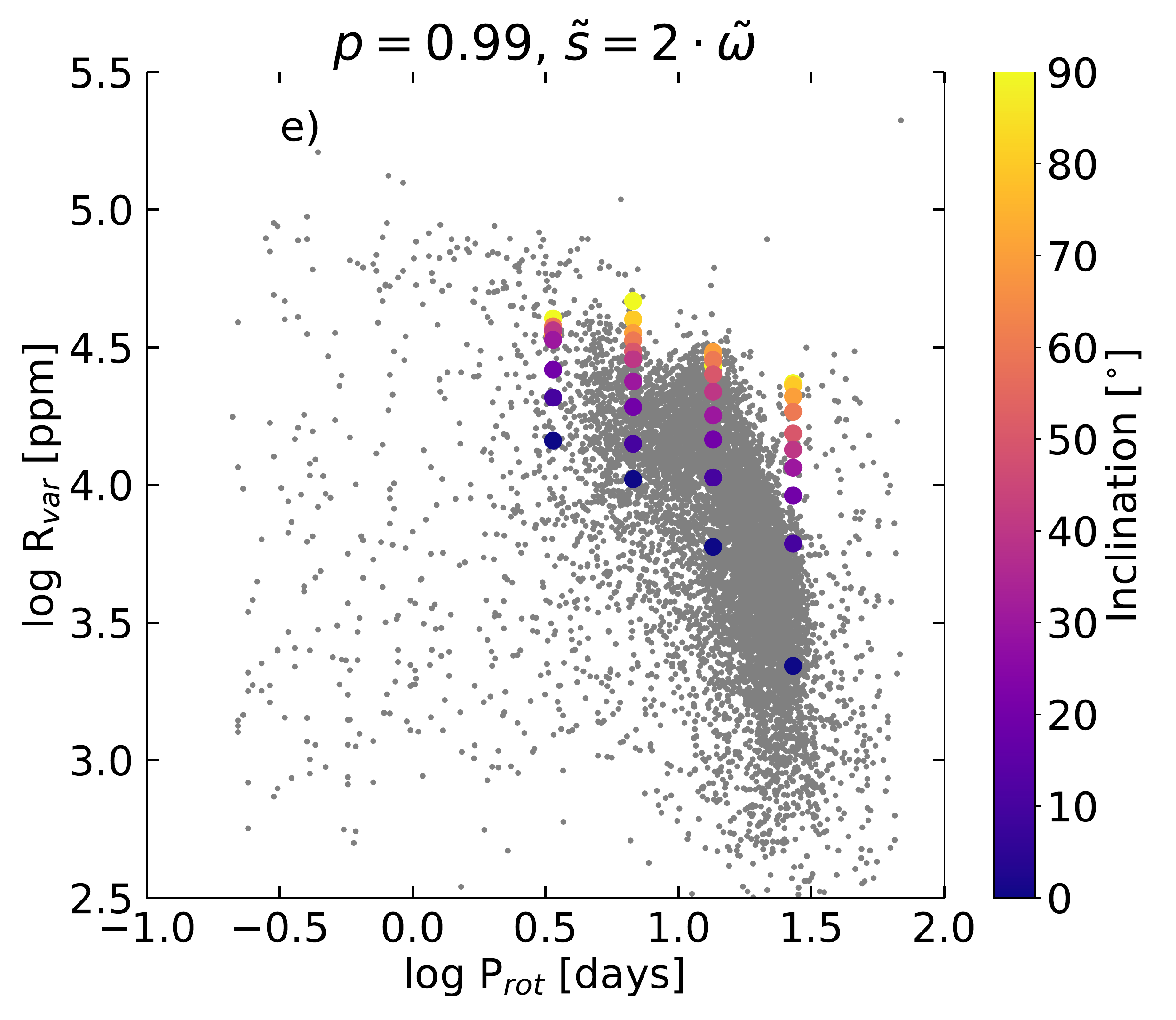}
%\caption{}\label{fig:fig_b}
\end{subfigure}
\begin{minipage}[b]{.48\textwidth}
\caption{Comparison of \rvar{} as a function of the rotation rate between \textit{Kepler} stars with effective temperatures between 5500--6000 K (grey dots) and the modelled stellar sample. Each panel includes different nesting probabilities $p$ and scalings between the number of active region emergences $\tilde{s}$ and the rotation $\tilde{\omega}$ in terms of the solar units. The different colours indicate the inclination of the stars of the modelled sample.}
\label{fig:comp_timo}
\end{minipage}

\end{figure}

We show the comparison between the calculated \rvar values of our simulated stellar samples and the sub-sample of \textit{Kepler} stars (grey dots) in Fig. \ref{fig:comp_timo}. Fig. \ref{fig:comp_timo} panel a shows the runs with $p=0$ and clearly, all of our calculated variabilities lie towards the lower edge of the distribution, especially for the faster rotators. With increasing nesting level (Fig. \ref{fig:comp_timo} panels b and c), \rvar{} is increasing and the values move towards the upper edge of the distribution. $p=0.99$ overestimates the variability of the solar case,  but is able to explain  the distributions of variabilities for the faster rotating stars. This implies that according to our modelling the degree of nesting should increase with decreasing rotation period. If the emergence rate is scaled twice as strong as cycle 22 \rvar{} is slightly increasing (Fig. \ref{fig:comp_timo} panels d and e), however, its effect is smaller than that of the nesting.

\begin{figure}[!h]
\centering
\includegraphics[width=.48\columnwidth]{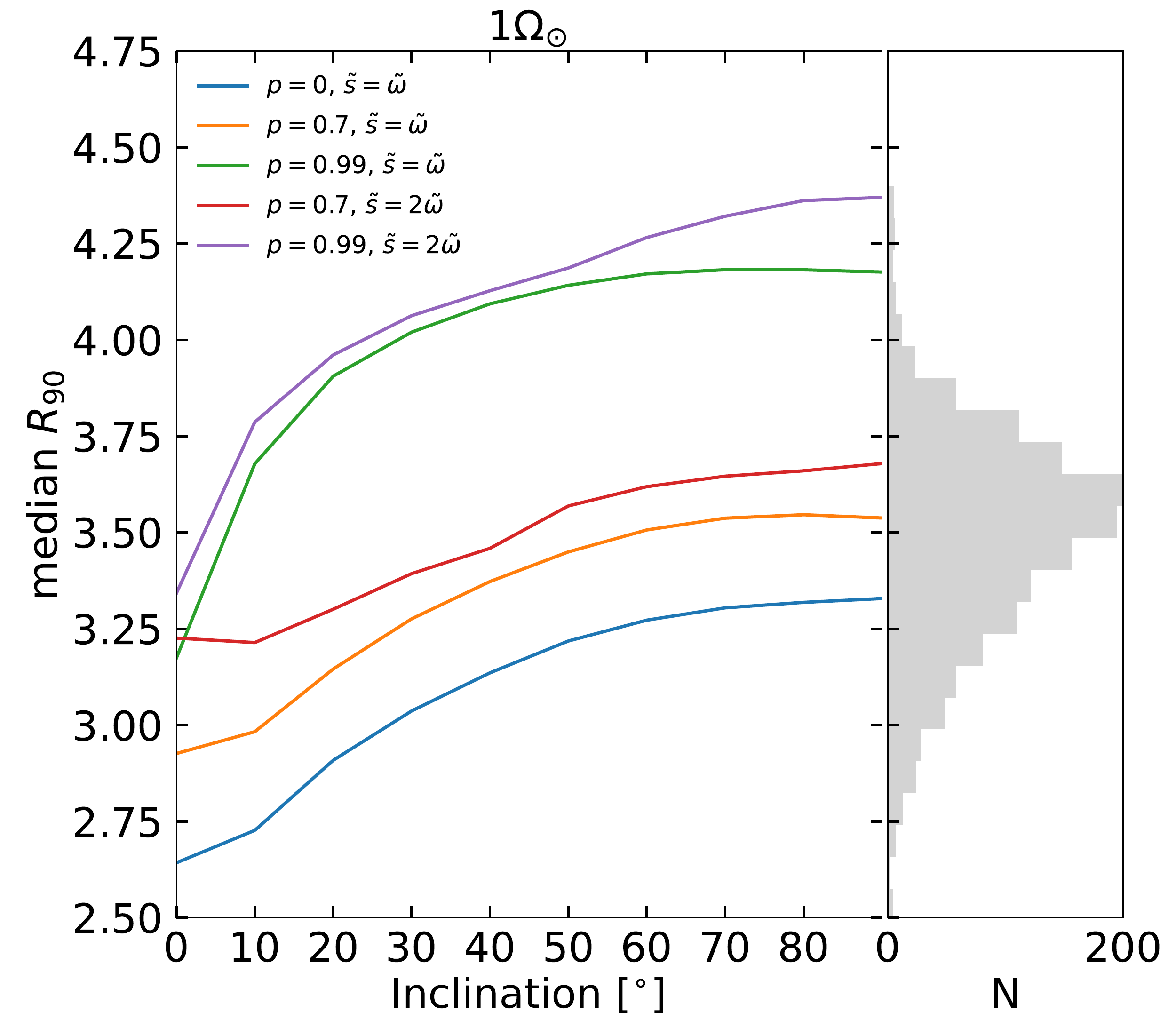}\quad
\includegraphics[width=.48\columnwidth]{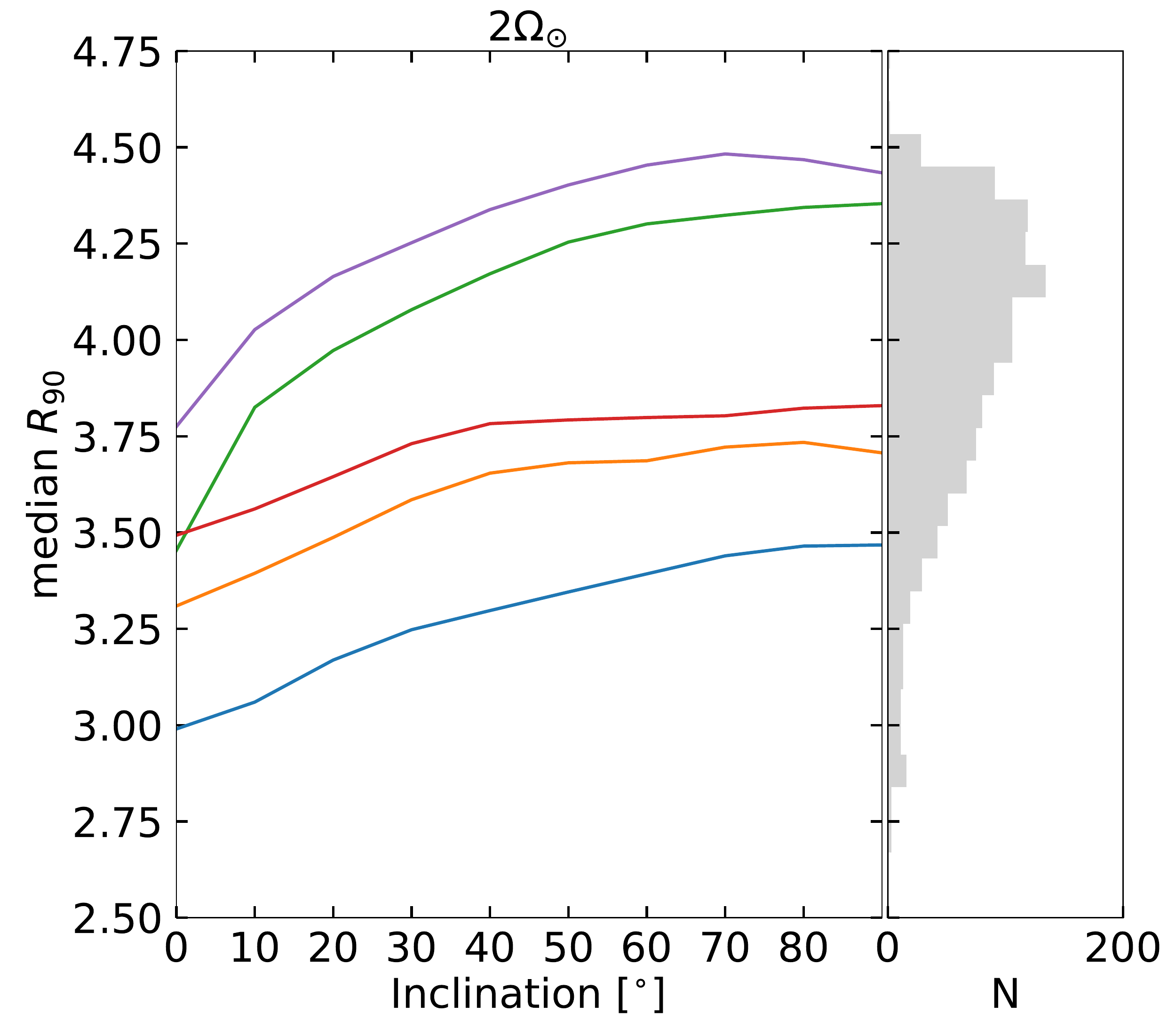}
\medskip
\includegraphics[width=.48\columnwidth]{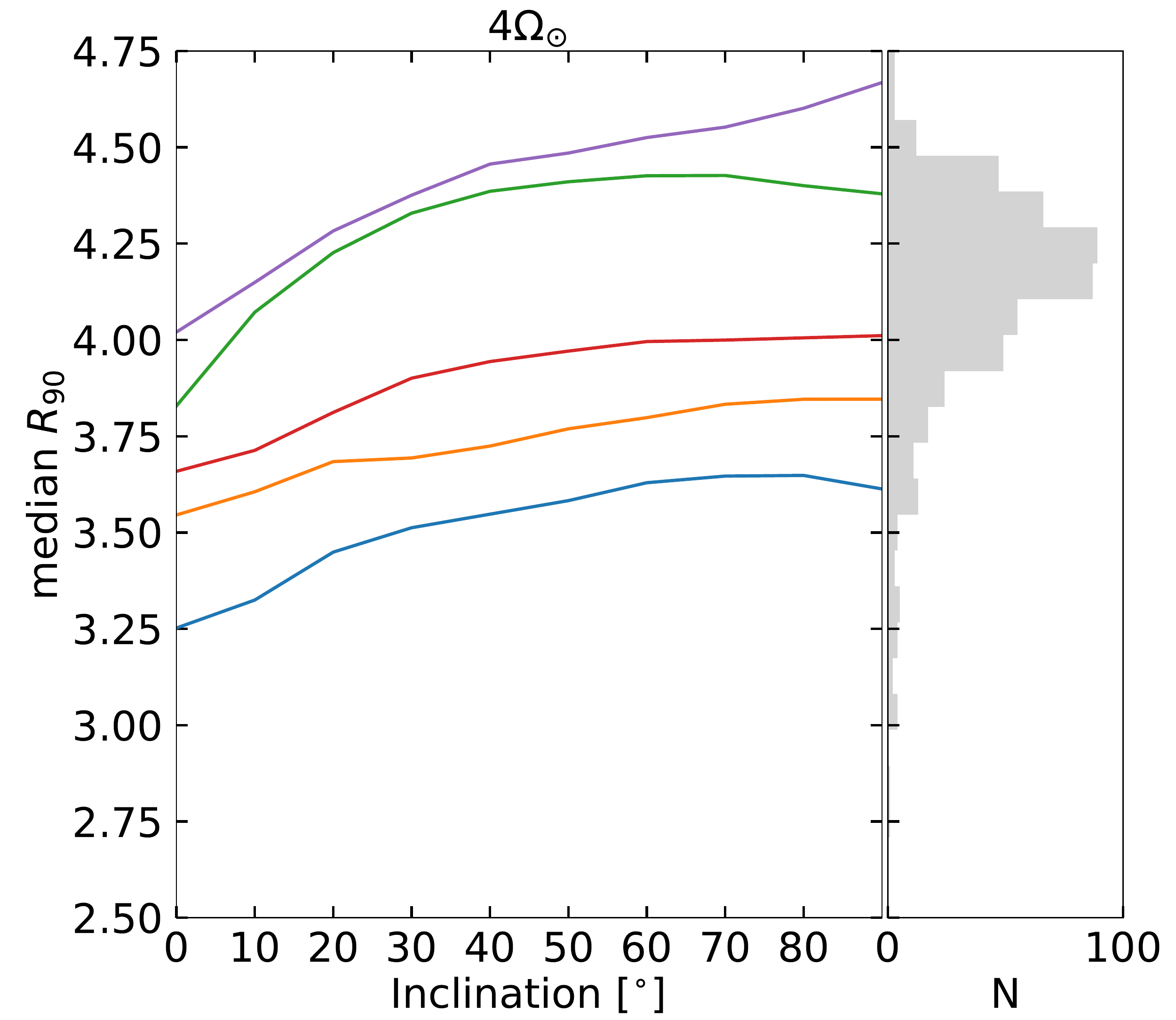}\quad
\includegraphics[width=.48\columnwidth]{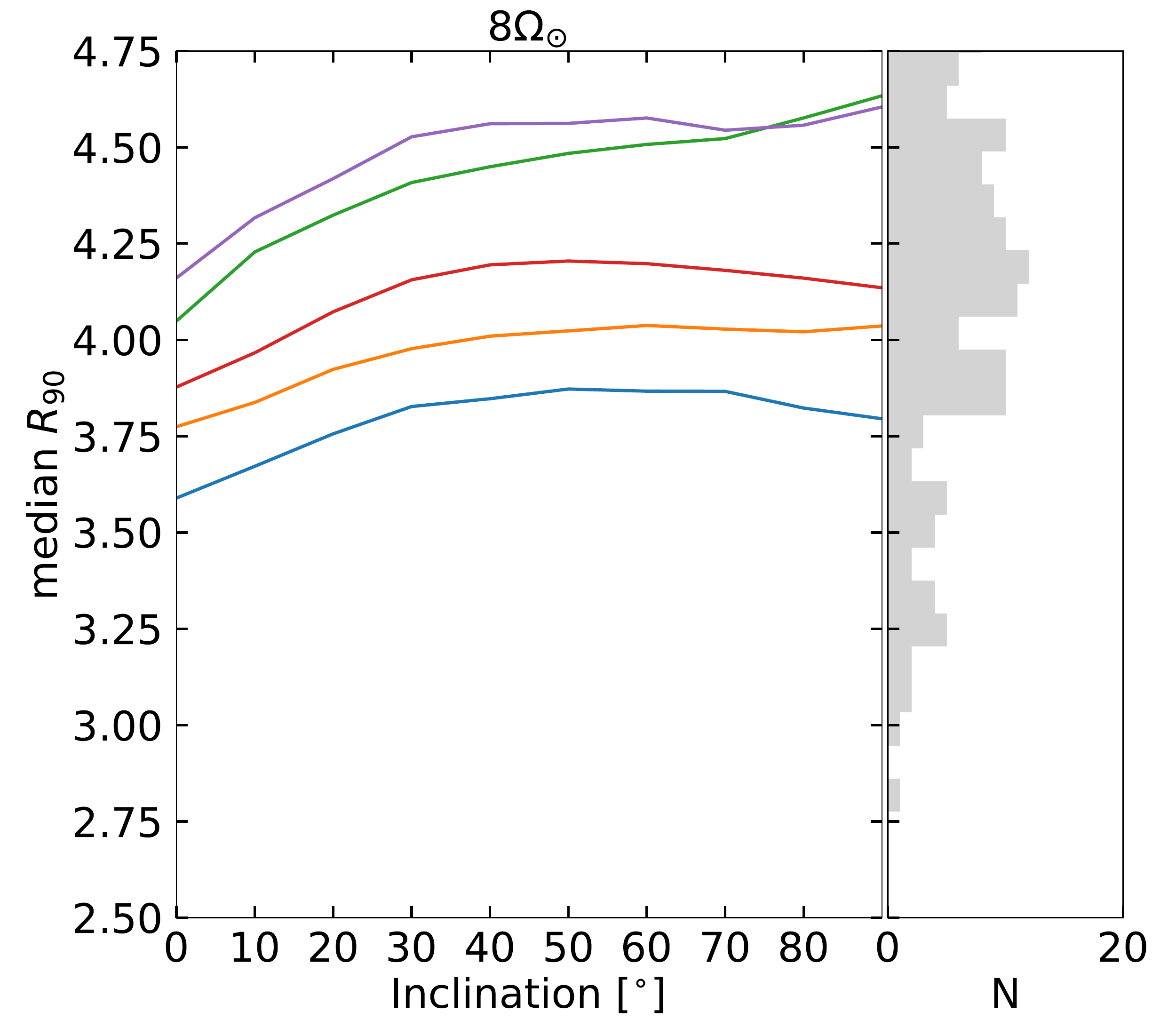}
\caption{Dependence of \rvar on the inclination. Each panel represents different realisations of the BMR emergence for a given rotation rate. The grey histograms are drawn from the \textit{Kepler} sample with the limitations of [23/X, 27/X] days for a given rotation rate X in units of the solar rotation.}
\label{fig:histograms}
\end{figure}

For a more detailed comparison between the simulations and the observations, we bin the distribution of observed variabilities. Namely, we compared the variabilities returned by our model for a star rotating X times faster than the Sun, with a sample of stars between [23/X, 27/X] days from the \textit{Kepler} sample. This comparison is shown in Fig. \ref{fig:histograms}. The histograms in grey display the range of variabilities within each of the rotation period bins. We note, that the number of stars used to draw the histograms from decreases from panel a to d. As seen already from Fig. \ref{fig:comp_timo}, the increase of the degree of nesting is of importance to explain the observed distribution of variabilities, the higher level of activity in terms of active regions for a given rotation rate is a lesser factor.

\section{Conclusions}

We coupled the model for emergence and surface transport of magnetic flux in Sun-like stars by \cite{Isik2018} with a model \citep[loosely based on the approach presented in][]{Nina1} for calculating stellar brightness variations. 
This allowed us to compute light curves of stars with rotation rates between 1 and 8\rotrate{} as they would be observed by \textit{Kepler} at different inclination angles. Following up on the findings of \cite{Isik2018} and \cite{Isik2020}, we investigated the impact of nesting of magnetic features emergences on the light curves and the amplitude of the variability.%
We compare the output of our model to the observed variabilities of \textit{Kepler} stars with temperatures between 5500--6000K. In particular, we aim at explaining the dependence of the amplitude of the variability on the rotation rate. Recently, \cite{Isik2020} showed that the solar model without nesting underestimates the variability of stars with known near-solar rotation periods \citep[see also][]{Timo2020}.  We found that the same is true for stars rotating faster than the Sun. Our runs without nesting dramatically underestimate the variability of all rotation rates.
We show that the observed dependence of {\it Kepler} stars variabilities on the rotation period can be explained by the increase of nesting degree on rotation rate.

%nesting of magnetic features increases with rotation rate.

%and scalings of the activity level in terms of active region emergence. 
%on stars with rotation rates from 

\chapter{Summary and prospects}\label{thesis_conclusio}

In this thesis I build a model for calculating the brightness variations of the Sun and Sun-like stars as they would be observed in different spectral passbands used in stellar instrumentation and at various inclinations. This model further allows applying lessons learnt from the solar paradigm to stars with higher activity levels and rotation rates.

In Sect.~\ref{sec:paper_1} I presented the model, which has enabled us to conduct not only the research presented in this thesis, but also opened up possibilities for further investigations beyond the photometric variability.  The model utilises a surface flux transport model, from which I obtain the area coverages of magnetic features,  which are then fed into the SATIRE model that attributes each of these features a brightness contrast depending on the position on the disc, the position of the observer and the wavelength. A strong advantage of this model over previous models is that it allows calculating variability not only on the activity timescale but also on the timescale of solar rotation. I have shown that the inclination effect enhances the variability on the activity timescale, in agreement with previous works \citep{Schatten1993,Knaack2001,Shapiro2017}. On the other hand, the effect of the inclination decreases the rotational variability.
The new model allowed me to disentangle the variability components arising from the evolution of magnetic features on the solar surface and from the solar rotation. It appears that solar brightness variations on timescales below 5 days are mainly attributed to the evolution of magnetic features and not to the rotation.

In Sect. \ref{sec:paper_2} I calculated the amplitude of the solar rotational variability as it would be observed through different filter systems regularly used in stellar observations.
I have shown that the effect of the inclination affects solar-stellar comparison. The mean variability in a sample of randomly inclined stars is on average 15\% lower than in stars observed from their equatorial planes. Interestingly, the amplitude of the TSI variability appeared to be an almost ideal representation of the solar variability for comparison to \textit{Kepler} stars with unknown inclinations.
I presented an explanation of why the coverages of stellar surfaces by faculae grow slower with activity than those by spots in Sect. \ref{sec:paper_3}. This tendency was observed for the Sun and it plays a crucial role in explaining the overturn from faculae- to spot-dominated regimes of long-term variability in stars more active than the Sun.
For this, I extended the model presented in Sect. \ref{sec:paper_1} to activity levels higher than solar. The results suggest that the cancellation of flux attributed to the faculae decreases with increasing activity level, due to flux cancellation due to encounters of flux patches with opposite polarity.
Finally, in Sect. \ref{sec:paper_4} I modelled brightness variations of stars rotating faster than the Sun. The calculations have revealed that the observed trends in stellar variability by \textit{Kepler} (e.g. the dependence of the amplitude of the variability on the rotation rate), can be explained if nesting of magnetic features on stellar surfaces increases with rotation rate.

All in all, I have investigated the impact of the inclination, the spectral passbands, and the distribution of the surface magnetic features on the variability of Sun-like stars. All these calculations have been, however, performed for stars with solar fundamental parameters.  %Stars however might also vary in their fundamental parameters.
\cite{witzkeetal2018} and \cite{Witzke2020} investigated the effect of the metallicity on the  variability of solar-like stars. They have found that  metallicity has a strong effect on facular contrasts relative to the quite Sun. \cite{Timo2020_2} combined the model I presented in Sect.~\ref{sec:paper_1} with the calculations of \cite{witzkeetal2018, Witzke2020} to calculate the variability of stars with solar rotation periods, but with various values of metallicity and observed at arbitrary inclinations. In particular, they found that metallicity has a crucial effect on the detectability of the rotation period.  On the Sun, the facular and spot contributions to the rotation signal almost fully compensate each other. This would significantly sophisticate the determination of solar rotation periods from the photometric time series if the Sun were observed by \textit{Kepler}. The contrast of faculae changes for stars with non-solar metallicity and stars move out of this aforementioned compensation regime. These results moreover allowed to explain the discrepancy between the number of stars with near-solar rotation rate in the \textit{Kepler} field, as predicted by galaxy evolution models \citep{VanSaders2019} and the actual number of stars with detected rotation rates \citep{McQuillan2014}. 

Stellar magnetic activity is often characterised by the emission in the Ca II H\&K lines. The S-index is one way of quantifying this emission for inter-comparison between stars. However, the S-index is contaminated by the photospheric contribution (e.g. the effective temperature of the star). Hence, other proxies, such as the  $R_{HK}'$ index \citep{Noyes1984, Rutten1984} have been introduced. $R_{HK}'$ is a derivative index of the S-index, where flux of the emission lines is corrected for the photospheric contribution.
\cite{Radick2018} used measurements of stellar $R_{HK}'$ to compare the chromospheric variabilities of the Sun with the other Sun-like stars and found that the Sun lies above the regression line representing their stellar sample, i.e. the Sun has a rather vigorous cycle in the chromospheric activity in comparison to stars with mean level of solar chromospheric activity. They noted that such a comparison is subject to a bias arising from the strength of the solar cycle.  \cite{Sowmya2020} have used the model presented in Sect. \ref{sec:paper_1} together with the calculations of the Ca II H\&K line profiles performed with the 
RH code \citep{Uitenbroek2001} to obtain time-series of the solar S-index as they would be observed at different  inclinations for the last 300 years. It was found that with decreasing inclination, the S-index variability decreases weakly on the activity cycle timescale, but strongly on the rotational timescale. When comparing the solar chromospheric variability over the last 23 cycles to the sample used in \cite{Radick2018}, it became clear that depending on the cycle strength, the Sun lies either below or above the regression presented by \cite{Radick2018}. Besides, the mean level of solar chromospheric variability during the last 23 cycles was shown to be in agreement with respect to other stars having near-solar activity level.

%astrometry- Sowmya

I have amply examined the impact of the transit and evolution of the magnetic features on the brightness of stars. The \textit{Gaia} mission provides not only photometry of stars, but also astrometry.  Astrometric measurements include those of stellar position, the parallax, and the proper motion. \cite{Sowmya2020_2} used the model presented in Sect. \ref{sec:paper_1} to calculate the astrometric signals in stars with solar rotation and activity level and found, that the astrometric signal is too small to be detected by \textit{Gaia}. Preliminary results of astrometric calculations employing the model of stellar variability of stars rotating faster than the Sun (as presented in Sect. \ref{sec:paper_4} have shown that the astrometric signal of these stars is higher and hence could be detected by \textit{Gaia}.

\bibliography{thesis}%if you have your bibliographic information in thesis.bib
\appendix
\cleardoublepage
\thispagestyle{empty}
\vspace*{2.2cm}
{\Huge\bf Appendix}
\chapter{Power spectra of solar brightness variations for TESS and Strömgren \textit{b} and \textit{y}}\label{app:paper1}

\begin{figure}[!h]
\centering
\includegraphics[width=0.75\textwidth]{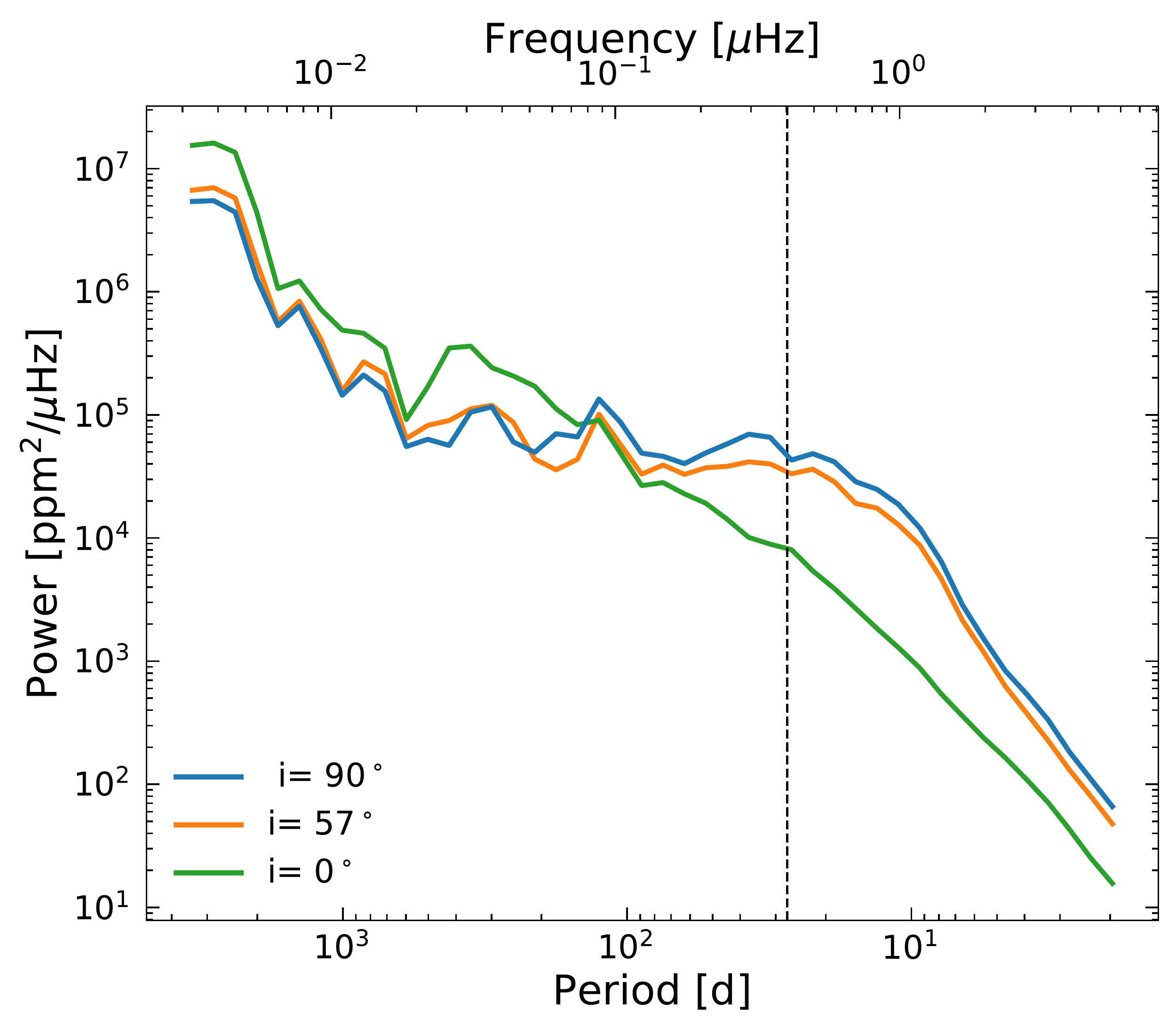}
\caption{Power spectra of the solar brightness variations for the TESS passbands for different inclinations.} 
\label{fig:PS_TESS}
\end{figure}

\begin{figure}[!h]
\centering
\includegraphics[width=0.75\textwidth]{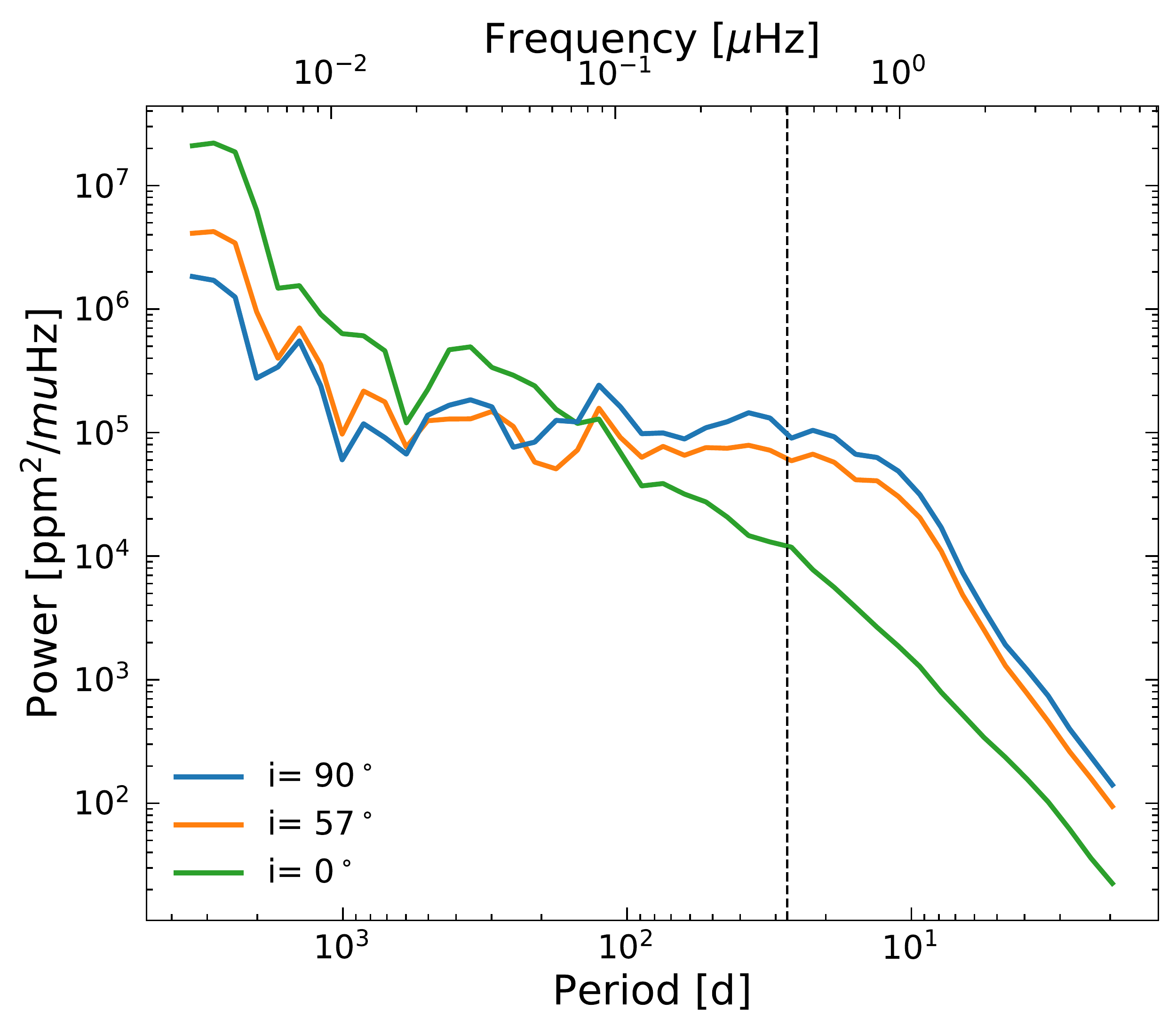}
\caption{Power spectra of the solar brightness variations for Strömgren \textit{b} for different inclinations.} 
\label{fig:PS_Str_b}
\end{figure}

\begin{figure}[!h]
\centering
\includegraphics[width=0.75\textwidth]{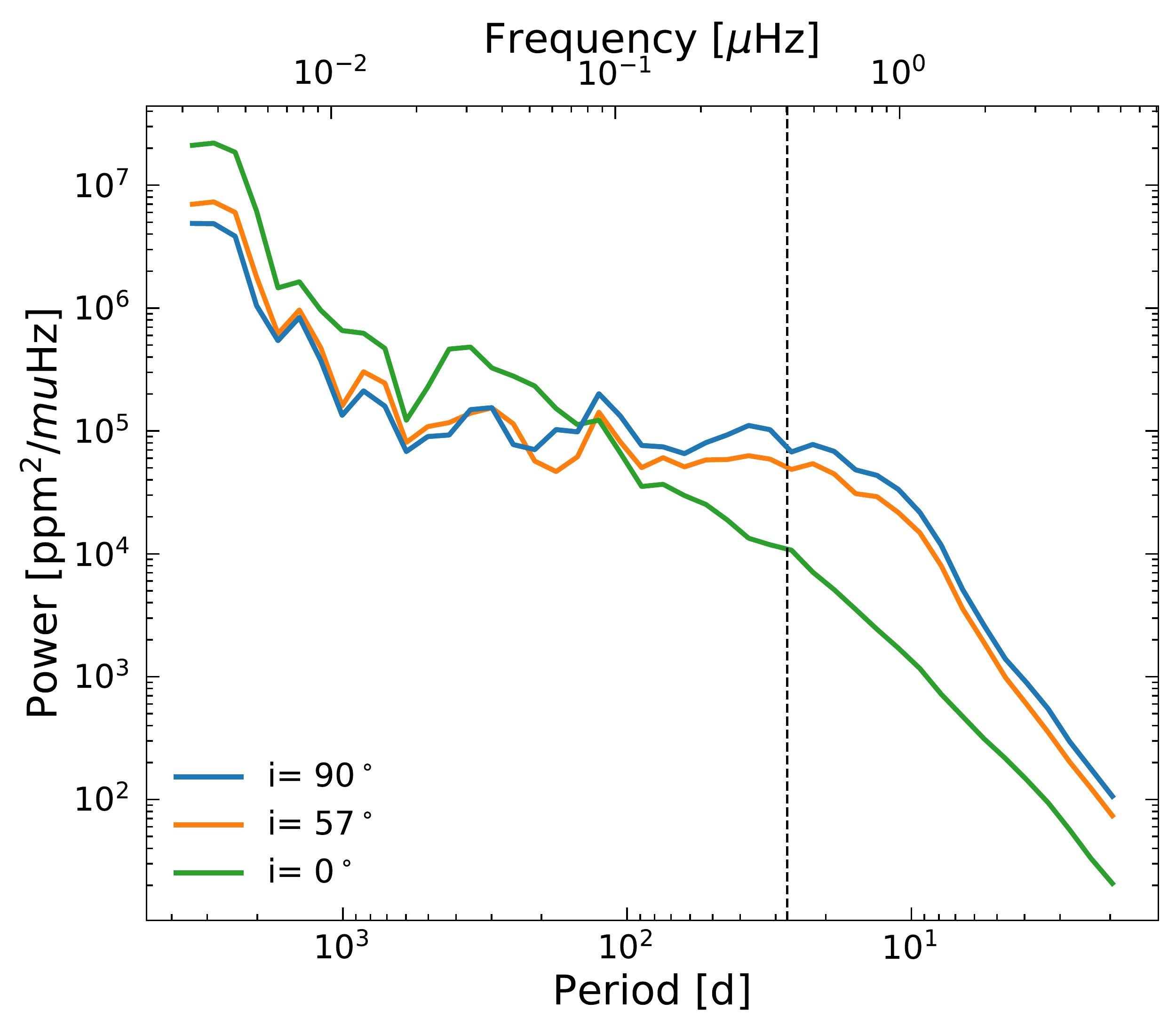}
\caption{Power spectra of the solar brightness variations for Strömgren \textit{y} for different inclinations.} 
\label{fig:PS_Str_y}
\end{figure}

\chapter{Obtaining the spot filling factors used in Sect. \ref{sec:paper_4}}\label{model_details}

\begin{figure}
\centering
\includegraphics[width=0.48\columnwidth]{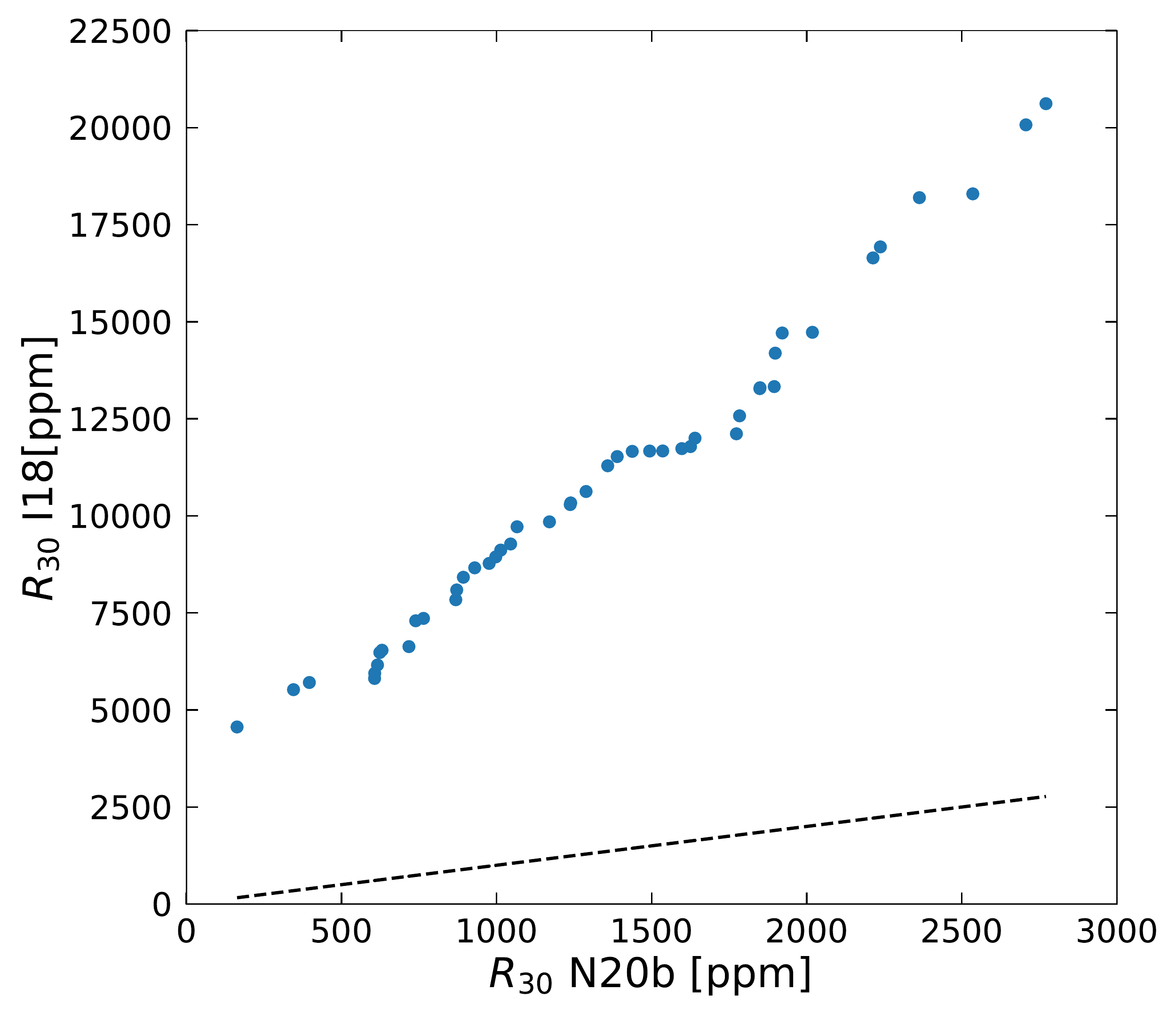}
\includegraphics[width=0.48\columnwidth]{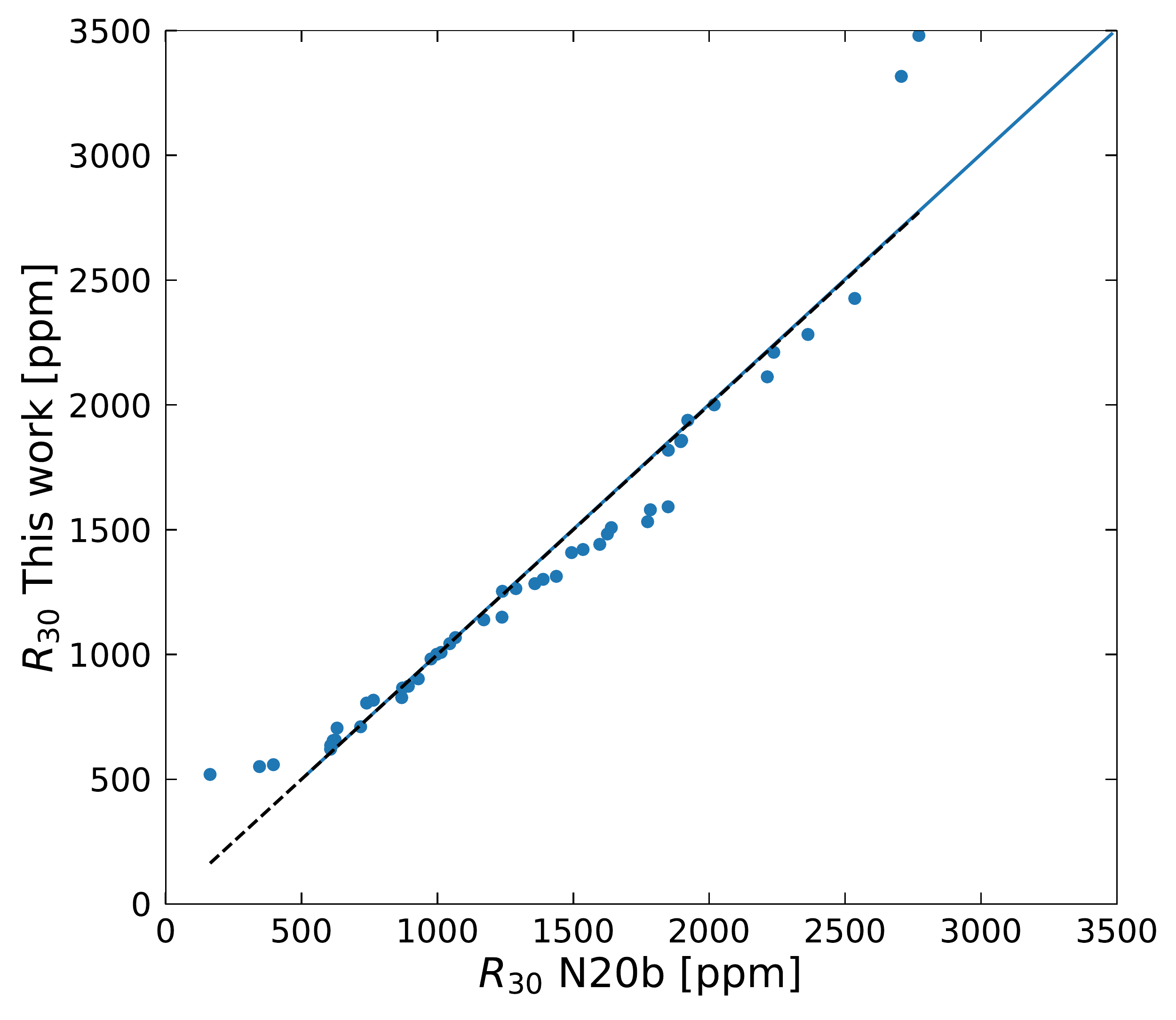}
\caption{Comparison between solar rotational variability represented via $R_{30}$ for different models. Left panel: Comparison between the \citetalias{Nina1} model and the approach of defining a single threshold for the spot filling factors as done by \citetalias{Isik2018}. Right panel: comparison between the \citetalias{Nina1} model and the two-threshold approach presented in this work. The black solid lines in both panels represents a one-to-one correspondence between the models, the blue solid line is the regression between the two models.}
\label{fig:params}
\end{figure}

\begin{figure}
\centering
\includegraphics[width=\columnwidth]{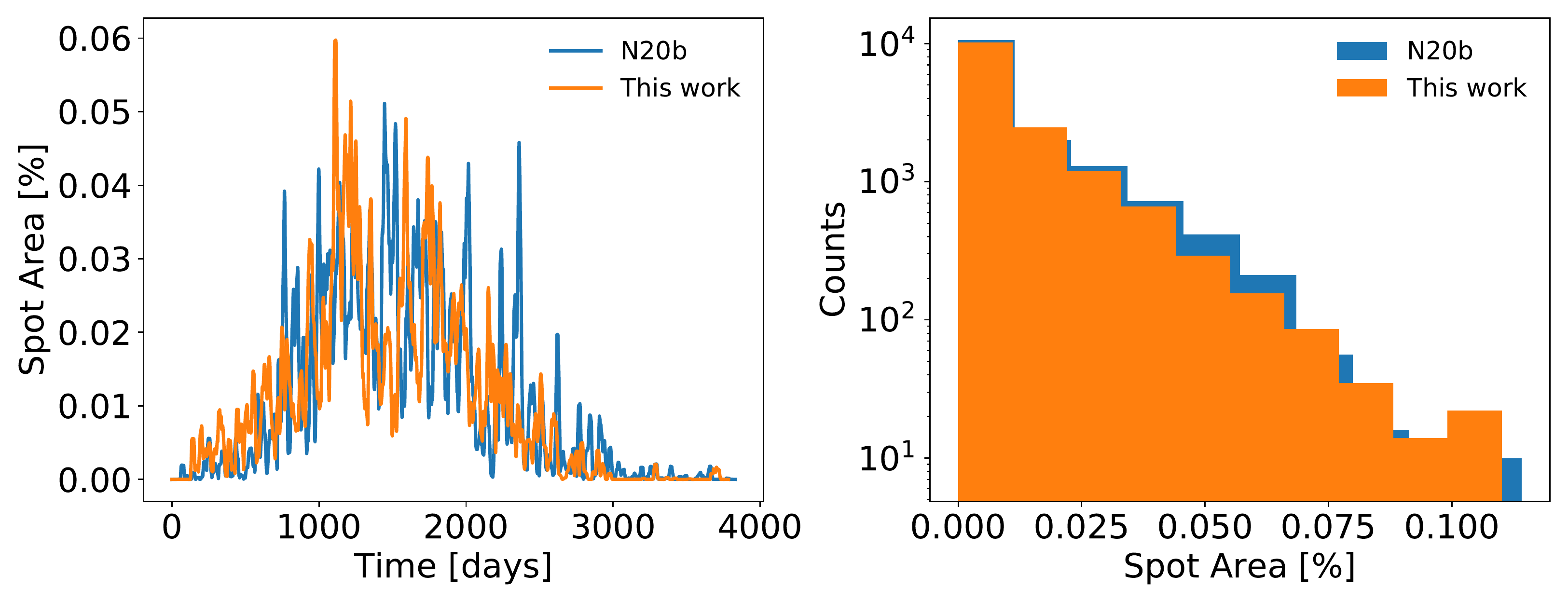}
\caption{Comparison between the spot disc areas of the \citetalias{Nina1} model and the approach outlined in this work. Left: 30-day running mean of the models, right: histograms of the spot area distribution.} 
\label{fig:spots}
\end{figure}

\begin{figure}[!h]
\centering
\includegraphics[width=0.8\textwidth]{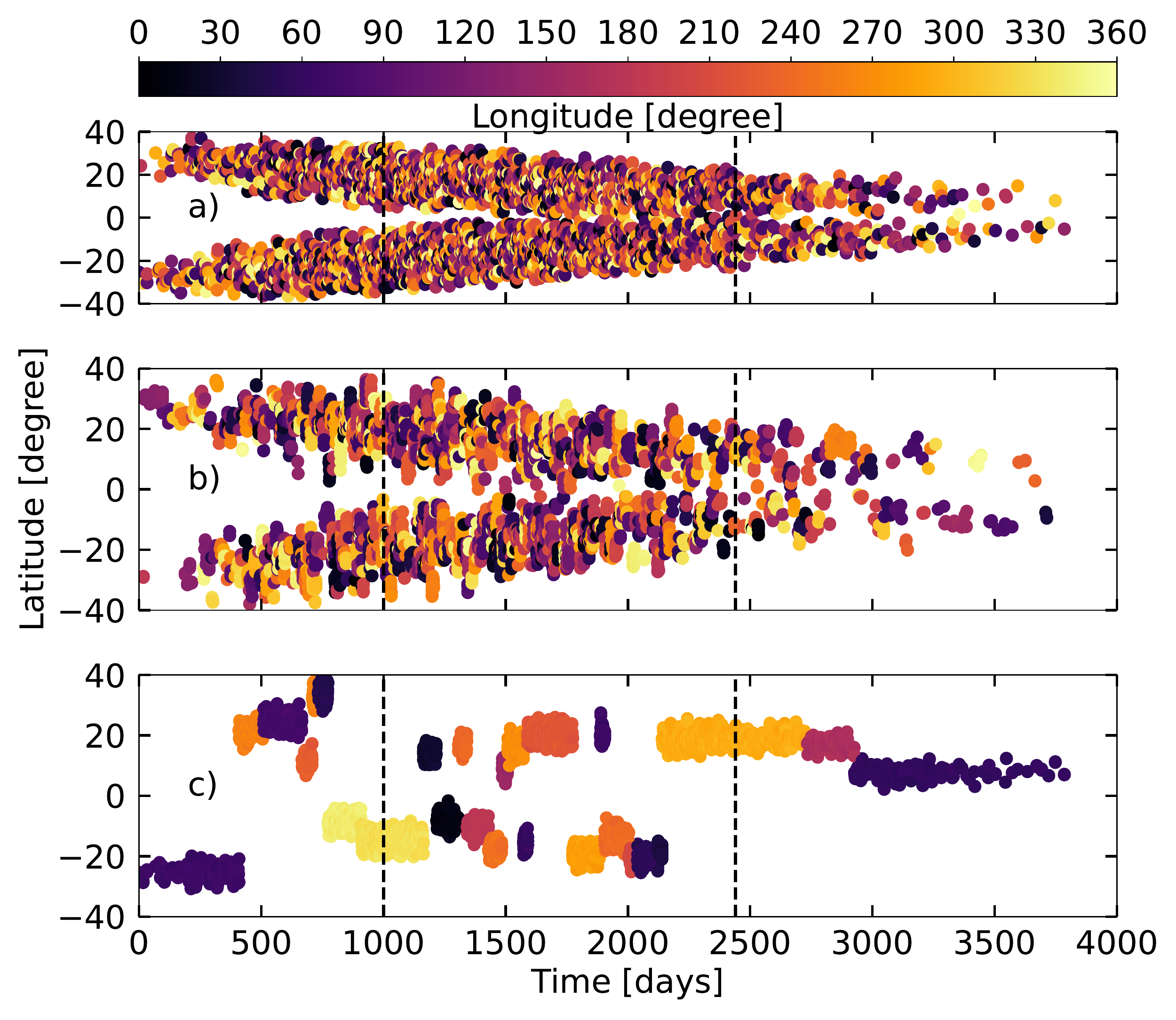}
\caption{Time-dependent emergence of bipolar magnetic regions (BMRs) in the surface flux transport model (SFTM) as a function of latitude for a star with 1\rotrate{} and different degrees of nesting, as expressed through $p$. Panel a) $p=0$,b) $p=0.7$, and c) $p=0.99$. The colours of the points represent the longitudinal position of the BMRs on the stellar surface. The vertical dashed black lines indicate the time-span considered for the brightness calculations in this work.}
\label{fig:butterfly}
\end{figure}

Attributing the magnetic flux from the SFTM to spot filling factors (as done in Sect. \ref{sec:paper_4}) has one major obstacle and that is the polar field that is present in the SFTM. An initial polar field is needed to compensate for the flux being transported towards the poles due to the meridional circulation. For the Sun, this polar field is about 10 G, which is consistent with observations. As we move to model more active stars, the polar field gets stronger to compensate for the larger amount of flux emerging on the surface of those stars \citep[see][]{Isik2018}. In order to not attribute all of the polar field to spots, as we do not observe polar spots on the Sun, we therefore set a lower threshold, $B_{min}$. We also define an upper threshold, $B_{max}$, that acts as a saturation threshold.
We varied $B_{min}$ and $B_{max}$ to find a set that matches the the rotational variability, $R_{30}$, caused by the spots of the \citetalias{Nina1} model. To calculate $R_{30}$, the obtained light curves are split into 30-day segments and within each segment, we calculate the difference between the extrema, and divide this value by the mean flux in the segment to get the relative variability. We directly consider the difference between the extrema instead of the differences between the 95th and 5th percentiles of sorted flux values, as is usually done in the literature with the more noisy \textit{Kepler} measurements  \citep[see e.g.][]{Basri2013}.  Additionally, we also only consider four years (corresponding to the time-scale of the \textit{Kepler} mission) during solar cycle maximum. The considered time-span is indicated by vertical dashed black lines in Fig. \ref{fig:butterfly}. We find that $B_{min} = 60$G and $B_{max}=700$G result in the best fit. We show a comparison between the models in Fig. \ref{fig:params} panel b). The slope of the linear regression (blue line) is 1.00196, with an intercept of -0.987 ppm and the Pearson-correlation coefficient is high (0.97). The mean rotational variability in the present model is comparable to that in the \citetalias{Nina1} model (1342.96 to 1341.31 ppm). We can also additionally compared the spot disc areas of the two models. Both yield a cycle averaged spot filling factors of 0.1\%, which is consistent with observations. We also show a direct comparison of the spot areas between the models in Fig. \ref{fig:spots}. The left panel shows the 30-day smoothed average of the spot filling factors, the right panel the histogram of the area distribution. We stress, that the daily and even the smoothed values are expected to be different between the \citetalias{Nina1} and the \citetalias{Isik2018} SFTM, as they used different source terms to compile the active region input record that is fed into the SFTM. However, as we have shown, both models return the same level of variability.

\chapter{Butterfly diagrams}\label{butterfly}

\begin{figure}[!h]
\centering
\includegraphics[width=0.8\textwidth]{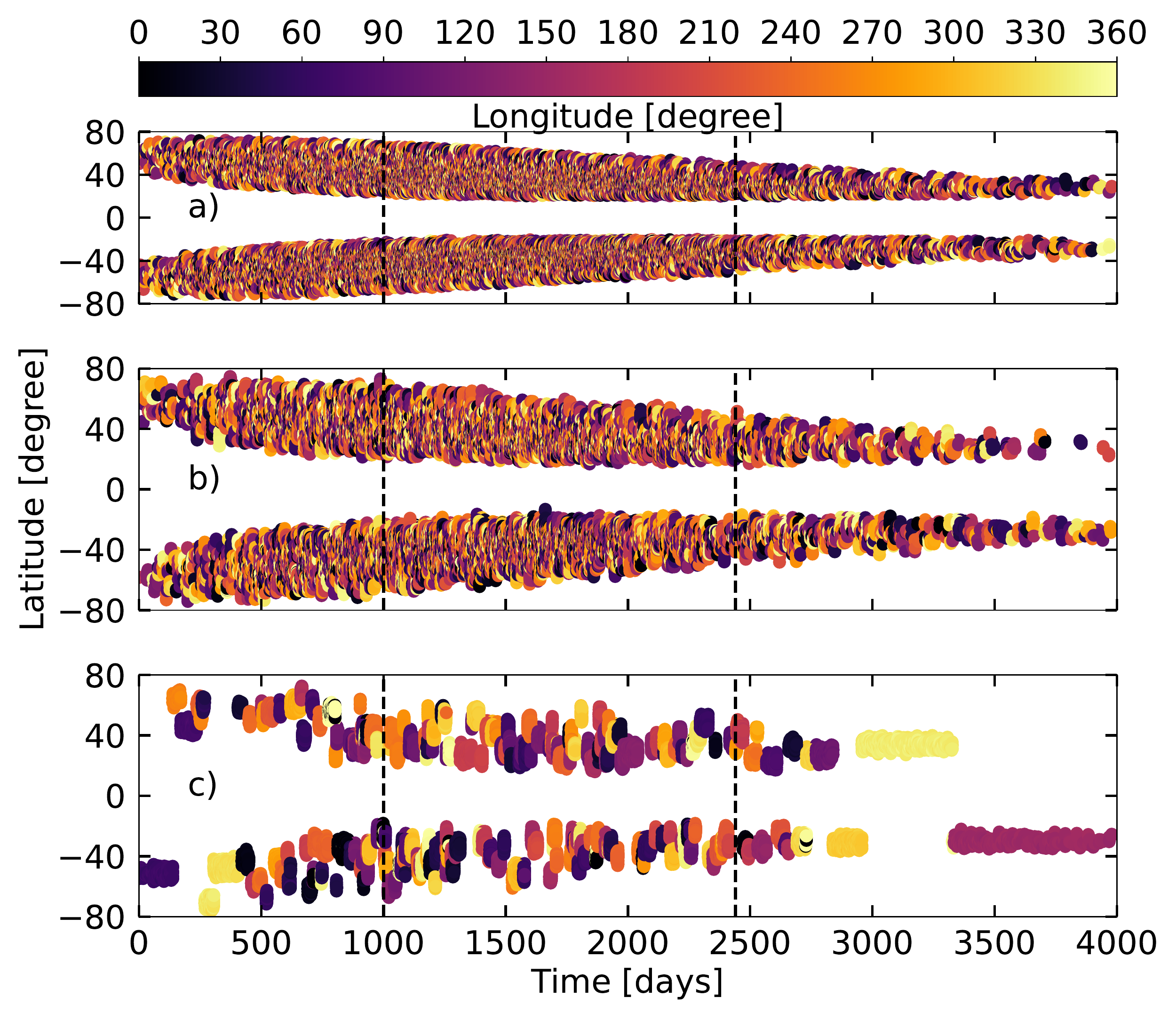}
\caption{Time-dependent emergence of bipolar magnetic regions (BMRs) in the surface flux transport model (SFTM) as a function of latitude for a star with 8\rotrate{} and different degrees of nesting, as expressed through $p$. Panel a) $p=0$,b) $p=0.7$, and c) $p=0.99$. The colours of the points represent the longitudinal position of the BMRs on the stellar surface. The vertical dashed black lines indicate the time-span considered for the brightness calculations in this work.}
\label{fig:butterfly2}
\end{figure}

\chapter{Light curves of the full 4--years}\label{full_LC}

\begin{figure}
\centering
\includegraphics[width=\textwidth]{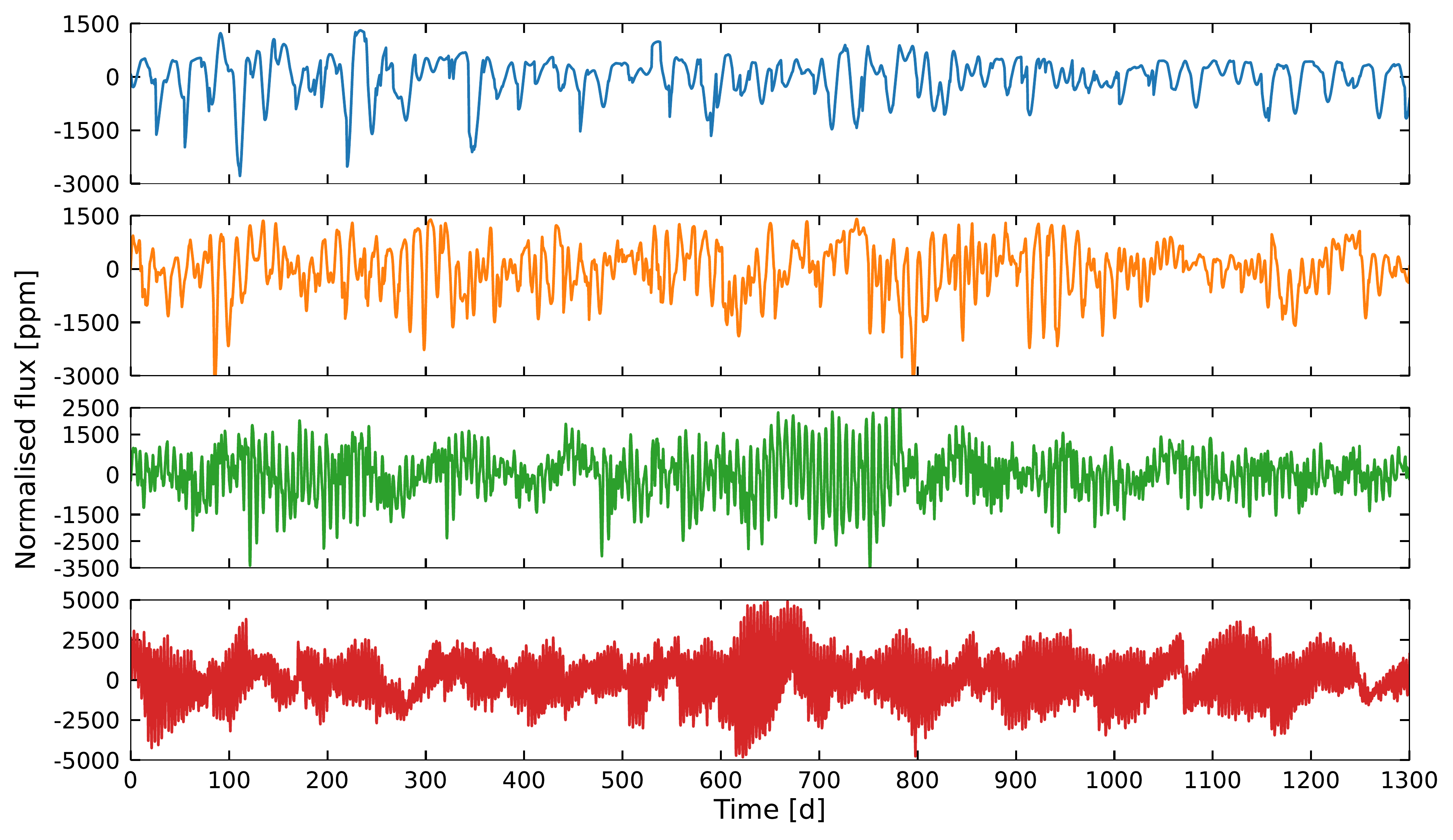}
\caption{Synthetic stellar light curves (LCs) for stars with different rotation rates as they would be observed in the \textit{Kepler} passband at an inclination of $i=90\degree$. Shown are rotation rates values of 1\rotrate{} (blue),  2\rotrate{} (orange), 4\rotrate{} (green), and  8\rotrate{} (red).  The degree of nesting is set to 0\%. Plotted are the 4--years timespan of 1000-2440 days as indicated by the vertical black dashed lines in Fig. \ref{fig:butterfly}.}
\label{fig:LCs_eq1}
\end{figure}

\begin{figure}
\centering
\includegraphics[width=\textwidth]{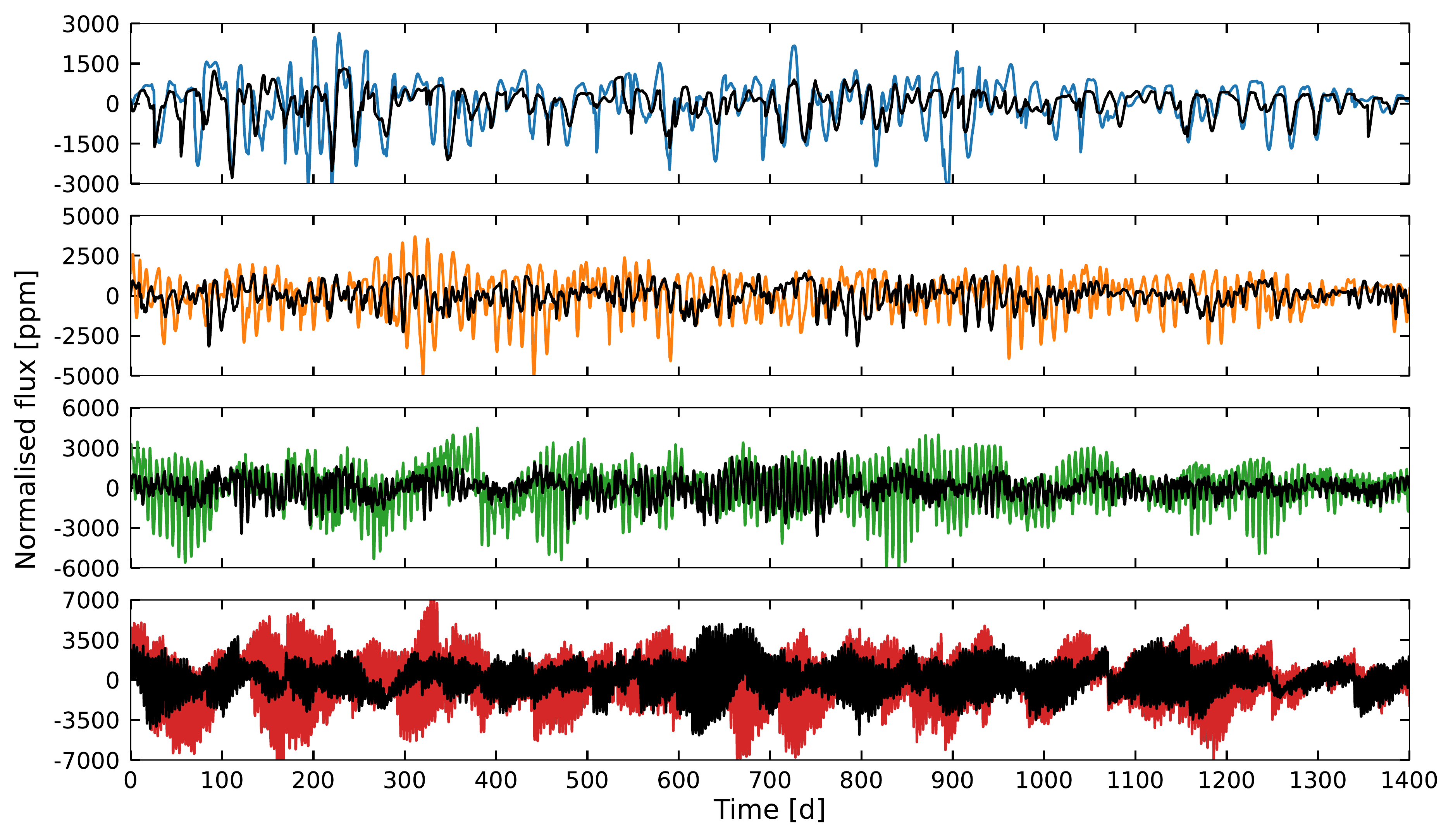}
\caption{The same as in Fig. \ref{fig:LCs_eq1} but the light curves are calculated with a nesting degree of 70\%. The light curves from Fig. \ref{fig:LCs_eq1}  are shown as black lines for comparison. Plotted are the 4--years timespan of 1000-2440 days as indicated by the vertical black dashed lines in Fig. \ref{fig:butterfly}.} 
\label{fig:LCs_eq2}
\end{figure}

\begin{figure}
\centering
\includegraphics[width=\textwidth]{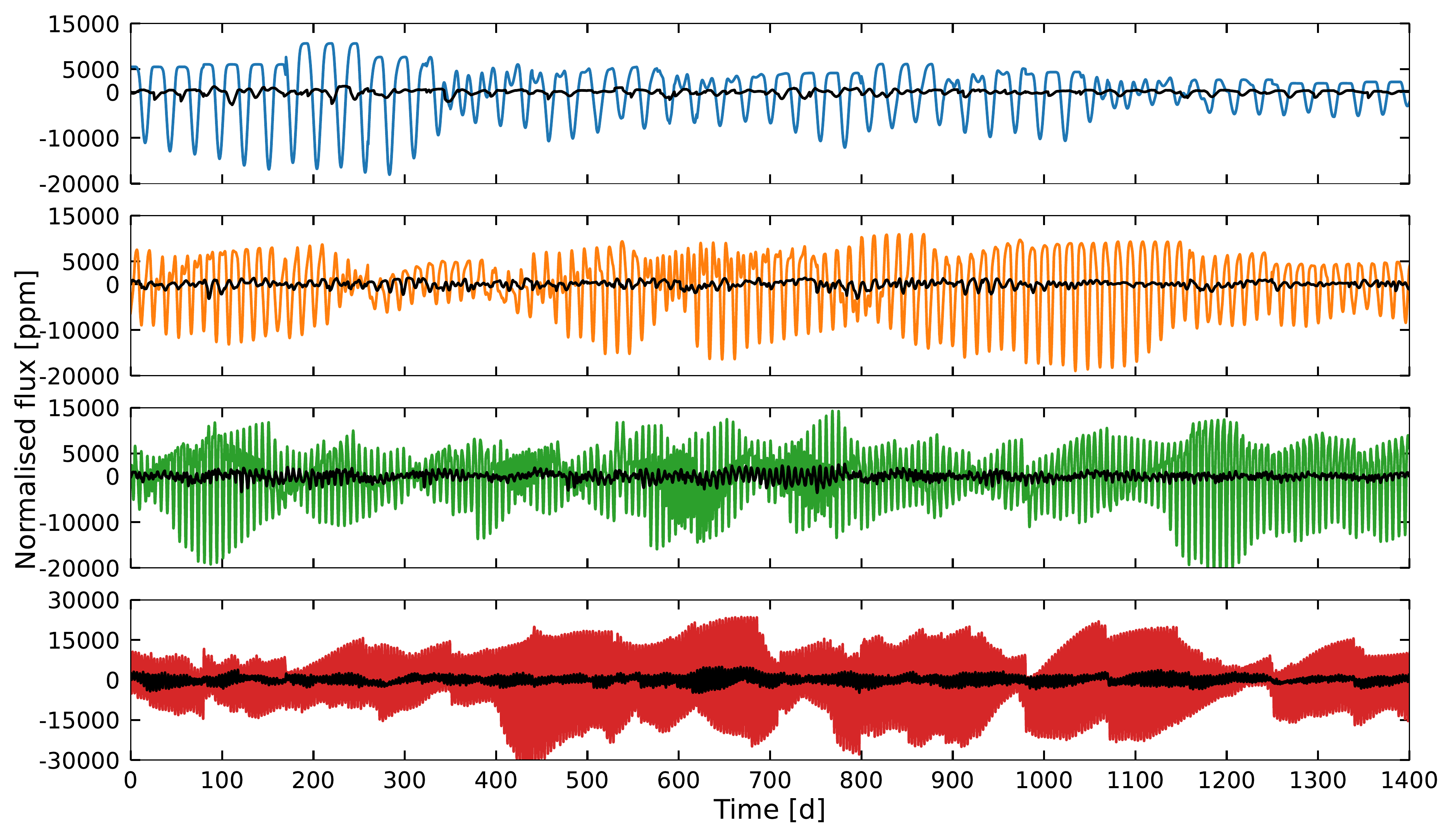}
\caption{The same as in Fig. \ref{fig:LCs_eq1} but the light curves are calculated with a nesting degree of 99\%. The light curves from Fig. \ref{fig:LCs_eq1}  are shown as black lines for comparison. Plotted are the 4--years timespan of 1000-2440 days as indicated by the vertical black dashed lines in Fig. \ref{fig:butterfly}.} 
\label{fig:LCs_eq3}
\end{figure}

\begin{figure}
\centering
\includegraphics[width=\textwidth]{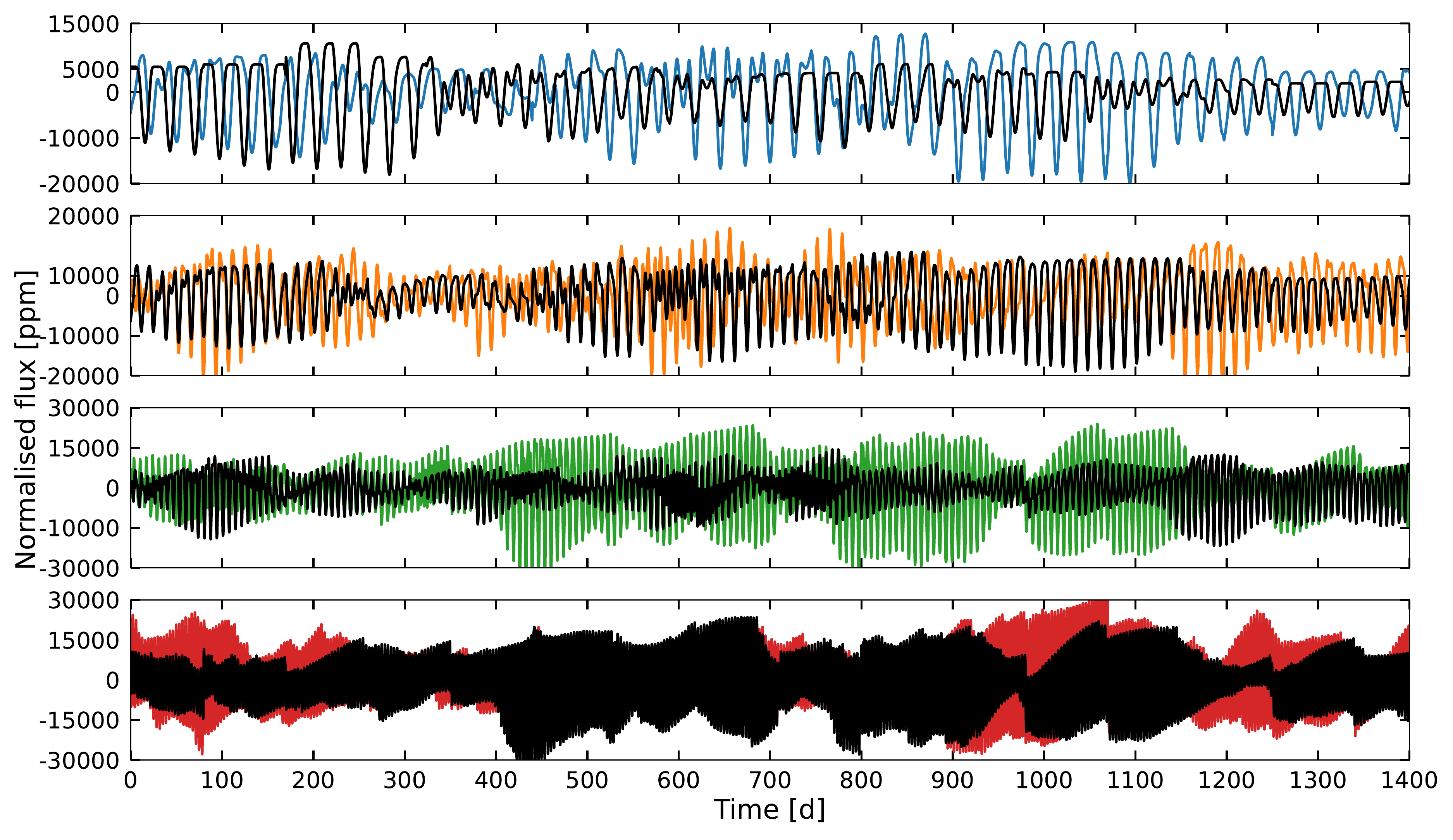}
\caption{The same as in Fig. \ref{fig:LCs_eq1} but the light curves are calculated with a nesting degree of 99\% and an activity scaling of $\tilde{s} = 2\cdot\tilde{\omega}$. The light curves from Fig. \ref{fig:LCs_eq1}  are shown as black lines for comparison. Plotted are the 4--years timespan of 1000-2440 days as indicated by the vertical black dashed lines in Fig. \ref{fig:butterfly}.} 
\label{fig:LC_eq_act}
\end{figure}

\chapter*{Publications\markboth{Publications}{Publications}}
\addcontentsline{toc}{chapter}{Publications}

{\bf\large Refereed publications}\\
\begin{itemize}
\item \textbf{N.-E. N\`{e}mec}, E. I\c{s}{\i}k, A. I. Shapiro, S. K. Solanki, N. A. Krivova, and Y. Unruh, "Connection measurements of solar and stellar brightness variations", 2020, A\&A, 638, A56
\item \textbf{N.-E. N\`{e}mec}, A. I. Shapiro, N. A. Krivova,
R.V. Tagirov, R. H. Cameron, S. K. Solanki and S. Dreizler, "Power spectra of solar brightness variations at different inclinations",  2020, A\&A, 636, A43
\item T. Reinhold, A. I. Shapiro,  V. Witzke, \textbf{N.-E. N\`{e}mec}, E. I\c{s}{\i}k, and  S.~K.~Solanki, {"Where have all the solar-like stars gone? Rotation period detectability of solar-like stars at various inclinations and metallicities"}, ApJ Letters, 908, 2, id.L21
\item D. Shulyak, L. M. Lara, M. Rengel, and \textbf{N.-E. N\`{e}mec}, "Stellar impact on disequilibrium chemistry and observed properties of hot Jupiter atmospheres", 2020, A\&A, 639, A48
\end{itemize}

{\bf\large Accepted publications}\\
\begin{itemize}
\item K.~Sowmya ,A. I. Shapiro, V. Witzke, \textbf{N.-E. N\`{e}mec}, T.~Chatzistergos, N.~A.~Krivova, and S.~K.~Solanki, {"Modeling stellar \mbox{Ca\,{\sc II} H \& K} emission variations. I. Effect of inclination on the S-index"}, accepted in ApJ
\end{itemize}

%{\bf\large Conference proceedings}\\
%\begin{itemize}
%\item Your 
%\item contributions
%\item to converence proceedings
%\end{itemize}

\chapter*{Acknowledgements\markboth{Acknowledgements}{Acknowledgements}}
\addcontentsline{toc}{chapter}{Acknowledgements}

First and foremost, I want to thank \foreignlanguage{russian}{Sasha}. I could not have been luckier to have you as my supervisor. When I was struggling, whether that had to do with my PhD life or otherwise, your door was always open and your advice has helped me a lot. With you, I was able to grow both as a young researcher, but also as a person. You are a good example as a group leader and supervisor, as you always treated your group members with respect and showed compassion, when necessary. Natalie, your kindness and openness have helped me as well to develop so much in this last three years. Thank you as well. Sami, I am thankful, for your ideas and inputs into the work and for sometimes being the voice of reason if Sasha and I had different ideas, 

I also want to thank Stefan for being in my TAC and for your suggestions to improve my work further and agreeing to be in my examination committee. I thank the other committee members, Prof. Dr. Wolfram Kollatschny, Prof. Dr. Ariane Frey and Hardi as well.

A big thank you to everyone in the SOLVe group, especially to Sowmya, Timo and Veronika. I enjoyed our discussions and the cakes. Emre, as an honorary group member, I also want to thank you for our collaboration and I am looking forward to work with you further in the future.

I want to thank Sonja Schuh, our IMPRS coordinator. You were always there to help to navigate administrative regulations and were extremely helpful during my tumultuous time as student representative.

What makes this particular IMPRS school so welcoming and enjoyable are its students and a couple of those deserve to be mentioned.
Juxhin, you were my coffee buddy for a majority of our time together at MPS. You never failed to make me laugh, whether that was intentional, or often rather unintentional. Cosima and Hanna (although you are technically not an IMPRS student, but I see you almost as an honorary member), I want to thank you for being my "boulder buddies". I enjoyed bouldering already a lot, but you made it even better. Philipp and Bernhard, you also deserve to be mentioned, as you helped me to fly the Austrian flag high, much to the dismay of the German students around us. Kok Leng, you also deserve to be mentioned here. You are still a student at heart and I am very thankful for all of our heartfelt and silly discussions.

Worte k\"onnen nicht beschreiben, wie dankbar ich meinen Eltern, Michaela und Franz, und meiner Schwester Sarah bin. Ihr habt mit mir zahlreiche kalte N\"achte im Garten mit "Stendarl schauen" verbracht und habt mir zugeh\"ort, als ich von fernen Planeten und Galaxien fantasierte. Ihr habt mich immer auf meinem Weg unterst\"utzt, obwohl ihr wusstest, dass er mich von zu Hause wegf\"uhren w\"urde. Ohne euch w\"are ich nie soweit gekommen. 

This work has been funded by the European Research Council under the European Union's Horizon 2020 research and innovation program (grant agreement No. 715947).

\chapter*{Curriculum vitae\markboth{Curriculum vitae}{Curriculum vitae}}
\addcontentsline{toc}{chapter}{Curriculum vitae}

\textbf{Personal data} \\
Name: Nina-Elisabeth N\`{e}mec \\
Day of birth: 16. October 1993 \\
Place of birth: Amstetten, Austria \\

\noindent \textbf{Education}
\begin{itemize}
    \item Sept. 2017-January 2021 PhD \\
    International Max Planck Research School (IMPRS) at the Max Planck Institute for Solar System Research (MPS) and the University of G\"ottingen under the supervision of Dr. Alexander I. Shapiro in the SOLVe group
    
    \item 2015-2017 MSc in Astronomy\\
    University of Vienna, Vienna, Austria, Thesis title: \textit{The XUV Sun in time} under the supervision of Prof. Dr. Manuel G\"udel
    
    \item 2012-2015 BSc in Astronomy\\
    University of Vienna, Vienna, Austria, Thesis title: \textit{The solar wind in time} under the supervision of Prof. Dr. Manuel G\"udel.
\end{itemize}

\end{document}